\documentclass[aps,prd,a4paper,amsfonts,amssymb,nofootinbib,onecolumn,longbibliography]{revtex4-2}
\usepackage[utf8]{inputenc}
\usepackage[a4paper,
            bindingoffset=0.2in,
            left=1.5cm,
            right=1.5cm,
            top=1in,
            bottom=1.5cm,
            footskip=.25in]{geometry}
\usepackage{amsmath,stackrel}
\usepackage{hyperref}
\hypersetup{
    colorlinks,
    citecolor=black,
    filecolor=black,
    linkcolor=black,
    urlcolor=black
}

\usepackage[parfill]{parskip}  
\usepackage{graphicx}
\usepackage{ulem}
\usepackage{amssymb}
\usepackage{braket}
\usepackage{epstopdf}
\usepackage{mathtools}
\usepackage{subfig}
\usepackage{tgtermes}
\usepackage[toc,page]{appendix}
\usepackage{cases}
\usepackage{tikz} 
\usetikzlibrary{arrows} 
\usetikzlibrary{shapes,arrows,positioning,automata,backgrounds,calc,er,patterns}
\usepackage{tikz-feynman}
\usepackage{empheq}
\usepackage{comment}
\usepackage{stackrel}
\usepackage{slashed}
\usepackage{changepage}
\usepackage{soul} 
\tikzset{
    ->-/.style={decoration={
  markings,
  mark=at position .5 with {\arrow{>}}},postaction={decorate}},
    -<-/.style={decoration={
  markings,
  mark=at position .5 with {\arrow{<}}},postaction={decorate}},
    ->/.style={decoration={
  markings,
  mark=at position .4 with {\arrow{>}}},postaction={decorate}},
}
\newcommand{\op}{\mathcal{O}}

\newcommand{\mF}{\mathcal{F}}
\newcommand{\mL}{\mathcal{L}}
\newcommand{\mO}{\mathcal{O}}

\renewcommand{\arraystretch}{1.5} 

\usepackage{color}

\definecolor{mygreen}{RGB}{0,128,0}

\begin{document}

\title{The flair of Higgsflare:\\ Distinguishing electroweak EFTs with   $W_LW_L \to n\times h$}

\author{Raquel G\'omez-Ambrosio, }
\affiliation{ Dipartimento di Fisica “G. Occhialini”, Universit\`a degli Studi di Milano-Bicocca, and
INFN, Sezione di Milano Bicocca, Piazza della Scienza 3, I – 20126 Milano, Italy
}

\author{Felipe J. Llanes-Estrada, Alexandre Salas-Bern\'ardez and Juan J. Sanz-Cillero}
\affiliation{
Univ. Complutense de Madrid, Dept. Física Teórica and IPARCOS, Plaza de las Ciencias 1, 28040 Madrid, Spain
}

\date{\today}

\begin{abstract}
 The electroweak symmetry-breaking sector is one of the most promising and uncharted parts of the Standard Model; but it seems likely that new electroweak physics may be out of reach of the present accelerator effort and the hope is to observe small deviations from the SM. Given that, Effective Field Theory becomes the logic method to use, and SMEFT has become the standard. However, the most general theory with the known particle content is HEFT, and whether SMEFT suffices should be investigated in future experimental efforts. Building on investigations by other groups that established geometric criteria to distinguish SMEFT from HEFT (useful for theorists examining specific beyond-SM completions), we seek more phenomenological understanding and present an analogous discussion aimed at a broader audience.  
 
We discuss various aspects of (multi-) Higgs boson production from longitudinal electroweak gauge bosons $W_LW_L\to n\times h$ in the TeV region as the necessary information to characterise the Flare function, $\mathcal{F}(h)$, that determines whether SMEFT or HEFT is needed.  We also present tree-level amplitudes including contact and exchange channels, as well as a short discussion on accessing $\mF$ from the statistical limit of many bosons. 
We also discuss the status of the coefficients of the series expansion of $\mathcal{F}(h)$, its validity, whether its complex-$h$ extension can be used to predict or not a tell-tale zero, and how they relate to the dimension-6 and -8 SMEFT operators in the electroweak sector. We derive a set of new correlations among BSM corrections to the HEFT coefficients that help decide, from experimental data, whether we have a viable SMEFT.
This analysis can be useful for machines beyond the LHC that could address the challenging final state with several Higgs bosons. \\

\end{abstract}

\maketitle

\tableofcontents
\setlength{\parindent}{0.3cm}
\newpage 
\section{Introduction}

Two effective theories (EFTs) have come to the fore in trying to extend the successful Standard Model in the ignorance of which new physics may be present at a higher energy (if there is any).
There are many aspects of the Standard Model that can be pursued at accelerators (the many-parameter flavor structure in both lepton and quark sectors, the Higgs couplings to the fermions, the CP violating phases, QCD processes...) but at the energy frontier the most important aspects of physics that is being clarified right now is the nature of the mechanism of Electroweak (EW) Symmetry Breaking: whether it happens as the well-known discussion in the Standard Model paradigm, or whether new particles or interactions influence the global $SU(2)\times SU(2)\to SU(2)$ breaking pattern that is at the crux of electroweak interactions.
Several extended works have dealt with many of these aspects, for example~\cite{Mariotti:2016owy,deBlas:2019rxi,DiMicco:2019ngk,Dawson:2018dcd,Passarino:2013bha,Chiesa:2020awd}.

Throughout most of this paper we will discuss these theories in a regime where the energies of the scattered particles are much higher than their masses $m_h\ll E\ll \Lambda $. 
A bit surprisingly perhaps for Standard Model practitioners, in such regime and in the presence of new physics that would yield derivative couplings with $\partial \sim E\gg m_h$, the much discussed Higgs potential $V(h)$ is actually a correction that does not play the pivotal role
it enjoys in the SM.

 For a while, it was often stated that the Standard Model Effective Field Theory (SMEFT)~\cite{Jenkins:2013zja,Alonso:2013hga,Brivio:2017vri,Ghezzi:2015vva,Alioli:2022fng,Marzocca:2020jze} and the Higgs Effective Field Theory (HEFT)~\cite{Brivio:2013pma,Eboli:2021unw,Kozow:2019txg,Dobado:2019fxe,LHCHiggsCrossSectionWorkingGroup:2016ypw,Carena:2021onl} must encode similar physics, just in different coordinates, with HEFT being perhaps a bit more general because it does not incorporate the Higgs boson $h$ into an $SU(2)$ multiplet.

 However, the work of the San Diego and Oregon
 groups~\cite{Alonso:2016oah,Alonso:2016btr,Alonso:2015fsp,Cohen:2020xca} has sharpened the differences between both formulations. Every Lagrangian in the shape in SMEFT form, such as that in Eq.~(\ref{SMEFTEWlagrangianV}) below, can be recast in HEFT form. An example in the relevant energy region is given in  Eq.~(\ref{FbosonLagrangianLO}). 
 
 The converse statement need not always  be true. From the work at San Diego it has become apparent that this can be achieved only if a certain function $\mathcal{F}(h)$ presented shortly in Eq.~(\ref{expandF}) has a zero for some real value of the classical field $h$, $\exists\ h_\ast\in\mathbb{R} \arrowvert \mathcal{F}(h_\ast)=0$. (The precise and complete conditions as presently understood for this SMEFT $\leftrightarrow$ HEFT equivalence are presented in Section \ref{HEFTinSMEFTsection}). This function controls high-energy processes with multiple Higgs bosons in the final state: achieving a good control over it requires measurements with an increasingly large number of them, that would look in a detector like a flare of Higgs bosons, whence the name of $\mF$.
 
 However, this is quite an obscure geometric requirement removed from accelerator physics. In recent work~\cite{Cohen:2021ucp}, an attempt has been made to bring up a phenomenological connection between an eventual breaking of unitarity at some scale and the Lagrangian to be chosen. Aware that such unitarity failure is a feature of perturbation theory in Taylor-series form, and that it does not occur if instead one expands the inverse partial-wave scattering amplitude~\cite{Dobado:1989gr}, so that no information distinguishing the effective Lagrangians can really be gained that easily,  we here continue exploring what relevant phenomenology there is.
 
 A summary of our present understanding is given by the following scheme:

 \begin{center}\boxed{
 \begin{tabular}{ccccc}
       &          &                    &          & Specific correlations\\
 Valid &          & Double zero        &          & between $a_i$ \\ 
 SMEFT &$\implies$& of $\mathcal{F}(h)$&$\implies$& coefficients \\
       &          & at some $h_\ast$   &          & from expanding $\mathcal{F}(h)$\\
       &          &                    &          & around $h=0$
 \end{tabular}}
 \end{center}

Both aspects, the double zero of $\mathcal{F}(h)$ (see subsection~\ref{subsec:noSMEFT} below, for example) and the possibility of using correlations between $a_i$ coefficients to distinguish SMEFT from HEFT from experimental data (see subsection~\ref{subsec:correlations}) have been separately discussed in the last years. We here provide an integrated discussion with full detail, putting less weight on geometric aspects and more in field-theory and particle physics ones, more familiar to the typical reader, and make several new contributions.  
Finally, it is also worth commenting that similar relations exist between the coefficients of the nonderivative $V_{\rm HEFT}$
potential, since a valid SMEFT description of $V$ must obey a series of conditions analogous to those for the flare function  $\mF(h)$~\cite{Cohen:2020xca}. Assuming SMEFT's validity would then also impose important correlations between the coefficients of the potential --trilinear, quartic, etc.-- (see e.g.~\cite{Gonzalez-Lopez:2020lpd} for a HEFT phenomenological analysis). This is mostly beyond the scope of this article and will certainly be studied in future work, but we present a slim discussion in Appendix~\ref{coeffspotential}. 

\newpage
\subsection{Key aspects of the SMEFT Lagrangian}

The first of those theories is the Standard Model EFT (SMEFT) electroweak Lagrangian. Its symmetry breaking sector is expressed in terms of the $SU(2)$ doublet $H=\frac{1}{\sqrt{2}}\begin{pmatrix} \phi_1+i\phi_2 \\ \phi_4+i\phi_3 \end{pmatrix}$ (that can also be collected as an $O(4)$ quartet  $\boldsymbol\phi=(\phi_1,\phi_2,\phi_3, \phi_4)$) and takes the general form
 \begin{equation}
     \mathcal{L}_{\rm SMEFT}= A(|H|^2)|\partial H|^2+\frac{1}{2}B(|H|^2)(\partial (|H|^2))^2-V(|H|^2)+\mathcal{O}(\partial ^4) 
     \label{SMEFTEWlagrangianV} \ .
 \end{equation}
Where  both functions $A(|H|^2)$ and $B(|H|^2)$ are real, and analytic around $|H|^2=0$. The SM is retrieved by choosing $A(|H|^2)=1$ and $B(|H|^2)=0$. 
 Thus, the organization of the theory is carried around an electroweak symmetric vacuum, instead of the energy minimum at $\langle \phi_4\rangle = v$.
  
SMEFT arranges the order of usage of operators in terms of their canonical-dimension counting, so that the leading corrections to the SM are composed of dimension six operators (each multiplied by a Wilson coefficient and divided by the new physics scale squared, $\Lambda^2$).

In the TeV region, the derivative terms multiplying $B$ become much larger than $V$, and we can neglect this potential. It will be shown that the piece of most importance for this article is that function $B$, that contains the electroweak symmetry breaking physics in the TeV region in the presence of new physics, particularly the dimension-6 operator 
 $\mathcal{O}_{H\Box} =   |H|^2 \Box |H|^2 $.

\subsection{ HEFT Lagrangian\\ (for the Electroweak Symmetry Breaking Sector in the TeV region)}
What has come to be called the Higgs EFT (HEFT) Lagrangian (the second to bear that name) is an evolution of the Electroweak chiral Lagrangian. Its degrees of freedom are built from a Cartesian to spherical-like change of coordinates %
\begin{equation}\label{eq:EWNGB-spherical-coord}
\boldsymbol\phi=(1+h/v)\boldsymbol n  \, ,
\end{equation}
where $\boldsymbol n=(\omega_1,\omega_2,\omega_3,\sqrt{v^2-\omega_1^2-\omega_2^2+\omega_3^2})$, so that $\boldsymbol n\cdot\boldsymbol n=v^2$. It couples the $\omega_i$ Goldstone bosons to an additional low-energy Higgs field singlet, $h$, that is not assumed to be part of the $SU(2)$ 
Goldstone triplet. 
At leading order in the chiral counting, the scalar sector of the HEFT Lagrangian  (in EW Goldstone spherical coodinates $\omega^i$ in~Eq.(\ref{eq:EWNGB-spherical-coord})) is given by  
 \begin{eqnarray} \label{FbosonLagrangianLO}
{\cal L}_{\text{LO HEFT}} &=& \frac{1}{2}\mathcal{F}(h)
\partial_\mu\omega^i\partial^\mu\omega^j\left(\delta_{ij}+\frac{\omega^i\omega^j}{v^2-\boldsymbol{\omega}^2}\right)
+\frac{1}{2}\partial_\mu h\partial^\mu h  \, ,
\end{eqnarray}
where the function $\mathcal{F}$ scales the scattering amplitudes involving two, four, and generically an even number of Goldstone bosons.  
Thus, the flare function $\mF(h)$ relates the EW Goldstone processes to amplitudes with an arbitrary  number of Higgs bosons: 
of the same order in the chiral counting appropriate for HEFT,
\begin{equation} \label{expandF}
    {\mathcal F}(h)=1+\sum_{n=1}^{\infty}{a_n}\Big(\frac{h}{v}\Big)^n\;.
\end{equation}
Usually, since only the first terms of the $\mathcal{F}$ function are known, the Lagrangian is expressed~\cite{Delgado:2015kxa,Delgado:2013hxa} in terms of $\mathcal{F}(h)\simeq \left[1+2a\frac{h}{v}+b\left(\frac{h}{v}\right)^2\right]$, with 
$a_1=2a$ and $ a_2=b\;.$
In the TeV regime, the leading corrections to this Lagrangian are not of order $m_W$ or $m_h$, both in the $100$ GeV range, but rather derivative couplings. This means that $V(h)$ is irrelevant and electroweak symmetry breaking in the TeV region is more naturally discussed in terms of the coefficients of the Higgs-flare function $\mathcal{F}(h)$.

At NLO, the Lagrangian relevant to study unitarity and resonances in the TeV regime 
acquires two further derivatives (so that amplitudes receive terms of order $s^2$) and
becomes
\begin{eqnarray} \label{bosonLagrangian}
{\cal L}_{\text{NLO HEFT}} &=& \frac{1}{2}\left[1+2a\frac{h}{v}+b\left(\frac{h}{v}\right)^2\right]
\partial_\mu\omega^i\partial^\mu\omega^j\left(\delta_{ij}+\frac{\omega^i\omega^j}{v^2-\boldsymbol{\omega}^2}\right)
+\frac{1}{2}\partial_\mu h\partial^\mu h %
\nonumber\\
 &+& \frac{4\alpha_4}{v^4}\partial_\mu \omega^i\partial_\nu \omega^i\partial^\mu\omega^j\partial^\nu\omega^j
+\frac{4\alpha_5}{v^4}\partial_\mu\omega^i\partial^\mu\omega^i\partial_\nu\omega^j\partial^\nu\omega^j
+\frac{g}{v^4}(\partial_\mu h\partial^\mu h)^2
\nonumber\\
 &+& \frac{2d}{v^4}\partial_\mu h\partial^\mu h\partial_\nu\omega^i\partial^\nu\omega^i
+\frac{2e}{v^4}\partial_\mu h\partial^\nu h\partial^\mu\omega^i\partial_\nu\omega^i
\ ,
\end{eqnarray}
that has been extensively studied in earlier work. Here we will concentrate on the LO Lagrangian (tree-level amplitudes $\propto s$) in  Eq.~(\ref{FbosonLagrangianLO}) with the Taylor series of ${\mathcal F}$ around the physical vacuum $h=0$ (with zero number of physical Higgs particles, that is, with $\phi_i = \langle \phi_4 \rangle\delta_{i4} = v\delta_{i4}$ in terms of the SMEFT coordinates) given by Eq.~(\ref{expandF}). The NLO coefficients of the second and third lines in Eq.~(\ref{bosonLagrangian}) should eventually encode similar physics to the $B$ function in Eq.~(\ref{SMEFTEWlagrangianV}), but we will leave exploring this connection for future work, and here concentrate on the comparison between $A(|H|^2)$, $B(|H|^2)$ and ${\mathcal F(h)}$.

\newpage
\section{TeV-scale relevant EW SMEFT in HEFT form}\label{HEFTinSMEFTsection}
\label{sec:convert}
We now show the explicit transformation to polar coordinates, and then to HEFT, of the
SMEFT electroweak Lagrangian of Eq.~(\ref{SMEFTEWlagrangianV}) in terms of the $SU(2)$ doublet $H$, neglecting the gauge couplings and $V(H)$ as is appropriate for TeV-scale physics $E\gg m_{h,W,Z}$ , following the discussion in~\cite{Cohen:2020xca}, with
$     \mathcal{L}_{\rm SMEFT}= A(|H|^2)| \partial 
     H|^2+\frac{1}{2}B(|H|^2)(\partial (|H|^2))^2 
  $. 
This is achieved by decomposing the doublet $H$ of the SMEFT framework in the spherical polar coordinates of Eq.~(\ref{eq:EWNGB-spherical-coord}),
\begin{eqnarray}
H &=& \left(1+\frac{h}{v}\right)\,  U(\omega) \,     \langle H\rangle    \, ,
\end{eqnarray}
 where $h$ denotes the radial Higgs-boson field in the SMEFT framework, $\langle H\rangle $ is  the chosen Higgs doublet vacuum, commonly taken to be 
$ \left( 0 \quad v/\sqrt{2} \right)^T $
$v=\sqrt{2\langle H^\dagger H \rangle}$ provides the modulus of the Higgs vacuum expectation value (vev), \and the $SU(2)$ matrix $U(\omega)$ contains the EW Goldstone bosons.
The vev modification due to higher-order corrections can always be later incorporated to the analysis by considering a  shift in the Higgs field~\footnote{This removes terms linear in $h$ and recenters the Higgs field expansion around the potential minimum.} $h\to h+\Delta$. 
Substituting $H^\dagger H= (v+h)^2/2$ in $A$ and $B$, this yields 
\begin{eqnarray}
\mathcal{L}_{\rm polar-SMEFT} &=& \frac{1}{2}(v+h)^2{A(h)}(\partial_\mu  \boldsymbol{n}\cdot \partial^\mu  \boldsymbol{n}) 
+ \frac{1}{2}\Big (A(h)+(v+h)^2 B(h)\Big )(\partial h)^2
\nonumber \\ 
&=&
\frac{1}{4} (v+h)^2 \, A(h) \, \langle \partial_\mu U^\dagger \partial^\mu U\rangle 
+ \frac{1}{2} (A(h)  + (v+h)^2 B(h)) (\partial h)^2  \, ,
\label{SMEFTtoHEFT1}
\end{eqnarray}
with $A((h+v)^2/2)\equiv \tilde{A}(h)\to A(h)$ now a function of $h$ to avoid cumbersome notation.
The SM, with $A=1$ and $B=0$, is the first and simplest of the family of SMEFT Lagrangians in Eq.~(\ref{SMEFTtoHEFT1}), and in this form it reads
\begin{eqnarray}
\mathcal{L}_{\rm SM} &=&   |\partial H|^2  
\, =\, 
\frac{1}{4} (v+h)^2 \,   \langle \partial_\mu U^\dagger \partial^\mu U\rangle  + \frac{1}{2}  (\partial h)^2  \, . 
\label{eq:SM-Lagr}
\end{eqnarray}

In the general case, even though
the coordinates of Eq.~(\ref{SMEFTtoHEFT1}) are now those of HEFT, the Lagrangian is not yet in its canonical form because, by convention, the HEFT Higgs field's $h$ kinetic term needs to be fixed to its free-wave standard expression 
\begin{eqnarray}
\mathcal{L}_{\rm HEFT} &=& 
\frac{v^2}{4} \mathcal{F}(h_1) \,   \langle \partial_\mu U^\dagger \partial^\mu U\rangle 
+
\frac{1}{2} (\partial h_1)^2  \, , 
\label{eq:HEFT-Lagr}
\end{eqnarray}
which requires a further change of the $h$ variable. Finding an $h_1$ field that absorbs the multiplicative factor in Eq.~(\ref{SMEFTtoHEFT1}) and that becomes the Higgs field in the HEFT framework, implies solving the differential condition~\cite{Giudice:2007fh} 
 \begin{equation}\label{htoh1}
     dh_1\, =\, \sqrt{A(h)+(v+h)^2 B(h)}\,\, dh \, ,
 \end{equation} 
that will collect all factors of $h_1$ to multiply only the Goldstone term to the right of Eq.~(\ref{SMEFTtoHEFT1}), from which the $\mathcal{F}$ of Eq.~(\ref{FbosonLagrangianLO}) can be read off,
\begin{equation}
     v^2\mathcal{F}(h_1)\, =\, (v+h(h_1))^2\, A(h(h_1))\ .
     \label{FfromSMEFT}
 \end{equation}
in terms of $h_1=h_1[A,B](h)$. In the next subsection~\ref{subsec:noA} we will show that non-trivial $A$ terms are unnecessary, so we can set $A=1$ and employ $B$ alone, which will determine the relation $h_1=h_1(h)$.
Once $h$ has been expressed in terms of $h_1$, the Lagrangian will have reached its HEFT form and the subindex in $h_1$ may be dropped~\footnote{Note we are dropping here the possible shift $h_1\to h_1 + \Delta$, required if there are modification to the SM Higgs potential. We are interested in high-energy effects and ignore non-derivative operators in $\mL_{\rm HEFT}$. This shift can easily be incorporated if needed.}. With this method, 
the coefficients of Eq.~(\ref{expandF}) expanding the generic HEFT Lagrangian radial function can be retrieved from the initial SMEFT.

To complete this discussion we will quickly digress, in the next subsection, to show that $A=1$ can be consistently taken, afterwards proceeding to carry out the transformation $h\to h_1$ for the relevant SMEFT pieces for the electroweak sector in the TeV energy regime where $m_h\ll E\sim \partial \ll \Lambda$.

\subsection{$A(H)$ is not really necessary for $\sqrt{s}\gg m_h$} \label{subsec:noA}

We here quickly show that it is possible, and can be more convenient, to eliminate the $n^{\rm th}$-power operators (for $n\geq 1$) obtained in an expansion of $A$, by means of a partial integration. For this, note that, up to a total divergence, 
\begin{eqnarray}
(H^\dagger H)^n |\partial H|^2 =
- \frac{n}{2}(H^\dagger H)^{n-1}   (\partial | H|^2)^2  -\frac{1}{2} (H^\dagger H)^n \left(  (\partial^2 H^\dagger)H + H^\dagger (\partial^2 H)\right) \,,
\label{eq:A-simpl}
\end{eqnarray}
obtained by using the relation $\partial^2 |H|^2 = 2 |\partial H|^2 + (\partial^2 H^\dagger)H + H^\dagger (\partial^2 H)$.  
This teaches us that we can always convert (by partial integration) any $n^{\rm}$-power operator of $A$-type into an $(n-1)$-power operator of $B$-type. 
The price to pay includes an irrelevant total derivative and a couple of terms proportional to $\partial^2 H$ and $\partial^2 H^\dagger$. However, the classical equations of motion of $H$  trade the derivative operators for $\partial^2 H $ (and its conjugate) by operators without derivatives, up to correction of higher dimension in $1/\Lambda^2$. In this way, the $A$-type of operators can be removed from the theory and transformed into $B$-operators at fixed dimension 6, 8, etc. 
Employing this freedom, we will set $\Delta A_{\rm BSM}=0$ and just keep the leading operator, $A=A_{\rm SM}=1$. Hence SMEFT can  be formulated in polar coordinates $(h,\omega^a)$ as     
\begin{eqnarray} \label{polarSMEFT1}
\mathcal{L}_{\rm polar-SMEFT} &=& 
\frac{v^2}{4} \bigg(1+\frac{h}{v}\bigg)^2 \,   \langle \partial_\mu U^\dagger \partial^\mu U\rangle 
+
\frac{1}{2} \bigg(1  + (v+h)^2 B(h) \bigg) (\partial h)^2  \, ,
\end{eqnarray}
instead of Eq.~(\ref{SMEFTtoHEFT1}).
The change of variables in the Higgs field, $h=h(h_1)$ of Eq.~(\ref{htoh1}) then becomes 
\begin{eqnarray}
dh_1 &=& \bigg(1  + (v+h)^2 B(h) \bigg) ^{1/2}  dh \, .
\end{eqnarray}
This change determines $\mathcal{F}$ in the form 
\begin{eqnarray}\label{cerodoble}
\mathcal{F}(h_1)   &=& \bigg(1+\frac{h(h_1)}{v}\bigg)^2 \, ,
\end{eqnarray}
with $h$ implicitly given~\cite{Giudice:2007fh} by the relation
\begin{eqnarray}
h_1 &=& \int_0^{h} \bigg(1  + (v+h)^2 B(h) \bigg) ^{1/2}  dh \, .
\end{eqnarray}

\subsection{Explicit computation with SMEFT's power expansion of $B(|H|^2)$}

\subsubsection{Order 6 in the SMEFT counting}

The SMEFT Lagrangian is an alternative parametrization of SM deviations, that assumes the SM symmetries and fields, and particularly assumes the traditional doublet structure for the Higgs field. The Higgs sector of this Lagrangian was introduced in Eq.~\eqref{SMEFTEWlagrangianV}, and it can be written more generally as, 

\begin{equation}
    \mathcal{L}_{\text{SMEFT}} =
    \mathcal{L}_{\text{SM}} + 
    \sum_{n=5}^{\infty}
    \sum_i
    \frac{c_i^{(n)}}{\Lambda^{n-4}} \op_i^{(n)}  \, .
\end{equation}

 At dimension 6, there are three operators of the SMEFT Warsaw basis~\cite{Grzadkowski:2010es} that directly distort the Standard Model's Electroweak Symmetry Breaking Lagrangian, which written in terms of the Higgs field doublet $H$ appropriate for SMEFT are ($\partial^2\equiv\Box$)
\begin{align} \label{Warsawbasis}
&  \op_H = (H^\dagger H)^3  \, ,
 & \op_{HD} = (H^\dagger D_{\mu} H)^*  (H^\dagger D^{\mu} H) \, ,  \nonumber \\
 & \op_{H \Box} = (H^\dagger H) \Box (H^\dagger H)\ .
\end{align}
They can of course be reexpressed in terms of the singlet field for the Higgs boson via $(H^\dagger H) = (h + v)^2/2$ (in polar coordinates this is manifestly gauge-independent).  
Those three operators are actually all that is needed for  Higgs-Goldstone boson scattering up to dimension 6 in the SMEFT counting. Moreover, $\op_{HD}$ breaks custodial symmetry so that it can be counted as higher order due to the small size of the corrections to Peskin-Takeuchi observables in the SM at LEP.

 We would like to remark that not only at dimension-6 but also at dimension-8 there is an additional operator with two derivatives acting only on a product of Higgs doublets. However, these terms violate custodial symmetry and they actually contribute to an independent type of HEFT operator, 
 Longhitano's $a_0$ Lagrangian term~\cite{Longhitano:1980iz,Longhitano:1980tm}. Consistently, this $a_0$ operator is related to the experimentally suppressed oblique $T$--parameter. Thus, we will no longer consider this type of custodial breaking operators in this article, although a similar study can be worked out if this kind of corrections needed to be included.
 
In turn, $\op_{H}$ is not a derivative operator, so that it does not contribute to the flare function that we are pursuing (though it does affect the Higgs self-coupling, namely the Higgs SM potential, and the vacuum expectation value, important near threshold, its impact in the TeV region is much smaller than that of the derivative operator).

In summary, only the $\mO_{H\Box}$ operator contributes to $\mF(h)$ at  order $\mathcal{O}(\Lambda^{-2})$.
Moreover, it has been shown in~\cite{Alonso:2021rac}, by geometric arguments, that only one operator is needed at this order, which is consistent with our discussion.
The rest of the electroweak operators of the Warsaw basis that the reader may be wondering about,
\begin{align}
&  \op_W = \epsilon_{ijk} W_{\mu }^{\nu i} W_{\nu}^{\rho j }
W^{\mu k}_{\rho}  \, ,
& \op_{HW} = (H^\dagger H) W_{\mu \nu}^i W^{\mu \nu i} \, , \nonumber \\
 & \op_{H B} = (H^\dagger H) B_{\mu \nu} B^{\mu \nu}    \, ,
 & \op_{H W B} = (H^\dagger \tau^i H) W_{\mu \nu}^i B^{\mu \nu}   \, ,
\end{align}
are necessary only if one intends to couple the transverse electroweak gauge bosons~\cite{Gonzalez-Lopez:2020lpd,Maas:2020kda}), but they are of no concern for our purposes of studying the TeV-region electroweak-symmetry breaking Lagrangian that requires only, by the equivalence theorem~\cite{Veltman:1989ud,Dobado:1993dg} in the TeV region, the Goldstone bosons $\simeq$ longitudinal $W_L$, $Z_L$ .
Further, a generic basis could also contain
an operator of the form
$\partial_{\mu}  (H^\dagger H) \partial^{\mu}  (H^\dagger H)$, but this is eliminated in the standard Warsaw treatment because it is equivalent to $\op_{H \Box}$ in Eq.~(\ref{Warsawbasis}) up to a total divergence, in analogy to Eq.~(\ref{eq:A-simpl}),
\begin{align}
 (X^2) \Box (X^2) = - \partial_{\mu} X^2 \partial^{\mu} X^2 + \underbrace{ \partial_{\mu} ( X^2  \partial^{\mu} X^2 )}_{\rm{surface \, \, term}} \, ,
\end{align}
or in terms of the $h$ singlet,
\begin{align}
  \op_{H \Box} & = (H^\dagger H) \Box (H^\dagger H) = - \partial_{\mu}  (H^\dagger H) \partial^{\mu}  (H^\dagger H) + \partial_\mu(...) = - (h +  v)^2 \partial_{\mu} h \partial^{\mu} h + \partial_\mu(...)\ .
\end{align}

Therefore, the only contributing dimension-six operator of the Warsaw basis that preserves custodial symmetry is
 \begin{align}
 \label{convierteops}
  \op_{H \Box} & = (H^\dagger H) \Box (H^\dagger H) = - \partial_{\mu}  (H^\dagger H) \partial^{\mu}  (H^\dagger H)\, , 
\end{align}
that in the Lagrangian appears multiplied by the Wilson coefficient $c_{H\Box}$ and is suppressed by two powers of the high-energy scale $\Lambda$ respect to the dimension-4 Lagrangian.
Comparing with Eq.~(\ref{SMEFTEWlagrangianV}) we read the (constant) values $A(|H|^2)=1$ and $B(|H|^2)=-2\frac{c_{H\Box}}{\Lambda^2}$, that contain no fields.

\subsubsection{The role of $c_{H\Box}$ in SMEFT and bounds on its size from experimental data}

Because SMEFT has been used for a few years now to analyze  LHC data, there already exist bounds on the coefficient $c_{H\Box}$ from Run 2 of the machine (even if the associated operator $\mathcal{O}_{H\Box}$ is quite elusive in LHC fits);  we now recall those bounds.

The best overall constraints on the dimension-6 basis arise from 
Higgs-sector observables (production and decay)~\cite{Ellis:2020unq,Ethier:2021bye}, but it is only when combined with other electroweak channels that this  $c_{H\Box}$ coefficient can be well constrained. 
The reason for this is the way that $\mO_{H \Box}$ enters in the $B$ piece of the Lagrangian in Eq.~(\ref{SMEFTEWlagrangianV}). Its effect is to change the
Higgs wave-function normalization 
\begin{equation} \label{eq:chboxshift}
  \mathcal{L}_{\text{SMEFT}} = 
 \frac{1}{2}  \left( 1  -  \frac{2 c_{H \Box} v^2}{\Lambda^2}  
  \right) \partial_{\mu} h \partial^{\mu} h \, +\, ...
\end{equation}
instead of the Higgs couplings to other particles, that are not directly affected. Hence, the contribution of this operator to any on-shell production or decay process of a single Higgs boson appears as a kinematics-independent shift, as evident from Eq.~\eqref{eq:chboxshift}.  In particular, for the reference value of $\Lambda = 1$ TeV used in most analysis, this overall shift for the several processes considered, whether decay width or production cross-section, becomes
 \begin{equation} \label{ecudel012}
\frac{\sigma_{H,\ \text{SMEFT}}}{\sigma_{H,\ \text{SM}}}   \propto    \frac{\Gamma_{H,\ \text{SMEFT}}}{\Gamma_{H,\ \text{SM}}}  \propto  
  1 + 2 \frac{ c_{H \Box} v^2}{\Lambda^2}
  = 1 + 0.12 c_{H \Box}\, ,
\end{equation}
which was already numerically observed by the ATLAS collaboration and reported, for example in Table 1 of~\cite{ATLAS:2019dhi}. 
It is obvious that the numbers there, between 0.115 and 0.125, just reflect the exact 0.12 factor of Eq.~(\ref{ecudel012}). This is true in any process involving only one on-shell Higgs boson (as will also be the case in our Eq.~(\ref{amp:h}) below); but events with two or multiple $h$ particles such as Eq.~(\ref{amp:2h}) and following have different dependences on $c_{H\Box}$ and will allow a cleaner separation within SMEFT.
Also, the cross sections above in $\frac{\sigma_{H,\ \text{SMEFT}}}{\sigma_{H,\ \text{SM}}} $ are implicitly understood as their narrow Higgs-width approximation, where one Higgs is produced on-shell and then cascades into the final products. For off-shell Higgs studies the dependency would be different.

In consequence, this kinematically not very exciting $\mathcal{O}_{H\Box}$ operator is often overlooked and few works actually constrain it. 
Still, the works of Ellis {\it et al.}~\cite{Ellis:2020unq} and Ethier {\it et al.}~\cite{Ethier:2021bye} offer quite interesting bounds on $c_{H\Box}$, that at 95\% confidence level, and rounding off to the precision of the leading digit of the uncertainty, read as follows,
\begin{eqnarray}
c_{H\Box} \simeq -0.3\pm 0.7 \ \ {\rm (individual)} \\
c_{H\Box} \simeq -1 \pm 2 \ \ {\rm (marginalized)} \ .
\end{eqnarray}
Both are compatible with the Standard Model value $c_{H\Box}=0$.

\subsubsection{Operator of order 8 in the SMEFT counting}
\label{subsec:dim8op}

Going one order further in the $1/\Lambda^2$ power counting makes the SMEFT parametrization more interesting~\cite{Passarino:2019yjx,Dawson:2021xei,Hays:2018zze,Henning:2015alf}. In particular, the full dimension-8 basis in SMEFT was recently published in \cite{Murphy:2020rsh,Li:2020gnx}. To find the dimension-8 operator that contributes $\mathcal{O}(\Lambda^{-4})$ corrections to the flare function $\mathcal{F}(h)$ we can return to 
the $B(H)$ term in Eq.~(\ref{SMEFTEWlagrangianV})  remembering that $A(H)$ is irrelevant as per section~\ref{subsec:noA},  $ \frac{1}{2}B(|H|^2)(\partial (|H|^2))^2$, and set $B(|H|^2)\propto |H|^2$ instead of 1.
Therefore, at dimension-8 we only find the following operator:
\begin{equation}
    \op_{H \Box}^{(8)}  = |H|^4 \Box  |H|^2 =   - 2 |H|^2 (\partial (|H|^2))^2 = - (h + v)^4 \partial_{\mu} h \partial^{\mu} h  \ .
\end{equation}
 We have chosen this operator's normalization for convenience and resemblance to $\mO_{H\Box}^{(6)}$ in Eq.~(\ref{convierteops}) when expressed in terms of the Higgs doublet modulus $(v+h)/\sqrt{2}$. Through partial integration, it can be easily rewritten in other forms considered in the bibliography (up to a total derivative): 
\begin{equation}
    \op_{H \Box}^{(8)}\,=\,  - 2 |H|^2 (\partial (|H|^2))^2   
    \,=\,  |H|^4 \Box  |H|^2 
    \,=\,  2 |H|^4 |\partial H|^2  +   |H|^4 ( (\Box H^\dagger) H+H^\dagger (\Box H)) \,,    
\end{equation}
with the last contribution $|H|^4 ( (\Box H^\dagger) H+H^\dagger (\Box H))$ being proportional to the Higgs equations of motion (EoM), so they can be removed from the effective action and transformed into operators including EW Goldstone bosons and also fermions, 
and $Q_{H^6}^{(1)}=|H|^4 |DH|^2 $ the operator 
chosen for the dimension-8 basis in Ref.~\cite{Murphy:2020rsh}. The other possible two-derivative dimension-8 operator $Q_{H^6}^{(2)}$ breaks  custodial symmetry and will not be discussed in this article.

In the next Section, we will first work out the precise change of Higgs variable $h\to h_1$ 
from SMEFT to HEFT up to dimension-6. 
Afterwards, in subsection~\ref{subsec:dim8}, we will proceed up to the next order and provide the NNLO modification induced by  
this dimension--8 operator. 

\subsection{Change of coordinates $h^{\rm SMEFT}\to h_1^{\rm HEFT}$}

\subsubsection{Derivation and  result at dimension-6}

The first step to put the dimension-6 relevant SMEFT Lagrangian into HEFT form is then to change the variable as in Eq.~(\ref{SMEFTtoHEFT1}) yielding
  \begin{equation}
     \mathcal{L}_{\rm SMEFT}= \frac{1}{2}\Big (1-2(v+h)^2\frac{c_{H\Box}}{\Lambda^2}\Big )(\partial_\mu h)^2+\frac{1}{2}(v+h)^2(\partial_\mu  \boldsymbol{n}\cdot \partial^\mu  \boldsymbol{n})\;.
 \end{equation}

We next have to take the Higgs kinetic energy to canonical form.  
This requires integrating~(\ref{htoh1}) 
($t$ being the integration variable taking the place of $h$):
\begin{equation}
  h_1=\int_0^h  \sqrt{1-(v+t)^2\frac{2 c_{H\Box}}{\Lambda^2}}dt=\int_v^{v+h} \sqrt{1-t^2\frac{2 c_{H\Box}}{\Lambda^2}}dt=\sqrt{\frac{\Lambda^2}{2c_{H\Box}}}\int_{\theta_0}^{\theta_1} |\cos{\theta}|\cos{\theta}d\theta\;,
  \label{integral1}
\end{equation}
where $\theta_0=\arcsin{\sqrt{\frac{2c_{H\Box}}{\Lambda^2}}v}$ and $\theta_1=\arcsin{\sqrt{\frac{2c_{H\Box}}{\Lambda^2}}(v+h)}$.
Now, since $0\leq\theta_0\leq\theta_1\ll\pi/2$ (because the EFT coefficient is very small by current bounds \cite{Buchalla:2015qju}) we can assume that the cosine in Eq. (\ref{integral1}) is positive and hence find

\begin{eqnarray}    h_1&=&\frac{1}{2}\sqrt{\frac{\Lambda^2}{2c_{H\Box}}}\Big( \theta +\frac{\sin{2\theta}}{2}\Big)\Big|_{\theta_0}^{\theta_1}=\frac{1}{2}\sqrt{\frac{\Lambda^2}{2 c_{H\Box}}}\Big( \theta + \sin{\theta}\sqrt{1-\sin^2\theta}\Big)\Big|_{\theta_0}^{\theta_1}
\nonumber=\\
&& =\frac{1}{2}\left((v+h)\sqrt{1-\frac{2 c_{H\Box}(h+v)^2}{\Lambda^2}} -v\sqrt{1-\frac{2 c_{H\Box}v^2}{\Lambda^2}}  \right)+\nonumber\\
&&+\frac{1}{2}\sqrt{\frac{\Lambda^2}{2c_{H\Box}}}\Big(\arcsin{\sqrt{\frac{2c_{H\Box}}{\Lambda^2}}(v+h)}-\arcsin{\sqrt{\frac{2c_{H\Box}}{\Lambda^2}}v}\Big)\, .\label{result1}
\end{eqnarray}
Such expression is not particularly useful, especially taking into account that we need to invert it to obtain $h(h_1)$, so we explore it by 
expanding Eq. (\ref{result1}) to leading order in $c_{H\Box}/\Lambda^2$, finding
\begin{equation}
    h_1=h - \frac{1}{3}\Big(\frac{c_{H\Box}}{\Lambda^2}\Big)(h^3+3h^2v+3hv^2)+\mathcal{O}\Big(\frac{c_{H\Box}^2}{\Lambda^4}\Big)\, ,
\label{eq:h1-from-h-NLO}
\end{equation}
which can be iteratively inverted, yielding
\begin{equation}
    h=h_1+\frac{1}{3}\Big(\frac{c_{H\Box}}{\Lambda^2}\Big)(h_1^3+3h
    _1^2v+3h_1v^2)+\mathcal{O}\Big(\frac{c_{H\Box}^2}{\Lambda^4}\Big)
    \;.
    \label{eq:h-from-h1-NLO}
\end{equation}
Finally, we can use Eq. (\ref{FfromSMEFT}) and (\ref{eq:h-from-h1-NLO}) to obtain ${\mathcal F}(h_1)$,
\begin{eqnarray} \label{F_SMEFT6}
    \mathcal{F}(h_1)&=&\left(1+\frac{h_1}{v}\right)^2 +  \frac{2v^3  c_{H\Box}}{\Lambda^2}\left(1+\frac{h_1}{v}\right)  \left(\frac{h_1^3}{3v^3}+\frac{h
    _1^2}{v^2}+\frac{h_1}{v}\right )  +\mathcal{O}\Big(\frac{c_{H\Box}^2}{\Lambda^4}\Big)= \nonumber \\ 
   &=&  1 +   \left( \frac{h_1}{v} \right)  
  \left( 2 + 2 \frac{ c_{H \Box} v^2}{\Lambda^2} \right)  \ 
  +  \left( \frac{h_1}{v} \right) ^2 \left( 1 + 4 \frac{ c_{H \Box} v^2}{\Lambda^2}  \right)+  \nonumber \\ 
  & &+   \left( \frac{h_1}{v} \right) ^3  \left(  8 \frac{ c_{H \Box} v^2}{3 \Lambda^2} \right)  +  \left( \frac{h_1}{v} \right) ^4 \left( 2 \frac{c_{H \Box} v^2}{3 \Lambda^2}  \right) 
 \, ,
\label{eq:SMEFT-F-Lambda2}
\end{eqnarray}
which expresses the expansion coefficients of $\mathcal{F}$ in terms of the SMEFT Wilson coefficient (in the philosophy of the appendix of~\cite{Buchalla:2018yce}),
\begin{eqnarray}\label{coefsSMEFTHEFT}
a_1 = 2a=2\left(1 + v^2\frac{c_{H\Box}}{\Lambda^2}\right)\; , \;\;\;
a_2 = b=1+{4v^2}\frac{c_{H\Box}}{\Lambda^2}\;, \;\;\;
a_3 = \frac{8v^2}{3}\frac{c_{H\Box}}{\Lambda^2}\;, \;\;\;
a_4 = \frac{2v^2}{3}\frac{c_{H\Box}}{\Lambda^2}\, .\label{eq:firstBoxcorrections}
\end{eqnarray}
These relations expose the inclusion of SMEFT into HEFT: the $a_i$ coefficients, independent parameters in the latter, are correlated in SMEFT up to a given order, as all of the first four $a_i$ are given in terms of only one Wilson coefficient $c_{H\Box}$.
This feature has been suggested as a handle to discern, from upcoming experimental data, whether SMEFT will be applicable later on (in the presence of any separation from the SM values $a=b=1$).  
Measurements of the $\omega\omega\to n h$ scattering process (see Section~\ref{nhproduction})
would allow the determination of the $a_i$ and can probe the SMEFT-predicted correlations~\cite{Agrawal:2019bpm} (or the SM ones~\cite{Arganda:2018ftn,Gonzalez-Lopez:2020lpd}). 
In the absence of such correlations, it is plausible that a HEFT formulation would be needed. 

However, this difference can be put into question in the presence of unnatural Wilson coefficients. 
If the dimension-8 operators contribute at an order similar to that of the dimension-6 operator, because the coefficients are not of order 1 or because $\Lambda$ is not large enough to significantly suppress them, additional SMEFT parameters would appear in the expressions of Eq.~(\ref{coefsSMEFTHEFT}), 
decorrelating the $a_i$ coefficients and voiding the analysis.
Therefore, though perhaps orientative, given that naturalness may have already failed as a safe guiding principle in view of the light Higgs boson mass, more robust criteria that helps systematize the correlations to distinguish SMEFT from HEFT have been provided, and to them we turn in the next section.

At last, the position of the symmetric point $h_\ast$ that satisfies $\mF(h_\ast)=0$ is always 
given by $h(h_1)=-v$ (with $|H|=(h+v)/\sqrt{2}$), as seen directly from Eq.~(\ref{cerodoble}). 
In turn, the position of the symmetric point in HEFT coordinates becomes displaced from its SM value $h_\ast^{\rm SM}=-v$  by an   $\mO(\Lambda^{-2})$ SMEFT correction,
 \begin{equation} 
\label{sympoint2}
\frac{h_\ast}{v}\, =\, \frac{h_1(h)}{v}\bigg|_{h=-v}
\,=\, -1 + \frac{c_{H \Box}  v^2 }{3\Lambda^2}   \,,
\end{equation}
where we use the relation $h_1=h_1(h)$ in Eq.~(\ref{eq:h1-from-h-NLO}). An alternative derivation of this result is presented in Appendix~\ref{appendixD}.

A simple way to obtain this result is to observe that Eq.~(\ref{cerodoble}) for the flare function has a double zero when the SMEFT field is at the symmetric point of the Standard Model, $h=-v\to -1$ (normalizing the field by $v$). Substitution in Eq.~(\ref{eq:h1-from-h-NLO}) immediately yields Eq.~(\ref{sympoint2}).
Curiously, if one would look at Eq.~(\ref{F_SMEFT6}) instead, the position of one of the zeroes would be $h_1=-1=-v$, as is obvious from the first line. This is an effect of the reexpansion in Eq.~(\ref{eq:h-from-h1-NLO}): none of the four zeroes of Eq.~(\ref{F_SMEFT6}), a fourth-order polynomial, seems to be double; in the limit $\Lambda\to \infty$, two of them escape to infinity and the other two merge to give the correct one of Eq.~(\ref{sympoint2}). This means that, at face value and inside a small interval of width suppressed by $1/\Lambda^2$, Eq.~(\ref{F_SMEFT6}) can violate positivity (see subsection~\ref{subsec:positivity} below). This is corrected by the next $\Lambda^{-4}$ order.

\subsubsection{Result at dimension-8} \label{subsec:dim8}

We now quote the result of adding the 
operator of dimension 8 in Eq.~(\ref{subsec:dim8op}); the calculation follows along the same lines so we only quote the combined result for the flare function $\mathcal{F}$, which reads 
\begin{align} \label{F_SMEFT8}
   \mF(h_1) =  1 &+ \left( \frac{h_1}{v} \right)  
  \left( 2 + 2 \frac{ c_{H \Box}^{(6)} v^2}{\Lambda^2} 
  + 3 \frac{(c_{H \Box}^{(6)})^2 v^4}{\Lambda^4}  + 
      2 \frac{c_{H \Box}^{(8)} v^4}{\Lambda^4}  \right) + \nonumber \\
  &+
   \left( \frac{h_1}{v} \right)^2 \left( 1 + 
   4 \frac{ c_{H \Box}^{(6)} v^2}{\Lambda^2} + 
   12 \frac{(c_{H \Box}^{(6)})^2 v^4}{\Lambda^4} + 
    6 \frac{c_{H \Box}^{(8)} v^4}{\Lambda^4}  \right) + 
 \nonumber \\ &+
    \left( \frac{h_1}{v} \right)^3 
    \left(  8 \frac{ c_{H \Box}^{(6)} v^2}{3 \Lambda^2} + 
    56 \frac{(c_{H \Box}^{(6)})^2 v^4}{3 \Lambda^4}  +
    8 \frac{ c_{H \Box}^{(8)} v^4}{\Lambda^4}   \right) +
\nonumber \\ &+
 \left( \frac{h_1}{v} \right)^4 \left( 
 2 \frac{c_{H \Box}^{(6)} v^2}{3 \Lambda^2} +
 44 \frac{(c_{H \Box}^{(6)})^2 v^4}{3 \Lambda^4} + 
 6 \frac{c_{H \Box}^{(8)} v^4}{\Lambda^4} \right) + 
  \nonumber \\ 
  &+
   \left( \frac{h_1}{v} \right)^5
   \left( 88 \frac{(c_{H \Box}^{(6)})^2 v^4}{15 \Lambda^4} +
    12 \frac{ c_{H \Box}^{(8)} v^4}{5 \Lambda^4}  \right) + 
  \nonumber \\
  &+ \left( \frac{h_1}{v} \right)^6 
  \left( 44 \frac{(c_{H \Box}^{(6)})^2 v^4}{45 \Lambda^4} + 2 \frac{c_{H \Box}^{(8)} v^4}{ 5 \Lambda^4}  \right) + \op{(\Lambda^{-6})} \, .
\end{align}
Note that the bracket in each line provides the corresponding $a_j$ up to $\mO(\Lambda^{-4})$.  
Also, to make the order manifest and avoid confusion, we have denoted $c_{H\Box}$ by $c^{(6)}_{H\Box}$ in this paragraph 
and below whenever there might be any confusion. 

As for the symmetric point around which SMEFT is built, Eq.~(\ref{sympoint2}) takes a further correction of $\mO(\Lambda^{-4})$ that may take it away from the Standard Model value. 
 Again, Eq.~(\ref{cerodoble}) shows that the $SU(2)\times SU(2)$ fixed-point point condition $\mF(h_\ast)=0$ has always the solution $h=-v$, which in the in HEFT coordinates up to $\mO(\Lambda^{-4})$ SMEFT corrections is given by  
\begin{equation}
\frac{h_\ast}{v}\, =\, \frac{h_1(h)}{v}\bigg|_{h=-v} 
\,=\, -1 + \frac{c_{H \Box}^{(6)} v^2 }{3\Lambda^2}   +    \left( (c_{H \Box}^{(6)})^2  + 2 c_{H \Box}^{(8)} \right)\frac{v^4}{10\Lambda^4} \, .
\label{eq:SMEFTd8-sym-point}
\end{equation} 
where we used the relation between SMEFT and HEFT coordinates at this order:  
\begin{equation}
h_1 \,=\, h 
\, +\, \frac{c_{H\Box}^{(6)}}{3\Lambda^2}\left(v^3-(v+h)^3\right) 
\, +\, \frac{((c_{H\Box}^{(6)})^2+2c_{H\Box}^{(8)})}{10\Lambda^4}\left(v^5-(v+h)^5\right)\,+\,\mO\left(\frac{1}{\Lambda^6}\right)\, . 
\end{equation}

\section{Geometric and analytic distinction between HEFT and SMEFT}\label{Geometricsection}

This section succinctly exposes the precise theoretical conditions allowing to discern between SMEFT and HEFT, compiling the main results of several articles~\cite{Dobado:2019fxe,Alonso:2016oah,Alonso:2016btr,Alonso:2015fsp,Cohen:2020xca} in their geometrical aspects and adding an extended analytical discussion about the function $\mathcal{F}$ of our own. 
Much of the past confusion between the two EFT formulations arises from the fact that there are two coordinate systems to describe the same system of fields, for this, the San Diego group advocated for employing a geometric perspective to be able to make coordinate-invariant statements. 

\subsection{Flat SM geometry}
The $O(4)$ components in the scalar field of Eq.~(\ref{eq:EWNGB-spherical-coord}) used for the SM Higgs sector, $\boldsymbol{\phi}=(\phi^1,\phi^2,\phi^3,\phi^4)$ are taken to represent coordinates in a (momentarily flat, later in the next subsection curved) geometric manifold $\mathcal{M}$. 
$\boldsymbol{\phi}$ contains the Higgs field and the three ``eaten'' Goldstone bosons and has a Lagrangian (turning off gauge fields)
\begin{equation}
    \mathcal{L}_{\rm SM}=\frac{1}{2}\partial_\mu\boldsymbol{\phi}\cdot\partial^\mu \boldsymbol{\phi}-\frac{\lambda}{4}(\boldsymbol{\phi}\cdot\boldsymbol{\phi}-v^2)^2\;.
\end{equation}
In these Cartesian coordinates the global $O(4)$ transformations should act \textit{linearly}
\begin{equation}
    \boldsymbol{\phi}\to O\boldsymbol{\phi}\;,\;\;\;\;\;\;O^TO=\boldsymbol{1}\;.\label{linearsym}
\end{equation}
The field breaks the global electroweak symmetry $O(4)$ down to $O(3)$ by acquiring a vacuum expectation value
$$\langle\boldsymbol{\phi\cdot\phi}\rangle=v^2\ , $$
where $v\simeq 246$ GeV. Usually, the vacuum expectation value is chosen to be $\langle\phi_4\rangle=v$ while $\langle\phi_1\rangle=\langle\phi_2\rangle=\langle\phi_3\rangle=0$ and the Higgs field $h$ in these Cartesian coordinates is defined through the relation $\phi_4=v+h$. \par

The alternative coordinate system in which HEFT is based expresses the SM Higgs sector Lagrangian in polar form 
\begin{equation}
    \boldsymbol{\phi}=\Big(1+\frac{h}{v}\Big)\boldsymbol{n}(\boldsymbol\omega), \;\;\;\;\;\;\;\; \boldsymbol n\cdot\boldsymbol n=v^2\;.
\end{equation}
Clearly, the constraint $\boldsymbol n\cdot\boldsymbol n=v^2$ makes the $O(4)$ symmetry to be realized in a \textit{non-linear} way.
This comes about because the four components of $\boldsymbol n=(\omega^1,\omega^2,\omega^3,\pm \sqrt{v^2-{\boldsymbol \omega}^2})$
rotate linearly with $O$ in Eq. (\ref{linearsym}),
imposing a non-linear transformation law on the polar-coordinate Goldstone bosons, $w^a$ ($a=1,2,3$). In these polar coordinates, the Higgs sector SM Lagrangian is~\footnote{Note the difference with Eq.(2.12) in~\cite{Alonso:2016oah} where the $\omega_i\omega_j$ piece is absent. It is unnecessary unless amplitudes with more than two Goldstone bosons are analyzed, which we leave for future investigation.}
\begin{equation}
    \mathcal{L}_{SM}=\frac{1}{2}\Big(1+\frac{h}{v}\Big)^2(\partial_\mu \omega^i\partial_\mu  \omega^j )\left(\delta_{ij}+\frac{\omega_i\omega_j}{v^2-\boldsymbol{\omega}^2}\right)+\frac{1}{2}(\partial_\mu h)^2-\frac{\lambda}{4} (h^2+2vh)^2\;.
\end{equation}
In the SM, we thus see that
\begin{equation}
    \mathcal{F}(h=h_1) = 
   \left(1+\frac{h}{v}\right)^2
    \label{SMF}
\end{equation}

This exercise enlightens the fact that that the SM Higgs sector $O(4)$ symmetry can be realized both in a linear or a non-linear manner. That is why, when studying EFTs that extend the SM, one should  concentrate on objects which are invariant under field redefinitions~\cite{Passarino:2016saj}.
Again, the important distinction between the SM, SMEFT and HEFT is not that the realization of the $O(4)$ symmetry is linear or not, since a field redefinition can turn one into the other. That is why the key aspect one must look for are the geometric invariants on the scalar manifold $\mathcal{M}$, which are by definition invariant under field redefinitions. 

\subsection{Beyond SM: curved $\boldsymbol{\phi}$ geometry}

The kinetic term of the Lagrangian (giving the classical equations of motion as the geodesics of that manifold) in Eq.~(\ref{FbosonLagrangianLO}) has the form
$$\frac{1}{2}g_{ij}(\phi)\partial_\mu \phi^i\partial^\mu \phi^j$$
and can be interpreted in terms of a metric tensor $g_{ij}$ that provides lengths in the geometrical space of the $\boldsymbol{\phi}$ fields,  $g_{ij}d\phi_id\phi_j$ in any of the different coordinate choices. 
This is the point of contact between physics experiments at colliders, \textit{i.e.}  production and scattering amplitudes derived from this Lagrangian, and the field geometry, where the former are represented in terms of curvature invariants. We now briefly recall the results presented in~\cite{Alonso:2016oah,Cohen:2020xca}.

In the SM, the metric  is just a Kronecker delta $g_{ij}(\boldsymbol{\phi})_{SM}=\delta_{ij}$, but the SMEFT of Eq.~(\ref{SMEFTEWlagrangianV}) extends it to a more general form
\begin{equation}
    g_{ij}\arrowvert_{\tiny SMEFT}=A\Big(\frac{\boldsymbol{\phi}\cdot\boldsymbol{\phi}}{\Lambda^2}\Big)\delta_{ij}+B\Big(\frac{\boldsymbol{\phi}\cdot\boldsymbol{\phi}}{\Lambda^2}\Big)\frac{\phi_i\phi_j}{\Lambda^2}
\end{equation}
where $\Lambda$ is the new physics scale.
 It is always possible to express the  SMEFT metric (and in particular its SM limit) in HEFT form by changing to polar coordinates followed by an additional field redefinition to make the kinetic term of $h$ canonical  as in section~\ref{sec:convert}. This makes the HEFT metric take the generic form
 \begin{equation}
     g_{ij}(h,\boldsymbol{\omega})_{HEFT}=\begin{bmatrix}
\mathcal{F}(h)g_{ab}(\boldsymbol w) & 0 \\
0 & 1
\end{bmatrix}\label{HEFTmetric}
 \end{equation}
 where $g_{ab}(\boldsymbol \omega )$ is the $O(3)$ invariant metric on the scalar submanifold $O(4)/O(3)=S^3$ described by the Goldstone bosons in angular coordinates. 
The SM is the special case with a flat scalar manifold $\mathcal{M}$  (since its metric is just $\delta_{ij}$ for all values of the field: there exist global Riemann coordinates). 
For both SMEFT and HEFT, $\mathcal{M}$ has curvature. But what makes SMEFT different from HEFT?

Eq.~(\ref{HEFTmetric}) allows to interpret the function $\mathcal{F}(h)$ in the manifold $\mathcal{M}\ni \boldsymbol{\phi}$ as 
a scale factor akin to the $a(t)$ one in the Friedmann-Robertson-Walker metric. For each value of $h$  (in units of $v$) there is an $S^3$ submanifold (parametrized by the $\omega_i$), away from the origin of $\mathcal{M}$ by an amount $\mathcal{F}(h)$, that acts as a radial distance.\par
One can in that way always write SMEFT in HEFT form, but as we will see, the converse is not always true. This means that
 $$\text{SM}\subset\text{SMEFT}\subset\text{ HEFT}\;.$$
 This is so because, in order to write a HEFT as a SMEFT, there must exist an invariant point in $\mathcal{M}$ under the $O(4)$ symmetry. This translates into the condition that the function $\mathcal{F}(h)$ must vanish for some $h_\ast$, $\mathcal{F}(h_\ast)=0$. Hence this $h_\ast$, an invariant point under the $O(4)$ symmetry,  plays the role of an origin for the Cartesian-like SMEFT coordinates on $\mathcal{M}$. 
 
This invariant point is a necessity to deploy a linear representation of the $O(4)$ group around it, just as in Eq.~(\ref{linearsym}), and hence to write down HEFT as a SMEFT.
It happens that the existence of such zero of $\mathcal{F}(h)$ is not a sufficient condition for SMEFT to be possible. It may happen that non-analyticities arise at the fixed point $h_\ast$ or between that $h_\ast$ and $h=0$ (the physical vacuum), spoiling the possibility of constructing a viable SMEFT around $h_\ast$ that is applicable in particle physics. This is why we must require that the metric and thus $\mathcal{F}(h)$ be analytic in a sufficient domain (see \cite{Cohen:2020xca} for further detail). 
In a lower energy regime $m_h\sim E$ where the potential $V$ makes a relevant contribution, the same considerations also apply to $V$.  

Instead of obtaining these results by further following the powerful yet intricate geometric methods just mentioned, we will continue with coordinate-dependent field theory in the spirit of staying close to the phenomenological formulation that can be brought to bear at accelerator experiments.

\subsection{Zero and analyticity of $\mathcal{F}$  upon
passing from HEFT to SMEFT}
\label{subsec:noSMEFT}

We are going to show in this subsection how to proceed from HEFT to SMEFT and under what condition this is possible in an analytical manner in terms of the field coordinates. 
For this we need to combine the Higgs $h_1$ and the EW Goldstone bosons $\omega^a$ appropriate for HEFT into the complex doublet $\Phi$ used in SMEFT. Problems can arise about whether the resulting Lagrangian will obey minimal physical requirements from the theoretical (e.g., analyticity) and the phenomenological point of view (e.g., perturbation theory convergence).  

The first step to reconstruct the SMEFT form 
from the HEFT Lagrangian in Eq.~(\ref{eq:HEFT-Lagr})
is to define a new Higgs variable $h$ from the HEFT one, $h_1$ in this subsection,  by the condition 
\begin{eqnarray}
\mathcal{F}(h_1(h)) = F^2(h_1(h)) \,=\, \bigg(1+\frac{h}{v}\bigg)^2\, , 
\qquad \qquad 
F(h_1(h)) \,=\,  1+\frac{h}{v} \, , 
\end{eqnarray}
implying the inverse relations
\begin{eqnarray} \label{Finverses}
h_1   \,=\, \mathcal{F}^{-1}\left((1+h/v)^2\right)\, , 
\qquad \qquad 
h_1 \, =\,  F^{-1}( 1+ h/v) \, . 
\end{eqnarray}

This change of variable unravels the standard HEFT normalization of the Higgs kinetic term in Eq.~(\ref{bosonLagrangian}), turning the Lagrangian in the ``polar-SMEFT'' form, 
\begin{eqnarray} \label{polarSMEFT2}
\mathcal{L}_{\text{polar-SMEFT}} &=& 
\frac{v^2}{4} \bigg(1+\frac{h}{v}\bigg)^2 \,   \langle \partial_\mu U^\dagger \partial^\mu U\rangle 
+
\frac{1}{2} \bigg(   \frac{1}{v} (F^{-1})'(1+h/v)  \bigg)^2 (\partial h)^2  \, .
\end{eqnarray}  
The second, $h$-kinetic term can also be expressed in terms of the square of $F$, that is, $\mathcal{F}$.
To achieve it, we can simply replace $(F^{-1})'(1+h/v)$  by  $  2 (1+h/v)\, (\mathcal{F}^{-1})'((1+h/v)^2)  $.  
We note that $F^{-1}$ and $\mF^{-1}$ are the inverse functions of $F$ and $\mF$, respectively, and $F(h_1)=1+h(h_1)/v$.

Up to this point there is no concerning issue; this polar-coordinate form half way between HEFT and SMEFT, that we also find when calculating in the opposite direction in Eq.~(\ref{polarSMEFT1}) is 
still a valid Lagrangian (if we postpone for a later moment the discussion on the convergence of the perturbative series) totally equivalent to HEFT.  

The possible obstacle to this conversion can however arise when trying to reconstruct the Higgs-doublet field $H$ from the EW Goldstone bosons $\omega^a$ in $U$ and the Higgs scalar field $h$,  making use~\cite{Cohen:2020xca} of
\begin{eqnarray}
\label{transformaSMEFT}
|H|^2 &=& \frac{(v+h)^2}{2}\, ,
\nonumber\\
|\partial H|^2 &=& \frac{(v+h)^2}{4}   \langle \partial_\mu U^\dagger \partial^\mu U\rangle +  \frac{1}{2}   (\partial h)^2 \, ,
\nonumber\\
(\partial|H|^2)^2 &=& (v+h)^2 \,   (\partial h)^2\,=\, 2 |H|^2 \,   (\partial h)^2 \, .
\end{eqnarray} 
The inversion of these linear equations to express Eq.~(\ref{polarSMEFT2}) in terms of $H$
brings about a possible $|H|^{-2}$ singularity in the SMEFT Lagrangian,
\begin{eqnarray}
\mathcal{L}_{\rm SMEFT} &=&  \underbrace{ |\partial H|^2}_{= \mathcal{L}_{\rm SM}} \quad +\quad 
\underbrace{ \frac{1}{2} \bigg[ \bigg(   \frac{1}{v} (F^{-1})'\left(\sqrt{2| H |^2/v^2}\right)  \bigg)^2 \,\,\,-\,\,\, 1\bigg] \, \frac{(\partial| H |^2)^2}{2 | H |^2}  }_{=\Delta \mathcal{L}_{\text{BSM}}}\, .
\label{eq:SMEFT-from-HEFT}
\end{eqnarray}

Such divergence is incompatible with a power-expansion as needed to deploy the SMEFT counting.
The only way that Eq.~(\ref{eq:SMEFT-from-HEFT}) can provide an analytical Lagrangian around $|H|=0$ to allow a valid SMEFT expansion in powers of $H$ is by restricting the $\mathcal{F}=F^2$ function of Eq.~(\ref{FbosonLagrangianLO}) to fulfill the condition 
\begin{eqnarray}   \bigg[
\frac{1}{v} (F^{-1})'\left(\sqrt{2| H |^2/v^2}\right)  \,\,\,-\,\,\, 1 \bigg] 
&\stackrel{| H |\to 0}{=}&  0 \,\,\, + \,\,\,   \mathcal{O}(| H |^2)  
\label{eq:analyticity-condition0}
\end{eqnarray}
so that this zero cancels the $|H|^{-2}$ denominator in Eq.~(\ref{eq:SMEFT-from-HEFT}).
Furthermore, even if the zero is cancelled, the analyticity of the SMEFT Lagrangian at any order implies that the $\mO(|H|^2)$ remnant must also have an analytic expansion in integer powers of $|H|^2$ (from the square-root) . Otherwise an expansion-breaking nonanalyticity is present and SMEFT becomes a theory that is not systematically improvable by furthering the expansion.  This remarkable fact can be traced to Eq.~(\ref{Finverses}), where the change of variables $h\to h_1$ happens at the level of individual singlet particles, whereas the doublet $H$ employed in SMEFT (and in the SM) needs to be squared to $|H|^2$ to produce an electroweak singlet, forcing the square root upon us.

The relation~(\ref{eq:analyticity-condition0}) is a differential equation for $(F^{-1})$ in the variable $z:=\sqrt{2|H|^2/v^2}$, whose integration leads to 
\begin{eqnarray}
 F^{-1}(z) 
&\stackrel{|H|\to 0}{=}&  F^{-1}(0) \,\, +\,\, v z \,\,  \,+\,\mathcal{O}(z^3)\,.  
\label{eq:Fm1}
\end{eqnarray}
 The analyticity of the SMEFT Lagrangian at all orders implies that the $\mO(z^3)$ remainder has an analytic expansion that only contains odd powers of $z$.
We solve for the $z$ variable around $z=0$ in terms of $F^{-1}(z)$, 
and invert to recover the original function
 $F=(F^{-1})^{-1}$ around the point $h_1^*\equiv F^{-1}(0)$, remembering from Eq.~(\ref{Finverses}) that $F^{-1}$ is a HEFT-Higgs $h_1$ value: 
\begin{eqnarray}
z = F(h_1) 
&\stackrel{| H |\to 0}{=}&   \frac{1}{v}(h_1\, -\,h_1^*) \,\, \,\,+\,\,\, \,\mathcal{O}((h_1-h_1^*)^3)\,,   
\label{eq:zeroF-cond}
\end{eqnarray}
where the $\mathcal{O}(z^2)$ remnant can be put in $\mathcal{O}((h_1-h_1^*)^2)$ form up to higher orders.  In terms of $\mathcal{F}$ the relation would be given by $(1+h/v)^2= 2|H|^2/v^2=z^2=F(h_1)^2=\mathcal{F}(h_1)$.   
 Moreover, the analyticity of the SMEFT Lagrangian at all orders implies that the solution of Eq.~(\ref{eq:Fm1}) for $z$ --shown in~(\ref{eq:zeroF-cond})-- has an analytical expansion around $h_1=h_1^*$ that only contains odd powers of $(h_1-h_1^*)$.   

The existence of that zero $h_1^*\equiv F^{-1}(0)$ of $F$ --and thus of its square $\mathcal{F}$--, and the analyticity required for a power series expansion (both of $\mathcal{F}$ and the Higgs potential $V$), broadly constitute the necessary and sufficient requirements for a given HEFT Lagrangian density  characterized by  $\mathcal{F}$ to be expressible as a SMEFT.
Let us summarize and make these findings, that agree with the ones presented in~\cite{Cohen:2020xca}, somewhat more precise:
\begin{enumerate}
    \item{} $F(h_1)$ must have at least a simple zero at some $h_1^*$, {\it i.e.},   $F(h_1^*)=0$. This implies that the function in the HEFT Lagrangian density of Eq.~(\ref{FbosonLagrangianLO}) $\boxed{\mathcal{F}(h_1^*)=F(h_1^*)^2=0}$ must have a double zero.  
    
    \item{} At that point $h_1^*$, $F$ must have the slope  $F'(h_1^*)=\frac{1}{v}$.  This translates into two conditions over $\mathcal{F}$, namely 
    $$\boxed{\mathcal{F}'(h_1^*)=0\ , \ \mathcal{F}''(h_1^*)=\frac{2}{v^2}}$$. 
    \item{} At that point $h_1^*$, $F$ must have zero curvature $F''(h_1^*)=0$, since the first correction must be $\mathcal{O}((h_1-h_1^*)^3)$. 
    From the point of view of $\mathcal{F}$ this translates as the constraint $\boxed{\mathcal{F}'''(h_1^*)=0}$. 
    
    \item{} 
 Finally, it is possible to exploit the analyticity of the SMEFT Lagrangian at higher orders, if the expansion is to be continued and be systematically improvable. In general, analyticity as an all-order requirement forces all even derivatives to vanish at the symmetric point: $F^{(\ell)}(h_1^\ast)=0$ for even $\ell$. From the point of view of $\mF$ this implies the vanishing of all odd derivatives,
    $\boxed{\mathcal{F}^{(2\ell+1)}(h_1^*)=0}$.  
\end{enumerate}

The first two conditions mean that the HEFT  $\mathcal{F}$ flare function must be an upward-bending parabola if an equivalent SMEFT is to exist. In the next subsection,  Figure~\ref{fig:Fstatus} shall expose that current knowledge is compatible with it, and allows to estimate how intensely one of the HEFT coefficients needs to separate from the SM or SMEFT for the latter not to be applicable.

Should Eq.~(\ref{eq:zeroF-cond}) fail, the SMEFT Lagrangian would not have an analytical expansion in powers of the doublet field $H$. 
Moreover, it is important to remark that in order to avoid a singularity, at least at dimension-6, the remnant in~(\ref{eq:analyticity-condition0}) must be at least  $\mathcal{O}(|H|^2)$, or equivalently, the remnant in~(\ref{eq:zeroF-cond}) must be at least $\mathcal{O}((h_1-h_1^*)^3)$. 

Various examples will be provided in subsection~\ref{subsec:examples} below.
 We will deal with the possibility of experimentally finding such zeroes $h_1^*$ in Section~\ref{ZeroesSection}.

\section{Generic properties of the flare function $\mathcal{F}(h)$}
\label{sec:generic-properties}

\subsection{Current knowledge of $\mathcal{F}(h)$}

In particle physics language, the appearance of the $\mathcal{F}(h)$ function in Eq.~(\ref{FbosonLagrangianLO}) controls 
the (derivative) coupling of a pair of $\omega\simeq W_L$ longitudinal gauge bosons to any number of Higgs bosons.

\begin{figure}[!t]   
\centering
\includegraphics[width=0.48\columnwidth]{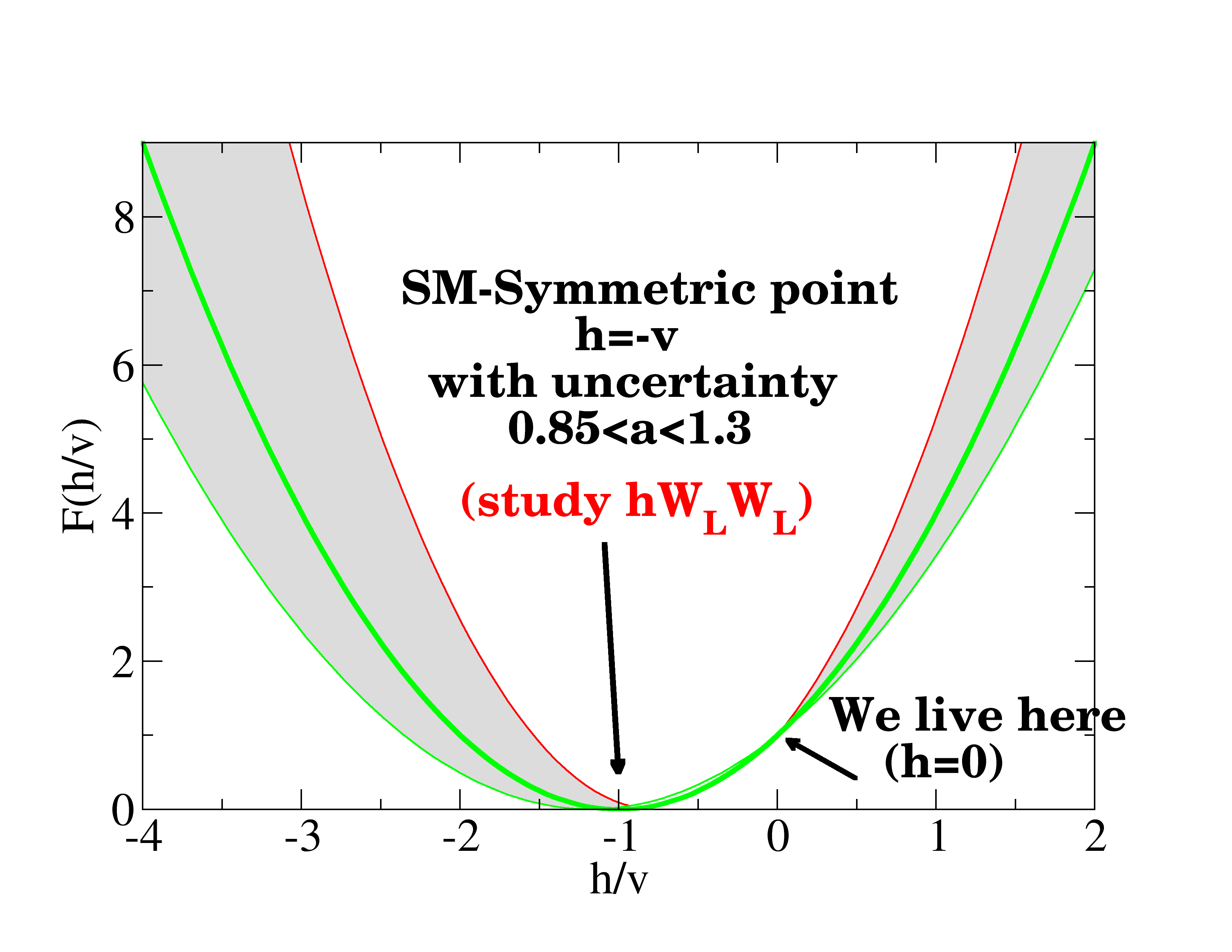}
\includegraphics[width=0.48\columnwidth]{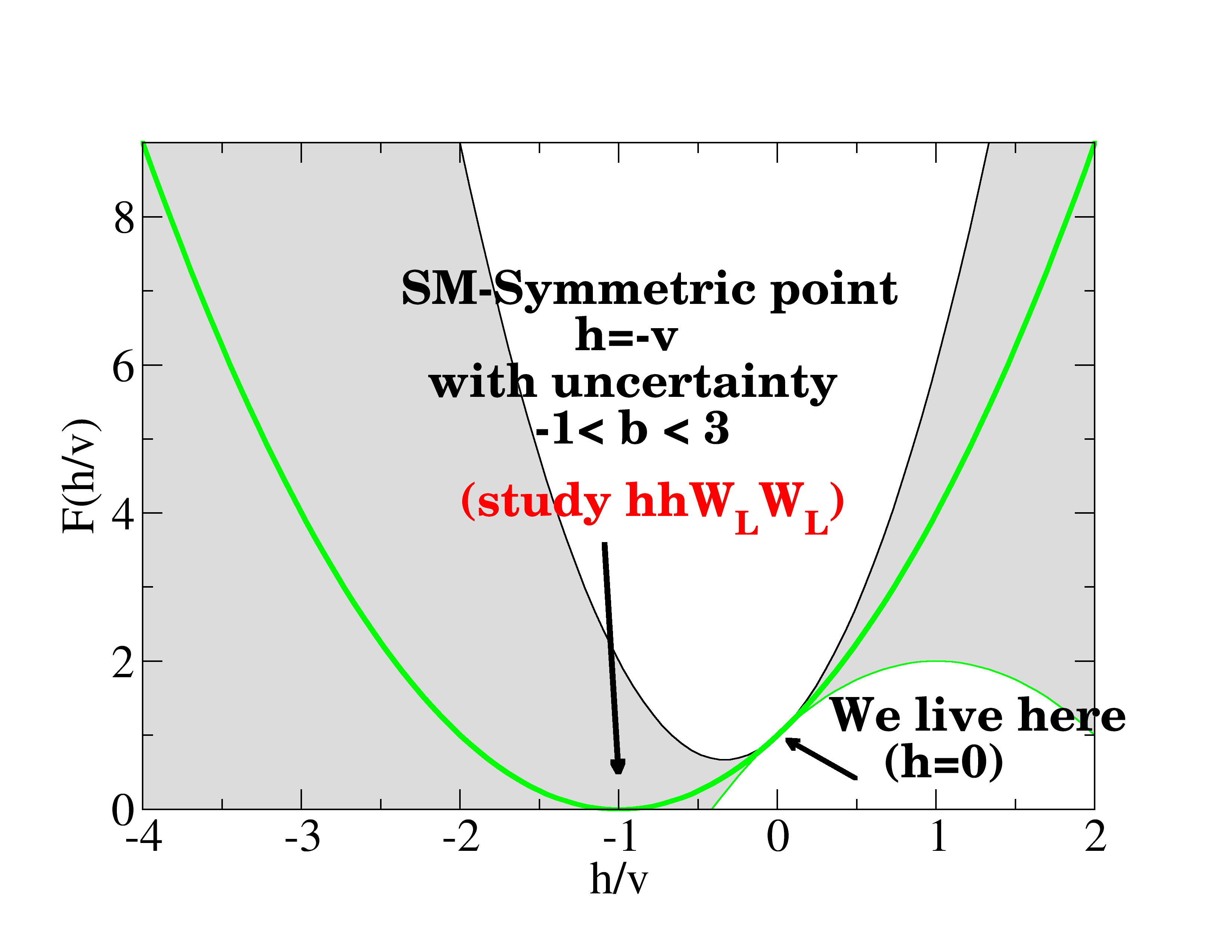}
\includegraphics[width=0.55\columnwidth]{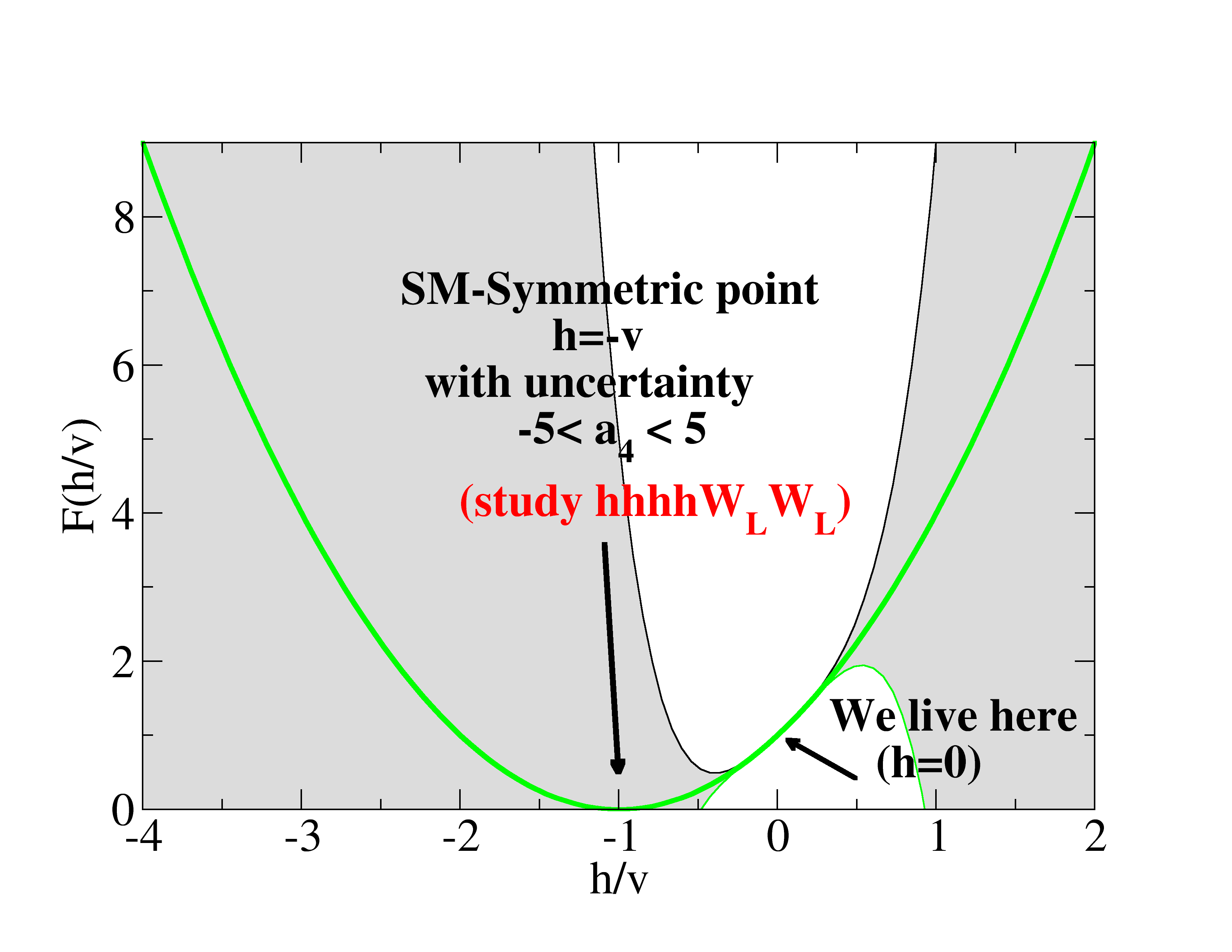}
\caption{\label{fig:Fstatus} {\small Sensitivity of $\mathcal{F}(h)$ to typical  parameter ranges.
The solid parabola (green online) is the SM prediction $1+2\frac{h}{v}+\left(\frac{h}{v}\right)^2$. The grey bands, in the order given,
show the uncertainty due to our present knowledge of the $a_i$ coefficients  that we vary one at a time around the SM value.
Respectively, the coefficients $a_1=2a$, $a_2=b$ and $a_4$, couple $\omega\omega$ to $h$ (the best constrained one, $a$), $hh$, and in view that a third order polynomial always has a zero, $hhhh$ with an even number of powers as the next most interesting one (and we ignore $a_3$ by itself).} 
}
\end{figure}

While this coefficient is the dominant Higgs production in the TeV region, multiboson processes at the LHC in the hundred GeV energy regime already constrain, although not tightly, the coefficients of the $\mathcal{F}(h_1)$ expansion. 
Since, for the rest of the main body of the article, we will be concentrating on HEFT and there can be no confusion with the SMEFT $h$ field, \textbf{we will drop the subindex $h_1\to h$} and make it explicit whenever a change of coordinates between SMEFT and HEFT is used.
We give a graphical representation of the present status of $\mathcal{F}$ in Figure~\ref{fig:Fstatus}.

At the time of writing there is no significant deviation from the Standard Model, which is a particular case of SMEFT, meaning that there is no reason to doubt the applicability of SMEFT: the uncertainty bands by no means exclude a zero of $\mathcal{F}$, possibly where the Standard Model requires it, at $h=-v$. 
The SM line, as a particular SMEFT case, is a parabola with vertex at $\mathcal{F}=0$, as discussed in subsection~\ref{subsec:noSMEFT}.
The reason why we have cut off values $\mathcal{F}<0$ will become clear in the next comment.

\subsubsection{Positivity from boundedness of the Hamiltonian} \label{subsec:positivity}

From the LO HEFT Lagrangian in Eq.~(\ref{FbosonLagrangianLO}) we can construct the Hamiltonian of the theory, finding 
\begin{equation}\label{HamiltonianHEFT}
    H_{\text{HEFT}}=\int d^3\boldsymbol x \; \frac{1}{2}\left[\partial_0 h\partial_0 h+\boldsymbol\nabla h  \cdot \boldsymbol \nabla h \, +\,      \mathcal{F}(h) \, \left( \partial_0 \omega^i\partial_0 \omega^j+\boldsymbol\nabla \omega^i \cdot  \boldsymbol \nabla \omega^j \right)\left(\delta_{ij}+\frac{\omega_i\omega_j}{v^2-\boldsymbol{\omega}^2}\right) \right]\;.
\end{equation}
 From the condition that the Hamiltonian $H_{\text{HEFT}}$ must be bounded from below for vacuum stability one obtains that $\mathcal{F}(h)\geq 0$. This justifies the common usage of the form $F^2(h)$ instead of simply $\mathcal{F}$. 
 While a matter of taste, it is not clear what in particle physics is the quantity $F$ being squared (a radial distance in the $\omega^a$ field space), so we prefer $\mathcal{F}$ for most of the discussion.

One can also often see that,
after expanding the function $\mathcal{F}(h)$ around our physical low-energy vacuum $h=0$ as in Eq.~(\ref{expandF}),  
the positivity condition on $\mathcal{F}(h)$ is forgotten or not explicitly mentioned, although in those approaches employing $F(h)$ with $\mathcal{F}=F^2(h)$ it is automatically incorporated. 

Therefore we are going to distinguish three cases. First let us mention that if 
$\mathcal{F}$ requires an infinite expansion, the information about positiveness is intricately hidden in the coefficients $a_i$. 

The second case that we next address corresponds to the treatment of experimental data within order by order EFT; $\mathcal{F}$ is truncated to a few terms and the customary assumption that $\mathcal{F}=F^2$ is accepted. However, because the most general polynomial of degree $n$ cannot be written as a square \cite{Polya}, we briefly discuss, as a third case, the possibility that $\mathcal{F}$ is well approximated by a polynomial, but this needs to be decomposed as  $\mathcal{F}=F_1^2+F_2^2$ that holds in all generality (because it corresponds to $|F|^2$, the modulus square of a complex function). 
For the sake of simplicity, we will express $h$ in units of $v$ in the discussion of this section~\ref{sec:generic-properties}. 
\vspace{1cm}

\paragraph{$\mathcal{F}=F^2$ is assumed.}

In this situation, there are restrictions among the coefficients $a_i$ that guarantee positivity of $\mathcal{F}$. We obtain them up to fourth order, by squaring the expansion of $F$.
If the expansion ended at first order, that is, normalizing $h$ by $v$,
\begin{equation}
    F(h)=1+\alpha h \implies \mathcal{F}(h)=F^2(h)=1+2 \alpha  h+h^2 \alpha ^2
\end{equation}
we obtain the relation 
\begin{equation}
a_2\, =\,  \frac{a_1^2}{4}\,,
\label{eq:posit-F-rel-h1}
\end{equation} 
or $b=a^2$ which is exactly the correct one for $a$ and $b$ in the Standard Model.

If that relation is experimentally found to be broken, then at least one more order is necessary in the expansion of $F$. This then implies that the degree of the flare function $\mF$ must be at least two orders higher,
\begin{equation}\label{poscorrels1}
F(h)=1+\alpha_1 h +\alpha_2 h^2
\implies 
\mathcal{F}(h)=1+2 \alpha_1  h+h^2 \left(\alpha_1 ^2+2 \alpha_2 \right)+ 2 \alpha_1  \alpha_2 h^3+
\alpha_2 ^2 h^4\, ,
\end{equation}
implying two correlations: 
\begin{equation}
    2a_3=a_1\left(a_2-\frac{a_1^2}{4} \right)\, ,  \ \ \ \ \ \ \  4a_4=  \left(a_2-\frac{a_1^2}{4} \right)^2\, ,
\label{eq:posit-F-rel-h2}
\end{equation}
These new correlations~(\ref{eq:posit-F-rel-h2}) would substitute~(\ref{eq:posit-F-rel-h1}), being a smoking gun of the presence of further BSM 3-Higgs and 4-Higgs vertices in the effective Lagrangian, which could then be measured in further collider experiments.

Iterating the analysis procedure, any eventual experimental deviations from  the relations in Eq.~(\ref{eq:posit-F-rel-h2}), 
immediately imply the presence of higher order coefficients in the function $F$, and therefore also in the flare function $\mF$ that provides the $WW\to n h$ effective vertices.

To  have enough freedom to accommodate the SMEFT values of the $a_i$ coefficients in Eq.~(\ref{coefsSMEFTHEFT}) to $\mathcal{O}(\Lambda^{-2})$, the expansion of $F$ needs to be run up to fourth order in $h$,
\begin{equation}
F(h)=1+\alpha_1 h +\alpha_2 h^2 +\alpha_3 h^3 +\alpha_4 h^4 
\dots
\end{equation} 
where the Greek names of the coefficients mimic those of the expansion of $\mathcal{F}$.
The positivity conditions on the coefficients $a_i$ of $\mathcal{F}$ can be obtained from
\begin{equation}
\mathcal{F}(h)=F^2(h)=1+2 \alpha_1  h+h^2 \left(\alpha_1^2+2 \alpha_2 \right)+h^3 (2 \alpha_1  \alpha_2 +2 \alpha_3 )+
h^4 \left(2 \alpha_1  \alpha_3 +\alpha_2^2+2 \alpha_4 \right)\dots
\label{eq:Fc=F2}
\end{equation}   
and essentially leave the first four $a_i$ undetermined, while those with $i=5\dots 8$ become dependent of those earlier four. Once more, an experiment that does not respect the corresponding correlations points to a higher term in the expansion of $F$ and so on. There is a tower of positivity correlations that should be experimentally tested as multiHiggs data in the correct kinematic region becomes available.

\vspace{1cm}
\paragraph{Most general non-negative polynomial
satisfying $\mathcal{F}=F_1^2+F_2^2$.} 

In this case, the flare function $\mathcal{F}(h)$ is the most general nonnegative polynomial $\mathcal{F}(h)\geq 0$  $\forall h\in\mathbb{R}$, and therefore, even of degree $2d$,
$\mF(h)=\sum^{2d}_{n=0} a_n h^n$.

We invoke the theorem \cite{Polya} that states that such a nonnegative polynomial can be decomposed as $\mF(h)=F_1^2(h)+F_2^2(h)$ 
in terms of two polynomials\footnote{
To demonstrate it, we will note that the most general form of a polynomial of order $2d$, that is $\mF(h)=a_n\prod_{i=1}^d(h-h^\ast_i)(h-\bar{h}^\ast_i)$, can be restricted because
positivity and real analyticity demand that both the real and complex roots must be double, 
 $$\mF(h):=| F(h)|^2\;\;\text{ with } F(h)=\sqrt{a_n}\prod_{i=1}^n(h-h^\ast_i)\ ;$$
the theorem then follows from taking the real and imaginary parts, $F_1(h)=\text{Re } [F(h)]$ and $F_2(h)=\text{Im }[F(h)]$. (That the real roots are double thwarts any sign change near them and guarantees positivity).}
$F_1(h)$ and $F_2(h)$ of degree $d$.
The equivalent of~Eq.(\ref{poscorrels1}) then becomes
\begin{equation}
\mF(h)=1 + 2 (\alpha_1 \cos\theta+\beta_1 \sin\theta) h + (\alpha_1^2+\beta_1^2+2 \alpha_2 \cos\theta+2 \beta_2 \sin\theta) h^2 + 2( \alpha_1 \alpha_2+\beta_1 \beta_2 ) h^3+ (\alpha_2^2+\beta_2^2) h^4  \,,
\end{equation}
having expanded $F_1$ up to second order  with coefficients $\alpha_i$, $F_2$ with coefficients $\beta_i$, and having noted that because of the normalization of the kinetic term, $\mathcal{F}(0)=F_1(0)^2+F_2(0)^2=1$, a practical parametrization is $F_1(0)=\cos\theta$, $F_2(0)=\sin\theta$ for some angle $\theta$.
The number of free parameters is now high enough
(five, $\theta,\, \alpha_{1,2},\, \beta_{1,2}$, for four orders that depend on them) so that, in an order by order expansion of the polynomial, the correlations are weaker than in the simplified case $\mathcal{F}=F^2$ with $F$ real.

Nevertheless, one should note that if, given an experimental situation, the highest given order $h^{2d}$ has a negative coefficient $a_{2d}$, then higher orders are necessary. This can be used in experiment to detect (further) new physics. Currently, the sign of $b=a_2$ is not known; discerning whether it is positive or negative is therefore an interesting experimental analysis exercise that, if it turned out to be negative, would immediately and of necessity point out to new coefficients $a_3$ and $a_4$ (and of course, indicate new physics, since in the Standard Model $b=1$).

\subsubsection{Restrictions on the coefficients $a_i$ from the invertibility guaranteeing a SMEFT}
\label{subsec:correlations}
The restrictions over $\mathcal{F}$ at the symmetric point $h_\ast$ that guarantees the existence of a SMEFT at the end of subsection~\ref{subsec:noSMEFT} translate into conditions over the $a_i$ at the physical vacuum $h=0$, that are constrained by experiment. Here we will write down the known ones. 
Let us express the series expansion around $h=0$ setting $a_0:=1$, and take $h$ normalized to $v$, so that $v=1$,
\begin{equation}
    \mathcal{F}(h) = \sum_{i=0}^n a_i h^i\ .
\end{equation}
 
Since the conditions over $\mathcal{F}$ are taken at $h_\ast$, we reexpand around that point, and in terms of $a^\ast_j=\mF^{(j)}(h_\ast)/j!$, find
\begin{equation}
    \mathcal{F}(h) = \sum_{j=0}^n a^\ast_j (h-h_\ast)^j\ .
\end{equation}
By matching the two expansions around the two different points it is easy to read off the coefficients $a^\ast_j$ (on which the conditions over $\mathcal{F}$ are expressed) in terms of the $a_i$ (more directly accessible to experiment).  The relation reads
\begin{equation}
    a^\ast_j = \sum_{k=0}^\infty a_{k+j} h_\ast^k \cdot b_{jk}\ .
\end{equation}
The coefficients of this expansion can be recursively 
obtained,
\begin{eqnarray}
    b_{0k} = 1\ \forall k \,, \nonumber \\
    b_{jk} = \sum_{l=0}^k b_{j-1\ l} \ .
\end{eqnarray}
The closed formula that solves this recursion 
\begin{equation}
        a_\ell^\ast = A_{\ell j} a_j\,,
\end{equation}
requires the following simple auxiliary matrix
\begin{eqnarray}
A_{\ell j}= \left\{
\begin{tabular}{ccc}
    $0$ & if & $j<\ell$\,, \\ 
    $\left(\begin{array}{ccc}   j \\ \ell  \end{array}\right)  \, h_*^{j-\ell}$ &  if &  $j\geq \ell$\, . 
\end{tabular}    
\right.   
\end{eqnarray}

We can now deploy the four straightforward conditions 
$\mathcal{F}(h_\ast)=\mathcal{F}'(h_\ast)=\mathcal{F}^{'''}(h_\ast)=0$, 
$\mathcal{F}^{''}(h_\ast)=2/v^2 \to 2$ as four linear constraints on the coefficients around the physical vacuum, namely
\begin{align} \label{matchcoeffs}
&\sum_{k=0}^\infty h_\ast^k a_k \cdot b_{0k} = a_0^\ast = 0\,, \nonumber & &\sum_{k=0}^\infty h_\ast^k a_{k+1} \cdot b_{1k} = a_1^\ast = 0\,, \nonumber \\
&\sum_{k=0}^\infty h_\ast^k a_{k+2} \cdot b_{2k} = a_2^\ast = 1\,, \qquad & &\sum_{k=0}^\infty h_\ast^k a_{k+3} \cdot b_{3k} = a_3^\ast = 0 \,.
\end{align}
\paragraph{Square-matrix four-coefficient truncation}
A first possible truncation of the series is to keep the terms with the first four  $(a_1,a_2,a_3,a_4)$  coefficients (the zeroth coefficient is identically $a_0=\mF(0)=1$ by construction of the HEFT formalism so we include it in the inhomogeneous term together with the $(a_1^\ast,a_2^\ast,a_3^\ast,a_4^\ast)$ values from the conditions on $\mathcal{F}$). These are the coefficients that collect dimension-6 SMEFT corrections to the SM as shown in Eq.~(\ref{F_SMEFT6}) or (\ref{eq:F-in-SMEFT}), and the linear system becomes
\begin{equation}\label{matrixforas}
\begin{pmatrix}
h_\ast & h_\ast^2 &h_\ast^3 &h_\ast^4 \\
1 & 2h_\ast & 3h_\ast^2 & 4h_\ast^3 \\
0 & 1 & 3 h_\ast & 6 h_\ast^2 \\
0 & 0 & 1 & 4h_\ast \\
\end{pmatrix} \begin{pmatrix} a_1\\ a_2 \\ a_3\\ a_4 \end{pmatrix}=
\begin{pmatrix} -1 \\ 0 \\ 1 \\ 0  \end{pmatrix} \ .
\end{equation}
The matrix has determinant $h_\ast^4$, so that barring the zero at
$h_\ast=0$ (the physical vacuum, where the coefficients $a_i$ and $a^\ast_i$ coincide), the system has a unique solution for each $h_\ast$.  
Such solution is that of Eq.~(\ref{F_SMEFT6}), with the symmetric point of SMEFT 
$h^*=-v+v \frac{c_{H\Box}^{(6)}v^2}{3\Lambda^2} $ that of Eq.~(\ref{sympoint2}), as can be easily checked by substitution. At this order in $1/\Lambda^2$, if $h_\ast$ would take a value different from that one, there would be a one-parameter family of coefficients that would yield a valid SMEFT. \\

\paragraph{Systematic order by order truncation}
Instead of that truncation, one could be more systematic and count powers of $h$ on the left and right of  Eq.~(\ref{matchcoeffs}), so that the respective left and the right hand sides of Eq.~(\ref{matrixforas}) are of the same order, say $N$. 
In that case the system  $A\, \vec{a}_\ast =\vec{a}$ has $N$ unknowns but $(N+1)$ equations and compatibility becomes an issue. The criterion of Rouche-Frobenius guaranteeing an algebraic solution then links possible values of the  $h_\ast$ with the unknown $a_{2n}^\ast$ that can appear on the right hand side of the equivalent system.

Taking $\mF(h)$ as a polynomial of order $h^4$, this compatibility condition is, 
\begin{equation}
1\,=\,\mF(0)\,=\, \frac{h_\ast^2}{v^2} \,  + \, a_4^\ast  \, \frac{h_\ast^4}{v^4} \,.
\label{eq:rel-polh4}
\end{equation}
Without extra work, the vanishing of $a_n^\ast$ for odd $n=1,3,5...$ yields the same relation even if $\mF(h)$ is a polynomial of order $h^5$. 
For a flare function $\mF(h)$, still polynomial but now of order $h^6$ (or even $h^7$) the constraint takes one more term,  
\begin{equation}
1\,=\,\mF(0)\,= \frac{h_\ast^2}{v^2} \,  + \, a_4^\ast  \, \frac{h_\ast^4}{v^4}\,  + \, a_6^\ast  \, \frac{h_\ast^6}{v^6} \,.
\label{eq:rel-polh6}
\end{equation} 
Let us recall that a non-singular SMEFT Lagrangian requires $a_0^\ast=0$, $a_2^\ast=1$  
and $a_n^{\ast}=0$ for odd $n=1,3,5...$, as shown earlier in subsection~\ref{subsec:noSMEFT}.
It is not difficult to check that SMEFT fulfills these relations~(\ref{eq:rel-polh4}) and (\ref{eq:rel-polh6}) at $\mO(\Lambda^{-2})$ and $\mO(\Lambda^{-4})$, respectively, as that effective theory predicts (see Eqs.~(\ref{F_SMEFT8}) and~(\ref{eq:SMEFTd8-sym-point})):
\begin{eqnarray}
 \frac{ h_\ast }{v} &=&  \frac{ \mF^{-1}(0) }{v} \,=\,   -1 + \frac{c_{H \Box}^{(6)} v^2 }{3\Lambda^2}   +    \left( (c_{H \Box}^{(6)})^2  + 2 c_{H \Box}^{(8)} \right)\frac{v^4}{10\Lambda^4} +\mO(\Lambda^{-6})\,,
 \nonumber\\
a_4^\ast &=& \frac{1}{4!}\mF^{(4)}(h_\ast)\,=\, \frac{2 c_{H \Box}^{(6)} v^2 }{3\Lambda^2} +\mO(\Lambda^{-6})\, ,
\nonumber\\
a_6^\ast &=& \frac{1}{6!}\mF^{(6)}(h_\ast)\,=\, \left(\frac{44 (c_{H \Box}^{(6)})^2  }{45} +\frac{2 c_{H \Box}^{(8)}   }{5}  \right)  \frac{ v^4 }{ \Lambda^4} +\mO(\Lambda^{-6})\, , 
\end{eqnarray}
with all these properties determined by the precise form of the flare function $\mF. $

If we extended the analysis to include $a_5$ and $a_6$, which is easily done and omitted for briefness, we would have two more Lagrangian parameters but only one further constraint over $\mathcal{F}$, namely the vanishing of its fifth derivative. This means that SMEFT would have a second parametric degree of freedom that could take any value. And in fact, this is precisely the case in Eq.~(\ref{F_SMEFT8}), that depends on the additional parameter $c^{(8)}_{H\Box}$ from the dimension-8 relevant Lagrangian. \\

\paragraph{Resulting testable correlations}
Table~\ref{tab:correlations}  collects the correlations between the $a_i$ coefficients of HEFT that we have worked out at order $\Lambda^{-2}$ and $\Lambda^{-4}$ (further correlations are possible from the higher odd derivatives of $\mathcal{F}$ vanishing, and all become a bit weaker numerically if yet higher orders in $1/\Lambda$ are studied, by the need of introducing further $a_i$ coefficients).

The correlation in the first row, second column of Table~\ref{tab:correlations} originates in a  quadratic one
 $2(\Delta a_2-2 \Delta a_1) -\frac{3}{4}\left(a_3-\frac{4}{3}\Delta a_1\right)=\left(-3\Delta a_1+\frac{5}{2}\Delta a_2-\frac{9}{8}a_3\right)^2$
 with two solutions for $a_3$, a small and a large one. In keeping near the SM value $a_3=0$ we take this second one and reexpand to linearize in $a_3$ so that it can be related to $a_1$ and $a_2$ in a straightforward manner; the difference is more suppressed than $\mathcal{O}(\Lambda^{-4})$ in the SMEFT 
 expansion.

 The remarkable property of these equations is that they are independent of the SMEFT parameters $c^{(n)}_{H\Box}$, that is, they are tests of the SMEFT theory framework itself, up to a given order in $1/\Lambda$, that cannot be rewritten away in terms of its parameters.

\begin{table}
    \caption{\small Correlations between the $a_i$ HEFT coefficients necessary for SMEFT to exist, at order $\Lambda^{-2}$ and $\Lambda^{-4}$. They are given in terms of $\Delta a_1:=a_1-2=2a-2$ and $\Delta a_2:=a_2-1=b-1$. This way, all the objects in the table vanish in the Standard Model, with all the equalities becoming $0=0$. 
    Notice that the r.h.s. of each identity in the second column shows the $\mO(\Lambda^{-4})$ corrections to the relations of the first column. The third one assumes the perturbativity of the SMEFT expansion.
    \label{tab:correlations}}
    \centering
    \begin{tabular}{|c|c|c|}\hline
        \textbf{Correlations}  & \textbf{Correlations}  & 
       ${\Lambda^{-4}}$ \textbf{Assuming}  
        \\ 
        \textbf{accurate at order} $\Lambda^{-2}$ & \textbf{accurate at order} $\Lambda^{-4}$ & \textbf{SMEFT perturbativity}
        \\ \hline
        $\Delta a_2=2\Delta a_1$ &  &   $|\Delta a_2| \leq 5 |\Delta a_1|$
        \\
        $a_3=\frac{4}{3} \Delta a_1$ & 
         $\left(a_3-\frac{4}{3}\Delta a_1\right) =\frac{8}{3}(\Delta a_2-2 \Delta a_1)   -\frac{1}{3}\left(\Delta a_1\right)^2$ 
        & 
        \\  
        $a_4=\frac{1}{3} \Delta a_1$ & 
        $\left(a_4-\frac{1}{3}\Delta a_1\right) = \frac{5}{3}\Delta a_1 - 2\Delta a_2 +\frac{7}{4} a_3=  $
        & 
        those for $a_3$, $a_4$, $a_5$, $a_6$
        \\
         & $ \phantom{\left(a_4-\frac{1}{3}\Delta a_1\right)}=\frac{8}{3}(\Delta a_2-2 \Delta a_1)-\frac{7}{12}\left(\Delta a_1\right)^2  $ &
        \\ 
        $a_5=0$ &  
        $a_5 = \frac{8}{5}\Delta a_1 -\frac{22}{ 15} \Delta a_2 +a_3=$ 
        & 
        all the same 
        \\ 
         &  $\phantom{a_5}= \frac{6}{5}  (\Delta a_2-2 \Delta a_1) -\frac{1}{3}\left(\Delta a_1\right)^2  $   & 
        \\ 
       $a_6=0$ & 
        $a_6=\frac{1}{6}a_5$
        &
         \\ 
        \hline
    \end{tabular}
\end{table}
 
These equations can be experimentally tested looking for the consistency of SMEFT.
Given tight experimental bounds on $a_1$, these relations (and those from $\mathcal{F}\geq 0 $) can already predict how the next HEFT coefficients will look like if SMEFT is valid.
This we will delay until subsection~\ref{subsec:numeritos} below.

The $1/\Lambda^2$ relations in the first column of Table~\ref{tab:correlations}, all hanging from $\Delta a_1$, are rather constraining given that one-Higgs production is well known. Those in the second column, as they depend also on $\Delta a_2$, which is much less well measured, are not very useful; but they can be further tightened by imposing perturbativity of the SMEFT expansion. \\

\paragraph{Perturbativity constraints} Perturbativity can be deployed by recalling that, at $\mathcal{O}(\Lambda^{-4})$,
\begin{eqnarray}
a_1 &=& 
  \left( 2 + 2 \frac{ c_{H \Box}^{(6)} v^2}{\Lambda^2} 
  + 3 \frac{(c_{H \Box}^{(6)})^2 v^4}{\Lambda^4}  + 
      2 \frac{c_{H \Box}^{(8)} v^4}{\Lambda^4}  \right)  \nonumber \\
a_2 &=&  \left( 1 + 
   4 \frac{ c_{H \Box}^{(6)} v^2}{\Lambda^2} + 
   12 \frac{(c_{H \Box}^{(6)})^2 v^4}{\Lambda^4} + 
    6 \frac{c_{H \Box}^{(8)} v^4}{\Lambda^4}  \right) \ .
\end{eqnarray}
For clarity, let us shorten notation for the rest of the paragraph, writing
\begin{eqnarray} \label{shorthanda1a2}
 \Delta a_1 &=& 2x +3x^2 +2y \nonumber \\
 \frac{\Delta a_2}{2} &=& 2x + 6x^2 +3y = \Delta a_1 +3x^2 + y\ . 
\end{eqnarray}
In general, there are two free parameters, $x$ and $y$. What perturbativity suggests is that each of the terms of the $\mathcal{O}(\Lambda^{-4})$ should not be larger than the $\mathcal{O}(\Lambda^{-2})$ term (this is akin to the Cauchy criterion for convergence of a sequence, but of course there is no guarantee that it will be satisfied at a fixed order; again, it is only a perturbativity argument, similar to the one in~\cite{Hays:2018zze}). 
Taking this at face value, it must be that $3x^2\leq 2|x|$  (by the way, this means that $|x|\leq 2/3$, that however is of little value as experimental constraints are much tighter) and that $|y|<|x|$. 

Returning to the first of Eq.~(\ref{shorthanda1a2}) and separately analyzing the positive and negative $x$ cases, we find
\begin{equation}
    |x|< {\rm max}\left(|\Delta a_1^-|,\frac{1}{2}|\Delta a_1^+|\right)
\end{equation}
and noting that half the upper 95\% uncertainty $\Delta a_1^+/2$ is larger than the lower one $\Delta a_1^-$ as discussed around Table~\ref{tab:correl-exp-bounds} below, leads us to 
\begin{equation} \label{corra2_c}
    |\Delta a_2| \leq 2|\Delta a_1| + 2\cdot \frac{3}{2} |\Delta a_1^+|\implies
    |\Delta a_2| \leq 5 |\Delta a_1^+|\ ,
\end{equation}
relation which we elevate to the third column of Table~\ref{tab:correlations}, in the understanding that the uncertainty there is the maximum ($+$) of the two asymmetric uncertainties, and where the absolute value bars have been at last dropped.

In the order in which experimental data can be used,

\begin{itemize}
\item  A nonzero measurement of $\Delta a_2$ signals new physics. SMEFT or HEFT are needed.
\item If additionally the stronger correlation $\Delta a_2 = 2 \Delta a_1$ is violated,  
severe corrections to $1/\Lambda^2$ SMEFT are suggested.
\item If the weaker correlation in Eq.~(\ref{corra2_c}), $\Delta a_2 \leq 5 \Delta a_1$  is violated,
those correlations make SMEFT unnatural and put its perturbative use into question but they do not necessarily rule it out as discussed in the next paragraph.
\item If the weakest correlation in the second column of Table~\ref{tab:correlations} is broken, the first two orders of SMEFT do not make much sense and the theory is falsified for all practical purposes.
\end{itemize}

To close this subsection, we note that the presence of the zero (and minimum) of $\mathcal{F}$ at $h^\ast$ is a distinguishing property in the TeV region, for near-threshold physics the Higgs potential $V(h)$ is also important. The SMEFT potential needs to be analytic too so that a power-expansion makes sense. The relevant theory regarding $V$ is  briefly discussed in Appendix~\ref{coeffspotential}.

\subsubsection{Unitarity imposes no constraint on the coefficients, causality may}
It has recently been proposed that unitarity violations in the effective theory could be used to describe the space of theories that can be characterized as HEFT but that, due to non-analyticities, can not be brought up to SMEFT form~\cite{Cohen:2021ucp}. 
While this may deserve further study, we are not very sure about that program. 

The reason is that the HEFT Lagrangian yields a properly Hermitian Hamiltonian, and therefore a unitary scattering matrix. Truncating an expansion of a partial wave amplitude in perturbation theory is indeed a procedure that violates unitarity, but this has nothing to do with the theory itself, but with the truncation. For example, in the well-known case of two-body elastic scattering one can, instead of the partial wave amplitude, expand first the inverse partial wave amplitude to one loop in the EFT
\begin{equation}
    \frac{1}{t^{IJ}} = \frac{1}{t^{IJ}_0+t^{IJ}_1}
\end{equation}
and then invert back to obtain 
\begin{equation}
    t^{IJ} \simeq \frac{(t_0^{IJ})^2}{t^{IJ}_0-t^{IJ}_1}\ .
\end{equation}
This expansion of the inverse amplitude, that can be carried out order by order, 
can also be derived from a dispersion relation, so it satisfies all analyticity properties expected from an elastic scattering amplitude. 
Additionally, elastic unitarity over the physical cut of the amplitude is exact,
no matter how strong the interaction, as long as the low-energy theory has the structure of HEFT (or Chiral Perturbation Theory or other similar theories with derivative couplings). 
This has been documented at length in the literature\cite{Dobado:1989gr,Delgado:2013loa,Espriu:2014jya,Delgado:2015kxa,Corbett:2015lfa,Salas-Bernardez:2022xk} so we will not delve any longer on the issue here. The point is that the failure (or not) of unitarity is not really about the theory, whether SMEFT, HEFT or another, but about the way to treat it to obtain observables. This is an ancient observation dating at least to the Effective Range Expansion~\cite{Bethe:1949yr} that needs to be discussed more often in  the context of high-energy physics.

On the other hand, causality does impose limits on the parameters of an effective Lagrangian, though they have not been very thoroughly studied and perhaps we will attempt this in future work. These come about because a scattered wave packet in the forward direction cannot precede the incoming wave packet (though this is possible at wide angles~\cite{Llanes-Estrada:2019ktp}). Perturbatively, Wigner's bound for the derivative of the phase shift of any partial wave $\delta_J$ respect to the centre-of-mass three-momentum $k$, in terms of the scatterer's radius $R$ is a well-known low-energy result~\cite{Pelaez:1996wk}, 
\begin{equation}
\frac{d\delta_{J}}{dk} \geq -R \,.
\end{equation}
However, what should be used
for $R$ in a relativistic scattering theory is less well understood.
Such a set of bounds on the scattering matrix (one for each of its partial wave projections) yields one-sided bounds on the $a_i$ coefficients.
Employing unitarized methods one can immediately set constraints by demanding that no poles of the amplitude lay on the first Riemann sheet~\cite{Espriu:2014jya,Delgado:2015kxa} of $s$, which also violate causality. But these poles typically fall in regions where the uncertainties of the unitarized amplitude~\cite{Salas-Bernardez:2020hua} are large. In all, we think that this deserves a separate investigation.

\subsection{Example functions to illustrate HEFT vs SMEFT differences}
\label{subsec:examples}

Let us illustrate the whole discussion with a few simple example functions 
(as opposed to the more ambitious construction of entire UV completions shown in \cite{Cohen:2020xca,Grinstein:2007mp}).

\subsubsection{Example flare functions $\mathcal{F}$ where SMEFT is applicable}

A couple of examples of HEFT flare functions that lead to regular SMEFT Lagrangians are:

\begin{itemize}
    \item The SM has $\mathcal{F}(h)=(1+h/v)^2$ that of course is analytic, possesses a zero at $h_\ast=-v$ and trivially fulfills all correlations in Table~\ref{tab:correlations} since $\Delta a_1=0=\Delta a_2$, $a_i=0\ \forall i>2$.
    \item The Minimally Composite Higgs Model with symmetry breaking pattern $SO(5)/SO(4)$(Ref.~\cite{Contino:2011np}), with 
    $\mathcal{F}(h)=\frac{f^2}{v^2}\sin^2\left(\frac{h}{f} +\arcsin{\frac{v}{f}}\right)$, which expanded to fourth order in $h/v$ and second in $v/f$ yields~\footnote{
    Up to $\mO(v^4/f^4)$, the flare function is given by the polynomial 
        $
    \mF(h)=1 + \left(2-\frac{v^2}{f^2}- \frac{v^4}{f^4}\right)\frac{h}{v} 
    +  \left(1-\frac{2 v^2}{f^2}\right)\frac{h^2}{v^2} 
    +  \left(-\frac{4 v^2}{3 f^2}+ \frac{2v^4}{3f^4}\right)  \frac{h^3}{v^3}  
    +  \left(-\frac{ v^2}{3 f^2}+ \frac{2v^4}{3f^4}\right) \frac{h^4}{v^4}  
    +  \left(\frac{4 v^4}{15 f^4}\right)  \frac{h^5}{v^5}  
    +  \left(\frac{ 2 v^4}{45 f^4}\right) \frac{h^6}{v^6}\, .$  
    This result is fully  consistent with  the $\mO(\Lambda^{-4})$ SMEFT flare function  in Eq.~(\ref{F_SMEFT8}) for the relations $c_{H\Box}^{(6)}= -\Lambda^2/(2f^2)$, 
    $c_{H\Box}^{(8)} = - \Lambda^4/(2f^4)$. 
    }:
 \begin{eqnarray}
    \mF(h) &=& 1 + \left(2-\frac{v^2}{f^2}\right)\frac{h}{v} 
    +  \left(1-\frac{2 v^2}{f^2}\right)\frac{h^2}{v^2} 
    -  \frac{4 v^2}{3 f^2} \frac{h^3}{v^3} 
    -  \frac{v^2}{3 f^2} \frac{h^4}{v^4} \, .
    \end{eqnarray}    
    It is easy to observe that this is a particular case of the SMEFT flare function at $\mO(\Lambda^{-2})$ in Eq.~(\ref{eq:SMEFT-F-Lambda2}) after the identification 
    $c_{H\Box}  =  -\Lambda^2/(2f^2)$. 
\end{itemize}

\subsubsection{Example flare functions $\mathcal{F}$ where SMEFT is not applicable}

 Examples of HEFT Lagrangians that transform to non-regular SMEFT Lagrangians are given by the models with $\mathcal{F}=e^{2h/v}$ or $\mathcal{F}= 1+\frac{1}{2}\sin(4h/v)$. Such models fail to have a zero of $\mathcal{F}$, in such a way that the behaviour of its root function $F[h]=(h-h^*)/v$ is fulfilled by no real value of $h^*$ (there is no zero).

However, as seen in Section \ref{Geometricsection}, there is more than this condition in order to have an appropriate SMEFT Lagrangian in terms of $H$: we illustrate this with the dilatonic model~\cite{Halyo:1991pc,Goldberger:2007zk,Hernandez-Leon:2017kea}, that has a HEFT function $\mathcal{F}=(1+ a h/v)^2 $ that does present a zero at $h^*= - v/a$. Nonetheless, the corresponding SMEFT Lagrangian~(\ref{eq:SMEFT-from-HEFT}) happens to be singular for $a\neq 1$, with a pole at $H^{\dagger} H=0$:
\begin{align}
\mathcal{L}_{\text{SMEFT}}&= \mathcal{L}_{\text{SM}} +  
\frac{1}{2} \bigg[ \bigg(   \frac{1}{v} (F^{-1})'(1+h/v)  \bigg)^2 -1\bigg] \, (\partial h)^2 
\nonumber=\\
&=\, \mathcal{L}_{\text{SM}} + \frac{1}{2} \frac{(1-a)}{a} (\partial h)^2 
= \, \mathcal{L}_{\text{SM}} + \frac{1}{2} \frac{(1-a)}{a}  \, \frac{(\partial|H|^2)^2}{2 |H|^2} \, .\label{eq:dilatonexample}
\end{align}

It might be tempting to consider that the divergence of the second term in the second line in Eq. (\ref{eq:dilatonexample}) could be cured and removed by an appropriate rescaling of $h$, but this would disarray the operators in $\mathcal{L}_{\rm SM}$, which would not come together anymore to conform $\mathcal{L}_{\rm SM}$. In this case, it happens that there is a zero in $F(h)=1 + \frac{a h}{v}$ at $h^*=-\frac{v}{a}$ but the slope of $F$ is not $\frac{1}{v}$ but rather $F'(h^*)=\frac{a}{v} \neq \frac{1}{v}$ for $a\neq 1$.  

From a completely different approach, based on the phenomenology of the effective couplings, we could observe that the dilaton is not compatible with the SMEFT expansion, since SMEFT --in Eq.~(\ref{coefsSMEFTHEFT})-- predicts $\Delta b= 4 \Delta a$ up to $1/\Lambda^4$ NNLO corrections \cite{Agrawal:2019bpm,Sanz-Cillero:2017jhb}, while the dilatonic model predicts that we should be observing $\Delta b= 2 \Delta a$~\cite{Halyo:1991pc,Goldberger:2007zk,Hernandez-Leon:2017kea}, 
with $\Delta a\equiv a-1,\, \Delta b\equiv b-1$.  
The only way SMEFT could be able to reproduce the ``dilatonic data'' is through a 100\% correction from operators of dimension-8 and greater, indicating a breakdown of the $1/\Lambda$ expansion.

\subsubsection{Example of potentials $V$ where SMEFT is applicable}
Next, we propose two Higgs-Higgs self-interaction potentials that lead to regular SMEFT Lagrangians, for example

\begin{itemize}
    \item The SM potential (with $\lambda, -\mu^2$ both positive)
    is given by  
    \begin{equation}
    V_{\text{SM}}(H)=\mu^2H^\dagger H+\lambda (H^\dagger H)^2\, ,
    \end{equation}
    
    which, in HEFT coordinates, becomes\footnote{In this case, the correlations of table~\ref{tab:corV} in appendix~\ref{coeffspotential} below are trivially satisfied, because the variables there defined $\Delta v_3=\Delta v_4=\dots 0$ all vanish.} 
    \begin{equation} 
    V_{\text{SM}}(h)=\frac{m_h^2}{2}\left({h^2} +\frac{h^3}{v}+\frac{h^4}{4v^2}\right) \,, 
    \end{equation}
    with $v^2=-\mu^2/\lambda$ and $m_h^2=2 |\mu|^2$.

    \item The typical potential with the correlations obtained there in Appendix~\ref{coeffspotential} will need to have an expansion which, up to $\mathcal{O}(\Lambda^{-2})$ in SMEFT needs to have the form
\begin{align}
V(h) = \frac{m_h^2}{2}&\bigg[h^2+\frac{h^3}{v} \left({1}+\epsilon\right)+\frac{h^4}{v^2} \left(\frac{1}{4}+\frac{3}{2}\epsilon \right)+\frac{3\epsilon}{4}\frac{h^5}{v^3}+\frac{\epsilon }{8}\frac{ h^6}{v^4}\bigg]\, .
\label{eq:SMEFT-V-correlations}
\end{align}
It is possible to see that including the custodial-invariant SMEFT operator without derivatives, $\mathcal{O}_H$, in Eq. (\ref{Warsawbasis}) one gets the potential,
\begin{equation} 
V_{\rm SMEFT}(H)=\mu^2H^\dagger H+\lambda (H^\dagger H)^2 -  \frac{c_H}{\Lambda^2} (H^\dagger H)^3\, ,
\label{eq:V_SMEFT}
\end{equation}
which reproduces the structure of the coefficients in Eq.~(\ref{eq:SMEFT-V-correlations}). By expanding $H$ around its minimum, the SMEFT potential in HEFT coordinates, 
finally produces the structure in  Eq.~(\ref{eq:SMEFT-V-correlations})
with  $m_h^2=\, -2\mu^2 \left(1+ 3\epsilon/4\right)$ and $2\langle |H|^2\rangle =v^2= v_0^2\left(1-3\epsilon/4\right) $, where we made use of the lowest order vev $v_0^2=\, -\mu^2/\lambda$ and the $\mO(\Lambda^{-2})$ correction $\epsilon= - 2c_H v^4/m_h^2\Lambda^2= \, \mu^2 c_H    /(\lambda^2  \Lambda^2)$. Notice that, for sake of clarity in the illustration, here we have taken $c_{H\Box}=0$, so there is no Higgs field renormalization. (Notice also that treating only terms in the potential, \textit{i.e.} non-derivative couplings implies, up to a constant shift, $h=h_1$) 
\end{itemize}

\subsubsection{Example of potentials $V$ where SMEFT is not applicable}

An example of a potential which can not be written as a SMEFT is
\begin{equation}
V(H) =V_{\text{SM}}(H)+ \frac{\varepsilon}{H^\dagger H}
\end{equation}
with $\varepsilon$ a constant small enough so as to avoid unsettling the potential away from $h=0$ by a finite fraction of $v$ 
now there is no symmetric $O(4)$ point where the function is analytic, there is a divergence at the origin. 
Consistently with the symmetric-point criterion, SMEFT cannot be used: this model does not reproduce Eq.~(\ref{eq:SMEFT-V-correlations}).

\section{$ww\to n\times h$ for all $n$ in HEFT as the telltale process: \\ extraction of $\mathcal{F}(h)$ expansion coefficients}
\label{nhproduction}

In this section we will indicate how to extract the coefficients of the flare function $\mathcal{F}$ in a process where $n$ Higgses are produced in the final state.

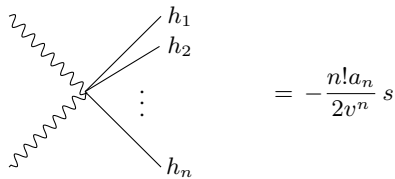
\begin{figure}[ht!]
\centering
     \begin{tikzpicture}[scale=1]
     \draw[decoration={aspect=0, segment length=1.8mm, amplitude=0.7mm,coil},decorate] (-1,1) -- (0,0)-- (-1,-1);
     \draw[] (0,0)-- (1,1);
     \draw[] (1.25,1) node {$h_1$};
          \draw[] (1.25,-1) node {$h_n$};
     \draw[] (0,0)-- (.98,0.6);
          \draw[] (1.25,0.6) node {$h_2$};
     \draw[] (.75,-.15) node {.};
     \draw[] (.75,0.) node {.};
     \draw[] (.75,-0.3) node {.};
     \draw[] (0,0)-- (1,-1);
       \draw[] (3.25,0) node {$\displaystyle{  \, =\,  -\frac{n!a_n}{2v^n} \,  s }$};  
\end{tikzpicture}
\caption{The $a_i$ coefficients of the flare function $\mathcal{F}$ control the contact piece of $\omega\omega\to nh$ processes. A large number $n$ of Higgs bosons in the final state would appear as a flare of them in the detector read out, whence the nickname of the function.}
\end{figure}

First we start by noticing that the measurement of the $\omega^+\omega^-\to h$ total cross section  gives us information the value of the first nontrivial coefficient of $\mF(h)$, $a_1=2a$. The value of $a$ is well constrained and hence we move on to identify the processes where the subsequent coefficients of the flare function can be measured.

Generalizing to $n>1$ Higgs bosons in the final state, the contributions to the amplitude will come from the contact diagram and the $t$-channel and $u$-channel diagrams. The contact diagram will give a contribution of $n! s a_n/(2v^n)$ whereas the $t/u$-channel will produce a string proportional to all the coefficients of $\mathcal{F}(h)$, $a_m$, for $1\leq m\leq n-1$. So that, for generic $n$, the amplitude will take the form
\begin{equation}
    T_{\omega\omega\to n \times h}=\frac{s}{v^n}\sum _{i=1}^{p(n)}\left(\psi_i(q_1,q_2,\{p_k\})\prod_{j=1}^{|\text{IP}[n]_{i}|}a_{\text{IP}[n]_{i}^j}\right)\;,
\end{equation}
where $\psi_i(q_1,q_2,\{p_k\})$ are functions depending on all four-momenta involved in the process (the two Goldstone bosons having momenta $q_1$ and $q_2$ and the $k$-th Higgs boson with momentum $p_k$) which will be made explicit below. These functions contribute to the angular integration used to obtain the total cross section of the process.  The symbol $\text{IP}[n]$ represents the integer partitions of $n$ and it is a collection of $p(n)$ vectors with length $|\text{IP}[n]_i|$ each, and components $\text{IP}[n]_i^j$. For example, for $n=4$
(see Eq.~(\ref{amp:4h}) given shortly), $\text{IP}[4]=\{\{4\},\{3,1\},\{2,2\},\{2,1,1\},\{1,1,1,1\}\}$ and hence $p(4)=4$, $|\text{IP}[4]_i|=\{1,2,2,3,4\}$ and $\text{IP}[4]_2^1=3$. In that case the amplitude takes the form
\begin{equation}
    T_{\omega\omega\to 4 \times h}=\frac{s}{v^4}\left(4!a_4+a_3a_1\psi_2(q_1,q_2,\{p_k\})+a_2^2\psi_3(q_1,q_2,\{p_k\})+a_1^4\psi_4(q_1,q_2,\{p_k\})\right)\;.
    \end{equation}

The strategy is to fit to data each $a_n$ with increasing $n$ starting form the one-Higgs boson production, then fit two-Higgs boson production, etc. We have developed a small program for the computation of the amplitudes $T_{\omega\omega\to n\times h}$ that can be provided by the authors on request. We present in the next subsection~\ref{subsec:amplitudes} the amplitudes for the production of one, two, three and four Higgs bosons.

\subsection{Amplitudes of $\omega\omega \to n\times h$ with $n=1,2,3,4$} 
\label{subsec:amplitudes}
Formally, the amplitude $\omega\omega \to h$ with the LO HEFT Lagrangian in Eq.~(\ref{FbosonLagrangianLO}) is 
given by
\begin{equation} \label{amp:h}
T_{\omega\omega\to h}=-\frac{a_1 s}{2 v}\;.
\end{equation}
There is no on-shell cross section associated to this amplitude (because of the impossibility to satisfy four-momentum conservation with three on-shell massless particles). The amplitude cannot be used off-shell because the Lagrangian of the EFT has been constructed on-shell.
Therefore we move on and quote the amplitude with two Higgs bosons in the final state, that is simply~\cite{Delgado:2015kxa}
\begin{equation} \label{amp:2h}
 T_{\omega\omega\to hh} \,=\, \frac{s}{v^2}(a^2-b) \, = \frac{s}{v^2}\left(\frac{a_1^2}{4}-a_2\right),
\end{equation}
but it will be useful to introduce some  notation to systematize what follows and give it in a more involved way: 
\begin{equation}
    T_{\omega\omega\to hh}=\frac{s}{v^2}\left(  a_1^2 \frac{( (z_1-2)f_1+(z_2-2)f_2 +2)}{4} -  a_2\right) 
\end{equation}
where we define, in the rest frame, the three-momentum fractions $f_i\equiv ||\vec{p}_i||/\sqrt{s}$ ($s=4||\vec{q}_1||^2$) for each Higgs boson; the angular functions $z_i\equiv 2\sin^2(\theta_i/2)$ with $\theta_i$ being the angle between the $i$-th Higgs boson and the first $\omega$ Goldstone boson momenta, ${\vec{q}}_1$ (that is, $z_1=1-\cos\theta$, $z_2=1+\cos\theta$ as usual in a two-body problem with $t$ and $u$ channels). 
We also define $z_{ij}\equiv 2\sin^2(\theta_{ij}/2)$,  $\theta_{ij}$ being the angle between the $i$-th and $j$-th Higgs bosons.

With this notation, the tree-level amplitude with a larger number of Higgs bosons
can be obtained (by automated means); the one with
three Higgs bosons in the final state is relatively manageable even when given in full,
\begin{align} \label{amp:3h}
    T&_{\omega\omega\to hhh}= -\frac{s}{8 v^3} \Bigg({\bf{a_{1}^3}} \Big[4 f_1 f_3^2 \left(\frac{z_{23} (f_1 z_{23}-1)}{f_3 (z_3-2 f_1 z_{23})+f_2 z_2}+\frac{z_{13} (f_1 z_{13}-1)}{f_1 (z_1-2 f_3 z_{13})+f_3 z_3}\right)\nonumber+\\
    &+2 f_3 \left(f_1 \left(\frac{z_{23}-2 f_2 z_{23}}{-2 f_1 f_3 z_{23}+f_2 z_2+f_3 z_3}+\frac{z_{13}-2 f_1 z_{13}}{-2 f_1 f_3 z_{13}+f_1 z_1+f_3 z_3}+z_{13}+z_{23}\right)+3 (z_3-2)\right)+\nonumber\\
    &+\frac{2 f_1 f_2 z_{12} (2 f_1 (f_2 z_{12}-1)-2 f_2+1)}{f_1 (z_1-2 f_2 z_{12})+f_2 z_2}+2 f_1 (f_2 z_{12}+3 z_1-6)+6 f_2 z_2-12 f_2+9\Big]+\nonumber\\
    &+{\bf{4 a_{1} a_{2}}} \Big[\frac{f_1^2 \left(2 z_1 (-2 f_2 z_{12}+f_3 (z_{13}+z_{23})-3)-4 f_2 z_{12} (f_3 (z_{13}+z_{23})-2)+3 z_1^2\right)}{2 f_1 f_2 z_{12}-f_1 z_1-f_2 z_2}+\nonumber\\
    &+\frac{2 f_1 f_2 (-2 f_2 z_{12} (z_2+1)+z_2 (f_3 (z_{13}+z_{23})+3 z_1-3)+z_{12})+3 f_2^2 z_2^2}{2 f_1 f_2 z_{12}-f_1 z_1-f_2 z_2}+6 (f_2+f_3-1)-\nonumber\\
    &-\frac{2 f_1 f_3 z_{23} (2 f_3 (f_1 z_{23}-1)-2 f_2+1)}{f_3 (z_3-2 f_1 z_{23})+f_2 z_2}-\frac{2 f_1 f_3 z_{13} (2 f_1 (f_3 z_{13}-1)-2 f_3+1)}{f_1 (z_1-2 f_3 z_{13})+f_3 z_3}-3 f_3 z_3\Big]+{\bf{24 a_{3}}}\Bigg)\;.
\end{align}

The amplitude with four Higgs bosons in the final state is complicated enough that it is worth to quote only one of the terms, corresponding to the ordering $(p_1,p_2,p_3,p_4)$
of the four $h$ momenta in the final state, with the other 23 permutations of these momenta not given.  This one term reads
\begin{align} \label{amp:4h}
   & T_{\omega\omega\to hhhh}=\frac{s}{16 v^4}\times \nonumber\\
   &\times\Bigg({\bf{a_{1}^4}}\bigg[\frac{  (2 f_{1} (z_{1}-f_{2} z_{12})+f_{2} z_{2})(2 f_{1} (-2 f_{2} z_{12}-f_{3} z_{13}+z_{1})+2 f_{2} (z_{2}-f_{3} z_{23})+f_{3} z_{3})}{(f_{1} (z_{1}-2 f_{2} z_{12})+f_{2} z_{2}) (f_{1} (z_{1}-2 (f_{2} z_{12}+f_{3} z_{13}))+f_{2} (z_{2}-2 f_{3} z_{23})+f_{3} z_{3})}\times\nonumber\\
   &\times (f_{1} (z_{1}-2 (2 f_{2} z_{12}+2 f_{3} z_{13}+f_{4} z_{14}))+f_{2} (-4 f_{3} z_{23}-2 f_{4} z_{24}+z_{2})-2 f_{3} f_{4} z_{34}+f_{3} z_{3}+1)\bigg]+\nonumber\\
   &+ {\bf{2a_1^2a_2}}\bigg[ \frac{(f_1 z_1+f_2 z_2) (2 f_1 (-2 f_2 z_{12}-f_3 z_{13}+z_1)+2 f_2 (z_2-f_3 z_{23})+f_3 z_3)}{(f_1 (z_1-2 f_2 z_{12})+f_2 z_2) (f_1 (z_1-2 (f_2 z_{12}+f_3 z_{13}))+f_2 (z_2-2 f_3 z_{23})+f_3 z_3)}\times\nonumber\\
   &\times (f_1 (z_1-2 (2 f_2 z_{12}+2 f_3 z_{13}+f_4 z_{14}))+f_2 (-4 f_3 z_{23}-2 f_4 z_{24}+z_2)-2 f_3 f_4 z_{34}+f_3 z_3+1) +\nonumber\\
   &+ \frac{(f_1 (z_1-2 (2 f_2 z_{12}+2 f_3 z_{13}+f_4 z_{14}))+f_2 (-4 f_3 z_{23}-2 f_4 z_{24}+z_2)-2 f_3 f_4 z_{34}+f_3 z_3+1)}{f_1 (z_1-2 (f_2 z_{12}+f_3 z_{13}))+f_2 (z_2-2 f_3 z_{23})+f_3 z_3}\times\nonumber\\
  & \times (2 f_1 (-f_2 z_{12}-f_3 z_{13}+z_1)+f_2 z_2+f_3 z_3)-\frac{(2 f_{1} (z_1-f_{2} z_{12})+f_{2} z_{2})}{f_{1} (z_1-2 f_{2} z_{12})+f_{2} z_{2}}\times\nonumber\\
  &\times(f_{1} (z_1-2 (2 f_{2} z_{12}+f_{3} z_{13}+f_{4} z_{14}))+f_{2} (-2 f_{3} z_{23}-2 f_{4} z_{24}+z_{2})+1) \bigg]+{\bf{4a_1a_3}}\times\nonumber \\
  &\times\bigg[(f_{1} (z_{1}-2 (f_{2} z_{12}+f_{3} z_{13}+f_{4} z_{14}))+1)+ \frac{(f_{1} z_{1}+f_{2} z_{2}+f_{3} z_{3}) }{f_{1} (z_{1}-2 (f_{2} z_{12}+f_{3} z_{13}))+f_{2} (z_{2}-2 f_{3} z_{23})+f_{3} z_{3}}\times\nonumber\\
 & \times(f_{1} (z_{1}-2 (2 f_{2} z_{12}+2 f_{3} z_{13}+f_{4} z_{14}))+f_{2} (-4 f_{3} z_{23}-2 f_{4} z_{24}+z_{2})-2 f_{3} f_{4} z_{34}+f_{3} z_{3}+1)\bigg]+\nonumber\\
 &+{\bf{4 a_{2}^2}}\frac{ s (f_{1} z_{1}+f_{2} z_{2}) (f_{1} (z_{1}-2 (2 f_{2} z_{12}+f_{3} z_{13}+f_{4} z_{14}))+f_{2} (-2 f_{3} z_{23}-2 f_{4} z_{24}+z_{2})+1)}{f_{1} (z_{1}-2 f_{2} z_{12})+f_{2} z_{2}}-\nonumber\\
 &-\bf{8a_4}\Bigg)
   +\text{perm.}
\end{align}
and the 23 permutations of the 4 final-state Higgs momenta ${{\bf p}_i}$ are to be taken in the computer code by invoking the amplitude with exchanged arguments. Permuting the $i$-th Higgs with the $j$-th Higgs will interchange $z_i\leftrightarrow z_j$, $f_i\leftrightarrow f_j$ and $z_{ik}\leftrightarrow z_{jk}$. (The indices of the $a_i$ coefficients are of course not to be permuted, as they correspond to the terms in the Lagrangian, not the external boson legs).

A check of these  amplitudes is to take the limit to the Standard Model
by setting the $a_i$  coefficients to the values 
$a_1=2$, $a_2=1$, $a_3=0$ and $a_4=0$.
Because the SM is renormalizable and unitary, these derivative terms must vanish, as indeed our computation reproduces,
having Eq.~(\ref{amp:2h}) and Eq.~(\ref{amp:3h}) above as well as Eq.~(\ref{amp:4h}) satisfy
\begin{equation}
 T^{SM}_{\omega\omega\to hh} 
 =0\; ;\ \ \
 T^{SM}_{\omega\omega\to hhh}    
 =0\; ;\ \ \ T^{SM}_{\omega\omega\to hhhh} 
 =0\; ,
\end{equation}
where conservation of momentum has been used.

\subsection{Cross-sections}

Equations~(\ref{amp:h})-(\ref{amp:4h}) and successive for an increasing number of Higgs bosons are what is needed for a phenomenological extraction of the $a_i$ coefficients in the TeV region.
From single Higgs production, through Eq.~(\ref{amp:h}),
$a_1$ is already constrained (see subsection~\ref{subsec:numeritos}), so current work focuses on two-Higgs processes which allows to address $a_2=b$ in Eq.~(\ref{amp:2h}). The $a_1$ appears squared (and is known to 10\% precision) and $a_2$ appears linearly, interference in this latter amplitude is possible and the sign of the deviations of $a_2$ from the SM value is at hand.

With $a_1$ and $a_2$ already constrained, it would become feasible to in turn constrain $a_3$ (null in the Standard Model) with Eq.~(\ref{amp:3h}) and so forth for higher coefficients with higher-point processes with more bosons in the final state. Since each successive amplitude is linear in the highest appearing coefficient, their signs can be determined if a separation from the SM value is found.

An important correlation that allows to ascertain whether SMEFT is at play comes from the observation that at order $1/\Lambda^{2}$ all the deviations from the SM in $a_1$ through $a_4$ stem from the same operator (see Eq.~(\ref{coefsSMEFTHEFT})). Note also that the amplitudes in subsec.~\ref{subsec:amplitudes} are the net deviations from the Standard Model in HEFT, since their SM prediction is zero.
Then, all those amplitudes are necessarily linear in the same Wilson coefficient
\begin{equation}
     T_{\omega\omega\to nh} \propto \left( \frac{s}{v^{n-2}\Lambda^2}\right) c_{H\Box} \ \ \ {\rm in\ SMEFT\ \ up\ \ to\ \ } \mathcal{O}\left( \Lambda^{-2}\right)\ .
\end{equation}
This means that taking ratios of cross-sections, the only parameter that encodes BSM physics in SMEFT in the relevant TeV energy region drops out, and what remains is a pure prediction, independent of the BSM physics scale, but dependent only on the structure of SMEFT
\begin{equation}
   \boxed{ \frac{\sigma(\omega\omega\to nh)}{\sigma(\omega\omega\to mh)} = {\text{ independent\ of }}\ c_{H\Box}}\ .
\end{equation}

The ratios become weakly dependent on the value of the parameter when order-$1/\Lambda^{4}$, dimension-8 terms are included, as seen in Eq.~(\ref{F_SMEFT8}). But if the SMEFT counting is sensible, this should be small and a reasonable prediction is possible. It will be explored numerically in a follow-up document. This can be a way of distinguishing whether SMEFT is  applicable or not, from ``low'' energy data, without access to the underlying UV completion of any new physics.

\section{Finding out whether the $\mathcal{F}(h)$ function has a zero}\label{ZeroesSection}

Among the precise conditions that allow to express a HEFT as a SMEFT,  thoroughly studied in \cite{Cohen:2020xca}, the first necessary requirement among those spelled out in subsection~\ref{subsec:noSMEFT}
is the existence of an $O(4)$ symmetric point $h=h_\ast$. 
This requires a zero, that recalling the Taylor expansion in Eq.~(\ref{expandF}) yields the relation
$$
\mathcal{F}(h_\ast)=1+\sum_{n=1}^{\infty}{a_n}\frac{h_\ast^n}{v^n}=0\;.
$$

In this section we will try to address what can be done, empirically and assuming that any UV physics is not known or understood (bottom-up approach) to improve the knowledge of whether such zero $h_\ast$ could be present.

\subsection{Finding the $O(4)$ fixed point candidate by looking at the polynomial approximation of $\mathcal{F}(h)$} 

\subsubsection{From one or two Higgs production}

Knowledge of the $a_i$ coefficients is rapidly evolving, as they directly correspond to the $\kappa_i$ scaling cross sections respect to the Standard Model ones.
A data-driven constraint for $a_1$  based on LHC run-I data can be found in~\cite{Brivio:2016fzo}; at 2$\sigma$, those authors conclude that $a\in [0.7,1.23]$.
A bound on $b$ was originally  obtained by examining the absence of a resonance in $W_L W_L$ scattering below 700 GeV \cite{Delgado:2014dxa} (according to \cite{Salas-Bernardez:2020hua}, the dispersive methods used for obtaining these bounds have a 10-20\% uncertainty on the position of the resonance).
Direct ATLAS and CMS work has improved those earlier limits, and
the latest bounds on the first two $a_i$ coefficients are discussed next in subsection~\ref{subsec:numeritos}; those coefficients $a_1=2a$ and $a_2=b$ remain the only ones  with current experimental constraints.

\begin{figure}[!t]
\centering
\makebox[\textwidth][c]{\includegraphics[width=90mm]{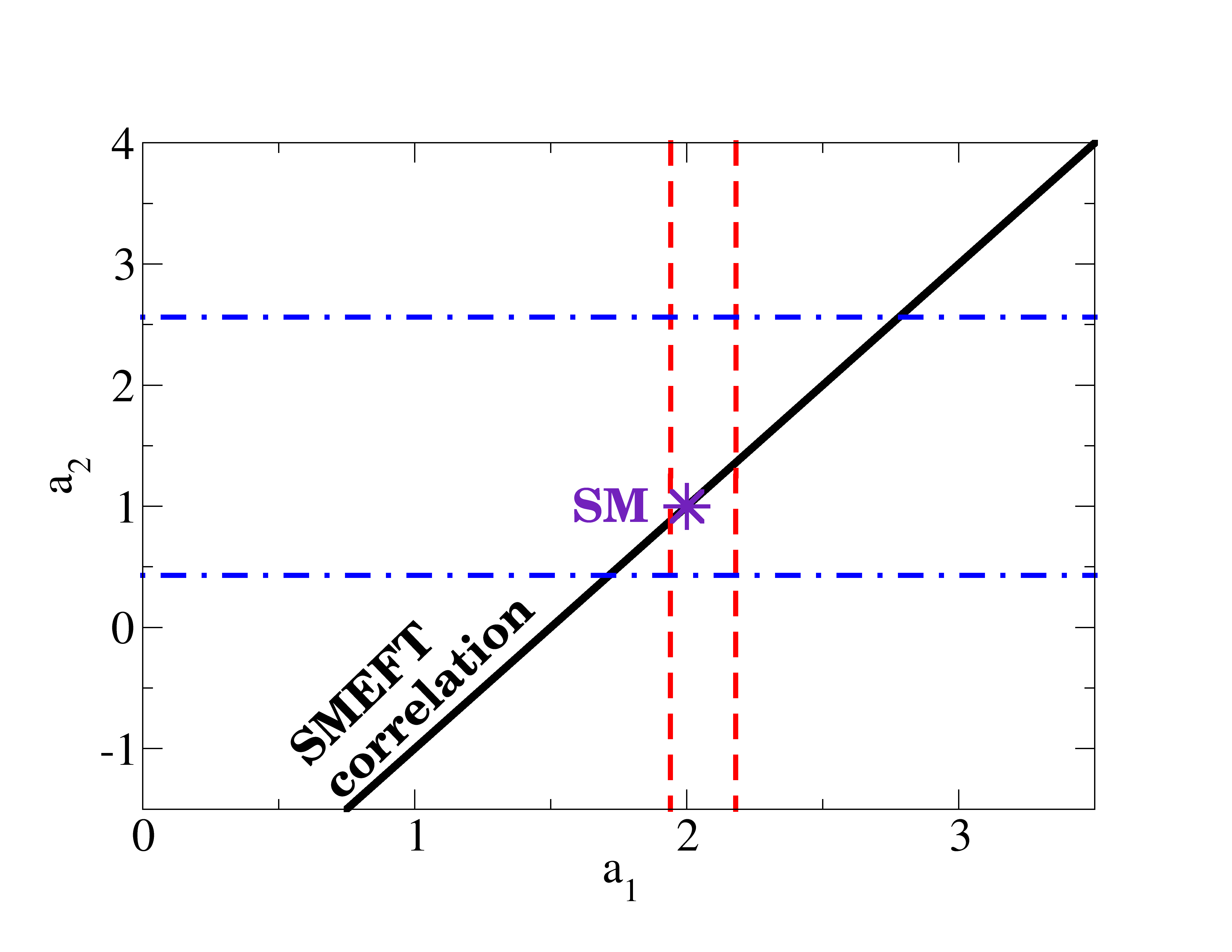}}
\caption{SMEFT at order $1/\Lambda^2$ predicts the correlation $a_2=2a_1-3$ from the first column in Table~\ref{tab:correlations}, which is plotted against the current 95\% confidence intervals for these two HEFT parameters~\cite{ATLAS:2020qdt,CMS:2022cpr}.}\label{SOFP}
\end{figure}

In Figure \ref{SOFP}, a straight line  showsthe SMEFT correlation obtained in the first column of Table \ref{tab:correlations}. The rest of the plane corresponds at most to HEFT theory. The SM is the point in the center of the figure. Finally the 95$\%$ confidence bands for the $a_1$ and $a_2$ parameters are presented as dashed lines with the numbers taken from the caption of Table \ref{tab:correl-exp-bounds}. 

\subsubsection{Multiple Higgs production: testing the SMEFT-induced correlations over the HEFT function $\mF$ coefficients} \label{subsec:numeritos}


We employ the correlations found earlier in Table~\ref{tab:correlations}, in conjunction with current direct experimental bounds on deviations of $a_1$ and $a_2$ from their Standard Model values, to propagate the information to other coefficients of $\mathcal{F}$ that are presently unconstrained {\it  provided SMEFT holds}.

These are then quoted in Table~\ref{tab:correl-exp-bounds}, an interesting new contribution of this article to the phenomenology of HEFT. If, for example, $a_3$ is measured to be different from zero, this would immediately establish new physics (which is known); but additionally, if it exceeds the bounds given in the table, it would mean that SMEFT correlations are being violated and the EFT has to be extended into HEFT.

The constrains in the first column assume the validity of SMEFT up to order $1/\Lambda^2$, $\mO(\Lambda^{-2})$; because of  the tight experimental bounds on the $WW\to h$ coupling $a_1=2a$, the remaining $a_n$ couplings are strongly limited.
If SMEFT is considered up to $1/\Lambda^4$,  $\mO(\Lambda^{-4})$ (as we do in the second column),
the $WW\to hh$ coupling $a_2=b$ becomes independent, as seen in Table~\ref{tab:correlations}; its experimental bounds are then also an input. Being poorly measured so far, it introduces a large uncertainty in the higher correlations. Thus, the bounds on the second column of Table~\ref{tab:correl-exp-bounds} are much looser.  
Those large uncertainties can be much reduced by improving the experimental knowledge of $a_2$: a decrease of its uncertainty by an order of magnitude scales almost linearly  and makes these errors roughly a factor $10$ smaller.

Notice that the values in third column in Table~\ref{tab:correl-exp-bounds} are similar for ATLAS and CMS. The reason is that when the experimental uncertainty of $\Delta a_2$ is very large, at the practical level, its only limitation comes from the constraint $|\Delta a_2|\leq 5 |\Delta a_1|$, this is, $\mbox{min}(5a_{1}^{\rm-},-5a_1^{\rm +}) \leq \Delta a_2\leq \mbox{max}(-5a_{1}^{\rm -},5a_1^{\rm +})$. 
Since effectively the bounds just depend on the allowed values for $a_1$ we are obtaining the same outcomes for ATLAS and CMS in the third column.

\begin{table}[!t]
    \caption{\small
    We input the 95\% confidence-level experimental bounds $a_1/2=a\in [0.97,1.09]$~\cite{ATLAS:2020qdt}  and, for the middle column,  $a_2=b=\kappa_{2V}\in[-0.43,2.56]$~\cite{ATLAS:2020jgy} (see the second erratum), by the ATLAS collaboration (top row) or the CMS collaboration (bottom row) interval of 
    $a_2=b=\kappa_{2V}\in[-0.1,2.2]$~\cite{CMS:2022cpr}.
    With them we have calculated and show here the expected corresponding 95\% CL intervals for several $W_LW_L\sim \omega\omega \to n h$ coupling, $a_n$, employing the relations of Table~\ref{tab:correlations}. Violations of the intervals in the first column would sow doubt on the SMEFT adequacy at $\mathcal{O}(\Lambda^{-2})$; surpassing any in the third column, on its perturbativity; and those of the middle column would void SMEFT of much significance as an EFT. They can be further tightened with improved experimental data for $\kappa_{2V}$.} 
    \label{tab:correl-exp-bounds}  
    \centering   
    \begin{tabular}{|c|c|c|}\hline
       \textbf{ Consistent SMEFT} &  \textbf{Consistent SMEFT} &  \textbf{Perturbativity of}\\
        \textbf{range at order} $\Lambda^{-2}$ & \textbf{range at order}  $\Lambda^{-4}$ & $\Lambda^{-4}$  \textbf{SMEFT} \\ \hline
        $ \Delta a_2\in [-0.12,0.36] $       
        & ATLAS  &  ATLAS 
        \\
        $a_3\in [-0.08,0.24] $  &  $a_3\in [-4.1,4.0]$  &  $a_3\in [-3.1,1.7]$
        \\  
        $a_4\in[-0.02,0.06]    $       & 
        $a_4\in [-4.2,3.9]$     & $a_4\in [-3.3,1.5]$ 
        \\
        $a_5=0$ &  $a_5 \in[-1.9,1.8]$ & $a_5 \in[-1.5,0.6]$ 
        \\
        $a_6=0$ &  $a_6=a_5 $ & $ a_6=a_5$
        \\ 
        \hline
& CMS & CMS  \\ 
& $a_3\in [-3.2,3.0]$
&$a_3\in [-3.1,1.7]$  \\
&$a_4\in [-3.3,3.0]$
&$a_4\in [-3.3,1.5]$   \\
&$a_5\in [-1.5,1.3]$
&$a_5\in [-1.5,0.6]$   \\
&$a_6=a_5$
&$a_6=a_5$   \\ \hline
\end{tabular}
\end{table}

\subsection{When Schwarz's Lemma guarantees a function's zero }
\label{subsec:Schwarz}
In this subsection we examine and adapt a known result from complex-variable analysis that guarantees the existence of a zero of a complex function: in the case of $\mathcal{F}(h)$ this would be an $O(4)$ fixed point candidate around which SMEFT could be built. 

The information that we would eventually need to have at hand to exploit the theorem would be a number of coefficients of the Taylor series,
depending on any future accelerators energy reach (subsec.~\ref{nhproduction}). 
To avoid too large a mathematical digression, Schwarz's Lemma and two of its corollaries are detailed in Appendix A. What can guarantee a zero of $\mathcal{F}$ is the second corollary.
The needed hypotheses are as follows:
\begin{itemize}
    \item{} First, the function $\mathcal F(h)$ (extended to be a complex function of a complex $h$ argument, \underline{in units of $v$} throughout this whole section) needs to be analytic inside a disc of radius $|h|=R$ around the vacuum $h=0$. This disk has to be large enough to reach the possible symmetric point  (\textit{i.e.}, $h=h_\ast$ or, in SMEFT, $|H|=0$) from the observed vacuum ({\it i.e.}, $\langle |H|\rangle=1/\sqrt{2}$, or in HEFT  $\langle h\rangle=0$), where one constructs the flare function $\mF(h)$.    
    
    \item{} Second, the image of that disc (the set of  possible values of $\mathcal{F}$) has to be contained inside another disc of radius $M$ (the maximum value of $|\mathcal{F}|$) centered at $\mathcal{F}(0)=1$. Finally, the derivative of the function is assumed to have been measured, so that  $\mathcal{F}'(0)=a_1$ is known.

\end{itemize}

The second corollary then guarantees  that a disc of radius $R_{\rm min }:= R^2a_1^2/
\gamma M$ centered at $\mathcal{F}=1$ is completely contained in the image of $\mathcal{F}$. Here $\gamma$ is 
\begin{equation}
    \gamma=\frac{(\sqrt{2}+2)(\sqrt2+1)}{\sqrt{2}}\simeq 5.83\;.
\end{equation}

Therefore, a zero of $\mathcal{F}(h)$ is ensured if that radius $R_{\rm min}$ is greater than one (so that $\mathcal{F}=0$ can be reached from $\mathcal{F}=1$),
\begin{equation} \label{conditionfor0}
R^2a_1^2/\gamma M > 1 \implies \exists h^\ast\ | \
\mathcal{F}(h^\ast)=0\ .
\end{equation}

Depending on how large the $a_i$ coefficients end up being, this lemma could provide a tool to extract a scale at which one is sure that there exists an $O(4)$ fixed point candidate. 

To use that second corollary in Appendix A, notice that by construction we have that ${\mathcal F}(0)=1$ and hence we can employ the auxiliary $g(h)\equiv \mathcal{F}(h)-1$ satisfying $g(0)=0$ and $g'(0)=\mathcal{F}'(0)=a_1$, which is the $g$ to which the corollary applies. This means that, if the function $\mathcal{F}(h)$ is analytic in the open disc of radius $R$, denoted as {$D(0,R)$}, then we will have that the condition for the existence of at least one (complex)
value of the Higgs field $h^\ast \in D(0,R)$ such that $\mathcal{F}(h^\ast)=0$ is (see Eq.~(\ref{2nddisk}) below)
\begin{equation} R^2 > \frac{\gamma M}{a_1^2}
\end{equation}
where $M$ is the maximum value that $|\mathcal{F}(h)|$ takes for $h\in D(0,R)$. For clarification see Figure \ref{fig:possiblezero}. Regrettably, the application of the lemma will give a definite positive answer to the existence of a zero if $M(R)$ is at most $R^2$, which means that we can only profit from the lemma for polynomials of order up to two (due to analyticity). This still leaves room for some cases that we explore below, saliently including the variations around the SM that are conceivable in the near future, with $\mathcal{F}$ up to order 4.

\begin{figure}[!t]
    \centering
    \begin{tikzpicture}[scale=1]
 \draw[dotted,fill=orange,opacity=0.5] (2,0) arc (0:360:2);
       \draw (1,2.2) node {$h\in \mathbb{C}$};
       
       \draw[blue] (.8,.9) node {$R$};
    \draw [very thin] (-2.3,0) -- (2.3,0);
    \draw [very thin] (0,2.3) -- (0,-2.3);
     \draw [color=blue,<->] (0,0) -- (1,1.7);

 \draw [color=blue,<->] (0,0) -- (1,1.7);
 
 \draw [thick, ->] (2.1,1) .. controls (2.5,1.2) and (3.6,1.2) .. (4,1);
 \draw (3,1.5) node {$\mathcal{F}(h)$};
\end{tikzpicture}
    \begin{tikzpicture}[scale=1]
 \draw[dotted,fill=cyan,opacity=0.1] (2.7,0) arc (0:360:1.7);
       \draw (-1,2.2) node {$\mathcal{F}(h)\in \mathbb{C}$};
       \draw (1,-.2) node {1};
       \draw[thick,red] (0,0) node {$\times$};
       \draw[thick,red] (0,0.2) node {\tiny possible fixed point};
       \draw[blue] (1.95,.8) node {$\frac{R^2c_1^2}{\gamma M}$};
    \draw [very thin] (-1.8,0) -- (2.3,0);
    \draw [very thin] (0,2.3) -- (0,-2.3);
     \draw [color=blue,<->] (1,0) -- (1.75,1.5);
      \draw[fill=blue,opacity=0.1] plot [smooth cycle] coordinates {(4,-1) (0.2,-2) (-.6,-1) (-1,0) (-.5,2) (2,2)};
      
      \draw [color=black,<->,opacity=0.6] (1.05,-0.05) -- (4,-1);
       \draw[black,opacity=0.6] (2.2,-.8) node {$M$};
      \draw[blue,opacity=0.4] (1,1.95) node  {$F(D(0,R))$};
\end{tikzpicture}
    \caption{{\small \textbf{Left-hand side:} the  disc of radius $R$, $D(0,R)$ (orange online), is the region where the Taylor approximation of the $\mathcal{F}(h)$ function is supposedly trusted, which can only be experimentally assessed. \textbf{Right-hand side:} the grayish outer region is the image of $D(0,R)$, namely $\mathcal{F}(D(0,R))$, and $M$ is the maximum distance of $\mathcal{F}(D(0,R))$ to 1 (thus, the maximum value of $|\mathcal{F}-1|$ for $|h|\leq R$). 
    Under the conditions of applicability for Schwarz's lemma, we can assure that the disc on the right (bluish online), $D(1,\frac{R^2c_1^2}{\gamma M})$, is contained in $\mathcal{F}(D(0,R))$, {\it i.e.} $D(1,\frac{R^2a_1^2}{\gamma M})\subset  \mathcal{F}(D(0,R))$.\\
    When it happens that $\frac{R^2c_1^2}{\gamma M}>1$, the radius of the disc in the image $\mathcal{F}$ plane around $\mathcal{F}(h=0)=1$, we are assured that $\mathcal{F}(h)$ has a zero for some $h^\ast\in D(0,R)$.}}
    \label{fig:possiblezero}
\end{figure}
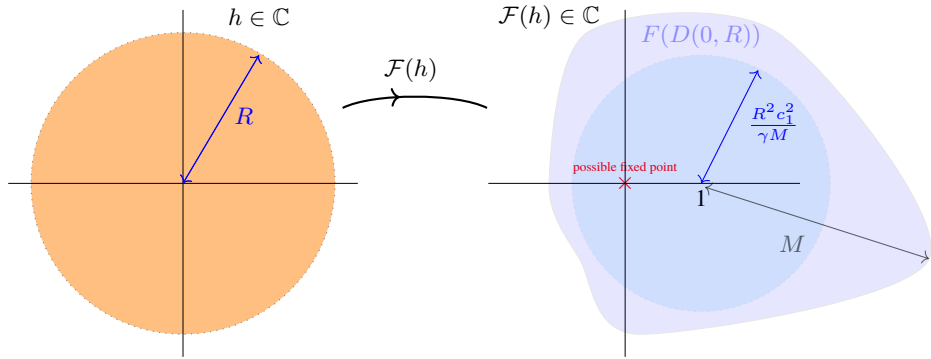$\,$
Experimentally, the full $\mathcal{F}(h)$ can not be measured. It is only its Taylor expansion that can be accessed in practice (unless the SM UV-completion is directly observed, of course). Hence, we must follow the logic:
\begin{enumerate}
\item First we must assume that $\mathcal{F}(h)$ is analytic in a neighborhood of the $h=0$ physical vacuum (and hence its Taylor expansion, and thus HEFT, makes sense). This region can be taken as the open disk $D(0,R)$.
\item Suppose we measure $k$ coefficients 
of the Taylor expansion of the function $\mathcal{F}(h)$ such that, for $h$ in units of $v$,
$$\mathcal{F}(h)=1+\sum_{i=1}^{k}a_i h^i+R_k(h)$$
where of course we trust the expansion up to an energy scale such that we know that, for $h\in D(0,R)$, $\mathcal{F}(h)$ is analytic~\footnote{The difficulty here is that, unlike in analyticity in Mandelstam  $s$ that ultimately follows from causality via Titchmarsh's theorem~\cite{Llanes-Estrada:2019ktp}, it is hard to find a guiding principle in $h$-- space that justifies assuming analyticity. At least we are exposing the necessary hypothesis, that is often taken for granted when writing down a SMEFT.}.
Here $R_k(h)$ is the remainder of the Taylor Expansion. We must neglect this Taylor remainder since, although it is bounded by $\{\max_{|h|=R} |\mathcal{F}(h)|\}\beta^{k+1}/(1-\beta)$ for $\beta\in[|h|/R,1)$, it cannot be experimentally accessed.
\item  Assign $$M=\max_{|h|=R} |\sum_{i=1}^{k}a_i h^i|.$$
The zeroth order coefficient is omitted because we must use the maximum of $g(h)\equiv \mathcal{F}(h)-1$, as described in the appendix.
Notice that the maximum modulus of $\sum_{i=1}^{k}a_i h^i_1$ is reached at the boundary of its domain thanks to the \textit{Maximum modulus principle}.
\item Using the second corollary we will have that we can assure the presence of an $O(4)$ fixed point if we reach a field intensity such that
\begin{equation}\label{criterionSchwarz}
{|h|^2}=R^2>  \frac{\gamma M}{a_1^2}
\end{equation}
\end{enumerate}

\paragraph{Standard Model case}
This discussion has been quite abstract, so let us try to apply Eq.~(\ref{conditionfor0}) in practice. The first obvious example is the Standard Model. 

We can  apply Schwarz's lemma  to either the $F(h)$ function, in the SM $F(h)=1+{h}$, or its square $\mathcal{F}$, the flare function. In the first case we see that $F(h)$ is analytic for all $h\in\mathbb{C}$, and hence we can take $R$ as big as we want. It is immediate to see that $M=R$ so that we can assure the presence of a zero of $F(h)$ whenever 
\begin{equation}
R^2 >  {\gamma}R\;,
\end{equation}
which can be met for $R=|h|>\gamma$ (in units of $v$).

If instead we apply Schwarz's lemma directly to the flare function $\mathcal{F}(h)=F^2(h)$, we find no useful information, as can be understood from the result in the next example. \\

\paragraph{Generic second-order polynomial}
Taking $\mathcal{F}(h)=1+a_1 {h}+a_2 {h^2}$, the condition to assure the presence of a fixed-point candidate becomes
\begin{equation}
    R^2>\frac{\gamma}{a_1^2}(|a_1|R+|a_2|R^2)\geq\frac{\gamma}{a_1^2}{{\max_{|h|=R}}\Big(\Big|a_1 {h}+a_2{h^2}\Big|\Big)}
\end{equation}
So that for $R$ sufficiently large, the condition will be met if $1>\frac{\gamma|a_2|}{a_1^2}$, i.e. $a_1^2/a_2>\gamma$ assures the presence of a fixed-point candidate.  For the known $a_1=2$ central value we obtain that
\begin{equation}
    \frac{4}{\gamma}>|a_2|\;\;\Rightarrow \;\;\text{if }a_2\in (-0.68,0.68) \;\Rightarrow\;\text{  zero of } \mathcal{F} \text{  assured}
\end{equation}
This result is in agreement with the condition of positivity on the discriminant of the polynomial which gives $\frac{a_1^2}{4}>a_2$ (SMEFT region in Fig. \ref{SOFP}) and hence guarantees a zero of $\mathcal{F}(h^\ast)=0$ for $h^\ast\in\mathbb{R}$.

Comparing to the interval for $a_2$ given by experiment and quoted in Table~\ref{tab:correl-exp-bounds}, we see that for negative $a_2$, the experimental bound is already inside the Schwarz's lemma limit; if the upper experimental limit also drops into the 0.68 boundary (which is not unthinkable, only a factor 3 better than the current LHC extraction), then Schwarz's lemma will tell us that a zero of $\mathcal{F}$ is at hand unless new discoveries of higher $a_i$ coefficients require further scrutiny.
Because a third-order polynomial always has a real zero, this takes us to a fourth order one, discussed in the next paragraph.

\paragraph{Perfect-square, fourth-order polynomial} 
Taking now a quadratic 
$F(h)=1+\alpha{h}+\beta {h^2}$ 
entails a quartic flare function 
\begin{equation}
    \mathcal{F}(h)=1+2\alpha  {h}+(\alpha^2+2\beta) {h^2}+2\alpha\beta {h^3}+\beta^2{h^4}\ .
\end{equation} 
In this case, the condition of Eq.~(\ref{criterionSchwarz}) that guarantees the presence of a symmetric point candidate $h_\ast$ becomes
$$\frac{\alpha^2}{\gamma}>\beta\,.$$
Squaring the above relation, for the central value $a_1=2$, we get the bound on the fourth order coefficient
$$\frac{a_1^4}{4 \gamma^2}>a_4\;\;\Rightarrow \;\;\text{if }a_4\in (-0.118,0.118) \;\Rightarrow\;\text{  zero of } \mathcal{F} \text{  assured}.$$
Notice of course that if $a_4$ is measured to be negative, higher order terms will be needed in the expansion of $\mathcal{F}$ (see Subsection \ref{subsec:positivity}) to guarantee its positivity.

\section{Far future: multiple Higgs production in extreme-$T$ collisions
to access the SM symmetric point}

The pion was first discovered in 1947~\cite{Perkins:1947mf,Lattes:1947mw} when precious few events from cosmic rays were obtained in photographic emulsions taken at high altitudes; nowadays, they are routinely produced by the thousands per event in central heavy-ion collisions at the LHC~\cite{ALICE:2021hkc}. Whereas currently multiple Higgs-boson events (or for that matter, multiple longitudinal gauge-boson ones) are not possible, one day they might come within reach. 
At that point, the entire $\mathcal{F}(h)$ function (or at least, to a very large order in the Taylor expansion) may become part of potential observables.
We wish to illustrate the possibility of accessing it with such future work in this subsection.

The idea of a large number of Higgs bosons (which inspires the title of this article) has been put forward before~\cite{Khoze:2017tjt,Khoze:2018mey,Khoze:2018qhz} although in a different context, in a proposal to solve the hierarchy problem. Here, we notice that the appearance of $\mathcal{F}(h)$ in Eq.~(\ref{FbosonLagrangianLO}) makes it that, in a thermal medium with temperatures of order the hundreds of GeV (over two orders of magnitude beyond what is possible today, but not an arbitrarily large scale, and within the validity of the EFT), the process $X\to n\times h + m\times W_L/Z_L$ with a thermal distribution becomes possible. In the next lines we propose a very schematic analysis chain that proceeds according to the following flow diagram:

 \begin{center}
\setlength{\arrayrulewidth}{0.3mm} 
\setlength{\tabcolsep}{0.1cm}  
\renewcommand{\arraystretch}{1.} 
 \boxed{
 \begin{tabular}{cccccccccccc}
         &          &                     &           &                 &         &Measure volume & & 
 Use                & & Compare & \\
 Measure &          & Fit to it           &            &Obtain $\mathcal{F}(h_1)$&         &$V$           & &  
 $V$, $T$,          &  & to \\ 
 $p_T$   &$\implies$& $T$ and             & $\implies$& from $E_{\boldsymbol{k}}$&$\implies$ & using HBT     & $\implies$         & $E_{\boldsymbol{k}}$& $\implies$ & measured \\
         &          & $E_{\boldsymbol{k}}$&           &               &         & interferometry  &  &
 to predict         & & $N$ \\
         &          &                      &            &              &         & & &
 $N$                & & as check 
 \end{tabular}}
 \end{center}
 \vspace{0.3cm}

\paragraph{Transverse-momentum distribution}
As pioneered by Hagedorn~\cite{Hagedorn:1966mqw,Hagedorn:1965st}, an observable revealing the statistical distribution is the {$p_T$-distribution of} the bosons produced. 

In the case of a free-boson gas with Lagrangian $\frac{1}{2}\left( (\partial h)^2 + (\partial \omega_i)^2)\right)$ this is given~\cite{Gupta:2020naz} by 
\begin{equation} \label{free_dist}
\left. \frac{d^2 N}{2\pi p_T dp_T d\eta} \right|_0 =
m_T \frac{gV}{(2\pi)^3} e^{\frac{\mu-E}{T}}\ .
\end{equation}
While $\mu$ is the chemical potential associated to a conserved particle number (which can be left out if events with different number of bosons are considered), $V$ is the volume of the source and $g=1$, 3 or 4 depending on what is measured ($h$, $V_L$ or both), we want to call attention to the transverse mass-like quantity $m_T= \sqrt{m^2+p_T^2}\simeq p_T$. 

Integrating Eq.~(\ref{free_dist}) over the longitudinal momentum (or rapidity) then yields a typical $p_T$ distribution
\begin{equation} \label{freeptspectrum}
    f_{\rm Bose}(p_T)dp_T=  {\rm constant}\ \times \ p_T dp_T
    \int_0^\infty dp_x \frac{1}{e^{\frac{1}{T} \sqrt{p_x^2+m_T^2}}-1}
\end{equation}

In the simplest free gas described by Eq.~(\ref{free_dist}), $f_{\rm Bose}(p_T)$ falls off as a simple exponential. This Boltzmann-like dependence is obtained from mean occupation numbers
\begin{equation}
    \bar{n}_{\alpha}(k)= X_{\alpha}(k) \frac{\partial \log Z}{\partial X_{\alpha}(k)}
\end{equation}
with $X_{\alpha}(k)= exp(-E_\alpha(k)/T)$ and the partition function expressed~\cite{Hagedorn:1966mqw} as
\begin{equation}
    Z = \sum_n \prod_{\alpha k} X_{\alpha}(k)^{\bar{n}_\alpha(k)}\ .
\end{equation}
Here $k=0,1,2,\dots \infty$ as corresponds to a boson occupation number. The momentum distribution can then be obtained from the density of states 
$\frac{V}{2\pi^2}d^3p$.

The $p_T$ dependence of Eq.~(\ref{freeptspectrum}) is modified in the interacting theory: this is what gives access to the function $\mathcal{F}(h)$. 

In case there is an interacting Hamiltonian, containing the  $\mF(h)$ function, 
it is possible, through the statistical distribution of bosons
to access it almost completely or at least to a very high degree in the $a_i$ expansion.  Such statistical distribution~\cite{Fetter} (see Eq. (26.6) in page 251 there, though in a nonrelativistic treatment) will amount to
\begin{equation} \label{ptspectrum}
    \frac{dN}{d^3 \boldsymbol{k}}=\frac{-gV}{(2\pi)^3} T\sum_{n\in\mathbb{Z}}\frac{1}{i\omega_n-\sqrt{E_{\boldsymbol{k}}^2+\Sigma(\boldsymbol k,i\omega_n)}}
\end{equation}
where $E_{\boldsymbol{k}}=\sqrt{\boldsymbol{k}^2+m^2}$, the summation is carried over Matsubara frequencies $\omega_n=2\pi n T$,  and
$\Sigma(\boldsymbol k,k_0)$ is the self energy of the (Higgs or Goldstone) bosons defined through
\begin{equation} \label{selfE}
    [G(\boldsymbol{k},k_0)]^{-1}=\left[ [G^0(\boldsymbol k,k_0)]^{-1} -\Sigma(\boldsymbol k,k_0)\right]^{-1}\ .
\end{equation}

The propagator $G(\boldsymbol{k},i\omega_n)$ (see \cite{Pawlowski:2015mia,Pawlowski:2017gxj} for a detailed discussion on analytic continuation of propagators) can be computed from the euclidean partition function, that has a path-integral representation,
\begin{align}
    Z&=\int \mathcal{D} h\mathcal{D}\boldsymbol{\omega}\exp{\Bigg\{-\int_0^\beta d\tau   \int d^3{x} \left[ \frac{1}{2}\mathcal{F}(h)
\partial_\mu\omega^i\partial_\mu\omega^j\left(\delta_{ij}+\frac{\omega^i\omega^j}{v^2-\boldsymbol{\omega}^2}\right)
+\frac{1}{2}\partial_\mu h\partial_\mu h  \right] \Bigg\}} \label{HEFTZ}
\end{align}
where summation in the Euclidean $\mu$ indices is assumed. We then  directly see how the $\mathcal{F}(h)$ function affects the statistical distribution of bosons through their self energy. The coordinate-space representation of this propagator for Higgs bosons will simply amount to
\begin{align}
    G(\boldsymbol{x},\tau)=&\int \mathcal{D} h\mathcal{D}\boldsymbol{\omega}\;h(\boldsymbol{x},\tau)h(\boldsymbol{0},0)\nonumber\times\\
    &\times\exp{\Bigg\{-\int_0^\beta d\tau   \int d^3{x}\left[ \frac{1}{2}\mathcal{F}(h)
\partial_\mu\omega^i\partial_\mu\omega^j\left(\delta_{ij}+\frac{\omega^i\omega^j}{v^2-\boldsymbol{\omega}^2}\right)
+\frac{1}{2}\partial_\mu h\partial_\mu h  \right] \Bigg\}}\;.
\end{align}
Once this path integration has been estimated on the lattice or by other means, the self energy from Eq.~(\ref{selfE}) can be extracted, and substituting it into Eq.~(\ref{ptspectrum}), a $p_T$ spectrum directly comparable with experiment can be obtained as a functional of $\mF$.

\paragraph{Number of Higgs bosons}

Additionally, we can try to get an idea of what is the number of Higgs bosons that should be produced in an experiment in order to access the SM $O(4)$ symmetric point $h_\ast=-v$. The SM Higgs potential is
\begin{equation}
    V_{SM}(\phi)=\frac{1}{2}\mu^2\boldsymbol\phi\cdot \boldsymbol{\phi}+\frac{1}{4}\lambda(\boldsymbol\phi\cdot \boldsymbol\phi)^2
\end{equation}
where $-\mu^2,\lambda>0$ and $\boldsymbol\phi\cdot \boldsymbol\phi=\phi_1^2+\phi_2^2+\phi_3^2+\phi_4^2$. Choosing the unitary gauge, the SM vacuum sits at $\phi_1=\phi_2=\phi_3=0$ and $\phi_4=v$ and $\phi=\phi_4=\sqrt{-\mu^2/\lambda}$. After redefining $\phi_4=v+h$ the physical Higgs mass amounts to $m_h=\sqrt{2\lambda v^2}$, which using $m_h=125.3\;\text{GeV}$ and $v=246\;\text{GeV}$ gives $\lambda\simeq 0.13$.

The invariant point under $O(4)$ in field space is the origin $\phi=0$. The difference of potential energy density between the SM vacuum and the SM $O(4)$ invariant point is
$$\Delta V=V(0)-V(v)=\frac{1}{4}\lambda v^4\simeq1.19\cdot 10^8\; \text{GeV}^4=1.49\times 10^{10}\;\text{GeV}/\text{fm}^3. $$\par
Now we wish to translate this energy density into a temperature, for doing so we look for $T$ such that
\begin{equation}
    \varepsilon(T)\equiv \frac{1}{(2\pi)^3}\int d^3\bold{k}\frac{E_\bold{k}}{e^{E_\bold{k}/k_B T}-1}=\Delta V\;.
\end{equation}
where $E_\bold{k}=\sqrt{c^2\hbar^2\bold{k}^2+m^2c^4}$ is the relativistic energy of a boson with three-momentum $\bold{k}$.
This gives a temperature of 
$k_B T=140\;\text{GeV}$ (which matches the EWSB second order phase transition critical temperature~\cite{Ramsey-Musolf:2019lsf}). Using this temperature we are ready to compute the number density of Higgs bosons at a temperature where the SM symmetric point is reached
\begin{equation}
    n(T)=\frac{N}{V}= \frac{1}{(2\pi)^3}\int d^3\bold{k}\frac{1}{e^{E_\bold{k}/k_B T}-1}=258\;\text{Higgs\ bosons}/(0.1\text{fm})^3.
\end{equation}
This is certainly a daunting concentration of energy and particle number that is not expected in a foreseeable future. But when/if it is achieved, the absolute number can serve as cross-check of the $p_T$ spectrum line shape to extract a temperature (hopefully the same) if the volume of the hot source, addressed next, is known.

\paragraph{Obtaining the volume of a multi-Higgs source}
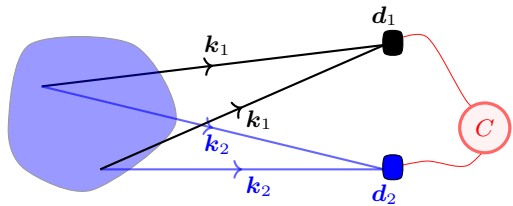
\begin{figure}[ht!]
   \begin{tikzpicture}[scale=1.1]
\draw[] (4,1.35) node  {$\boldsymbol{d}_1$};
\draw[] (2,1) node  {$\boldsymbol{k}_1$};
\draw[color=blue] (2,-.2) node  {$\boldsymbol{k}_2$};
\draw[] (2.5,0.1) node  {$\boldsymbol{k}_1$};
\draw[color=blue] (2.5,0.1-.8) node  {$\boldsymbol{k}_2$};
\draw[color=blue] (4,-.8) node  {$\boldsymbol{d}_2$};
\draw[color=red,opacity=0.9] plot [smooth] coordinates {(4,1) (4.5,1.1) (5,.3) (5.4,0)};
\draw[color=red,opacity=0.9] plot [smooth] coordinates {(4,-0.5) (4.5,1.1-1.5) (5,.3-.75) (5.4,0)};
     \draw [color=black,thick,->-] (-.1,0.5) -- (4,1);
      \draw[scale=0.5,fill=blue,opacity=0.4] plot [smooth cycle] coordinates {(3,.1) (1,-1.5) (-.6,-1) (-1,0) (-.5,2) (2,2)};
      
      \draw [color=blue,opacity=0.6,thick,->-] (0.6,-0.5) -- (4,-0.5);
      \draw [color=blue,opacity=0.6,thick,->-](-.1,0.5) -- (4,-0.5);
      
      \draw [color=black,thick,->-] (0.6,-0.5) -- (4,1);
 \draw[scale=0.5,fill=blue] plot [smooth cycle] coordinates {(8,-1.2) (8,.5-1.2) (8.4,.5-1.2) (8.4,-1.2)};
  \draw[scale=0.5,fill=black] plot [smooth cycle] coordinates {(8,+1.8) (8,.5+1.8) (8.4,.5+1.8) (8.4,+1.8)};
  \filldraw[color=red!60, fill=red!5, very thick] (5.2,0) circle (.3);
  \draw[color=red] (5.2,0) node  {$C$};
\end{tikzpicture}
    \caption{HBT/GGLP Interferometry: Detecting two particles with momenta $\boldsymbol{k}_1$ and $\boldsymbol{k}_2$ at the respective  detection points $\boldsymbol{d}_1$ and $\boldsymbol{d}_2$ and studying their correlation gives information about the dimensions of the homogeneity region of source in bluish.}
    \label{fig:HBT}
\end{figure}
It remains to guess what would be the hot source volume that a
future multi-Higgs factory, ({\it i.e.} a collider capable of producing statistically significant numbers of Higgs bosons) could achieve. 
This type of machine would allow us to explore the existence of such a symmetric point by directly heating the electroweak sector to populate it, and to explore the properties of the EW phase transition. 

With the data in hand, the volume could
be obtained by using Hanbury-Brown-Twiss (HBT) or its particle physics analogue Goldhaber-Goldhaber-Lee-Pais (GGLP) interferometry: the Higgs bosons exiting the collision would retain memory (by interference) of the radius of the source that emitted them. The technique is routinely used in astrophysics to establish the size of astrophysical objects from the emitted photons, and in nuclear collisions by analyzing pions. That future electroweak collider could likewise obtain the radius of a hot electroweak ball from the Goldstone and Higgs bosons emitted.

Let us state schematically how this interferometry works (a comprehensive review can be found in \cite{Weiner:1999th}). 
Suppose that a source has emission points continuously distributed in a  space-time volume $V_4$ with an emission probability amplitude, $\Pi({r};\boldsymbol{k})$, of emitting a particle with momentum $\boldsymbol{k}$ (on the mass-shell with plane-wave wave-function $\psi_{\boldsymbol{k}}(\boldsymbol{r})\propto e^{i\boldsymbol{k}\cdot \boldsymbol{r}}$) at the space-time point $r$. Hence, the total probability of observing the emission of one particle with momentum $\boldsymbol k$ from the source is $P(\boldsymbol{k})=\int_{V_4} d^4{r} | \Pi({r};\boldsymbol{k})|^2$.\\
Likewise, the total probability of measuring two particles with momenta $\boldsymbol{k}_1$ and $\boldsymbol{k}_2$, assuming the two emissions are uncorrelated, \textit{i.e.} $\Pi({r}_1,{r}_2;\boldsymbol{k}_1,\boldsymbol{k}_2)=\Pi({r}_1;\boldsymbol{k}_1)\Pi({r}_2;\boldsymbol{k}_2)$, amounts to
\begin{align}
P(\boldsymbol{k}_1,\boldsymbol{k}_2)&=\int_{V_4} d^4{r}_1d^4{r}_2 \left|
\frac{\psi_{\boldsymbol{k}_1}({ \boldsymbol{r}}_1)\psi_{{\boldsymbol{k}}_2}({\boldsymbol{r}}_2)+\psi_{{\boldsymbol{k}}_1}({\boldsymbol{r}}_2)\psi_{{\boldsymbol{k}}_2}({\boldsymbol{r}}_1)}{\sqrt{2}}
\right|^2 |\Pi({r_1};\boldsymbol{k}_1)|^2 |\Pi({r_2};\boldsymbol{k}_2)|^2\nonumber\\
&=P(\boldsymbol{k}_1)P(\boldsymbol{k}_2)+\int_{V_4} d^4{r}_1d^4{r}_2\cos{\big[(\boldsymbol{r}_1-\boldsymbol{r}_2)\cdot(\boldsymbol{k}_1-\boldsymbol{k}_2)\big]}|\Pi({r_1};\boldsymbol{k}_1)|^2 |\Pi({r_2};\boldsymbol{k}_2)|^2\;.
\end{align}

The GGLP experiment could be adapted to measuring, at two detection points $\boldsymbol{d}_1$ and $\boldsymbol{d}_2$, two Higgs bosons with precise momentum $\boldsymbol{k}_1$ and $\boldsymbol{k}_2$ respectively (see Fig. \ref{fig:HBT}). The correlation function among the two momenta is
\begin{equation}
C(\boldsymbol{k}_1,\boldsymbol{k}_2):=\frac{P(\boldsymbol{k}_1,\boldsymbol{k}_2)}{P(\boldsymbol{k}_1)P(\boldsymbol{k}_2)}=1+\frac{\int_{V_4} d^4{r}_1d^4{r}_2\cos{\big[(\boldsymbol{r}_1-\boldsymbol{r}_2)\cdot(\boldsymbol{k}_1-\boldsymbol{k}_2)\big]}|\Pi({r_1};\boldsymbol{k}_1)|^2 |\Pi({r_2};\boldsymbol{k}_2)|^2}{P(\boldsymbol{k}_1)P(\boldsymbol{k}_2)}.\label{eq:correlationHBT}
\end{equation}
Under the assumptions explained thoroughly in \cite{Lisa:2005dd} (neglection of higher order symmetrization, smoothness and equal time approximations, useful for large
 (RHIC-like) sources) the correlation function in Eq. (\ref{eq:correlationHBT}) simplifies to 
\begin{align}
    C(\boldsymbol{k}_1,\boldsymbol{k}_2)-1&=\int d^3\boldsymbol{r}'\mathcal{S}_{\boldsymbol{K}}(\boldsymbol{r}')\cos{\big[(\boldsymbol{r}_1-\boldsymbol{r}_2)\cdot(\boldsymbol{k}_1-\boldsymbol{k}_2)\big]}\nonumber\\
    \mathcal{S}_{\boldsymbol{K}}(\boldsymbol{r}'):&=\frac{\int_{V_4} d^4{r}_1d^4{r}_2|\Pi({r_1};\boldsymbol{k}_1)|^2 |\Pi({r_2};\boldsymbol{k}_2)|^2\delta(\boldsymbol{r}'-\boldsymbol{r}_1+\boldsymbol{r}_2)}{P(\boldsymbol{k}_1)P(\boldsymbol{k}_2)},
\end{align}
where $\boldsymbol{K}=\boldsymbol{k}_2+\boldsymbol{k}_1$ is the total momentum of the pair of outgoing particles. The function $\mathcal{S}_{\boldsymbol{K}}(\boldsymbol{r}')$ encodes \textit{``the distribution of relative positions of particles with identical velocities and total momentum} $\boldsymbol{K}$'' \cite{Lisa:2005dd} and it gives information about the size of the region of homogeneity of a source (\textit{i.e.} the region where the equilibrium assumptions can be taken). The curvature of $C(\boldsymbol{k}_1,\boldsymbol{k}_2)$ at $\boldsymbol q:=\boldsymbol{k}_1-\boldsymbol{k}_2=0$ is related to the mean-square separation of the three-dimensional quadrupolar moments \cite{Lisa:2005dd}
\begin{equation}
    -\frac{C(\boldsymbol{k}_1,\boldsymbol{k}_2)}{dq_i dq_j}\Bigg|_{\boldsymbol q=0}=\int d^3\boldsymbol{r}S_{\boldsymbol{K}}(\boldsymbol{r})r_i r_j\;.
\end{equation}
In this way we can obtain the volume of the region where the source can be considered homogeneous and the equilibrium conditions apply.

\section{Conclusions}

In this work we have bridged between the SMEFT and HEFT formalisms following the work of two other groups \cite{Alonso:2015fsp,Alonso:2016btr,Alonso:2016oah,Cohen:2020xca,Cohen:2021ucp}. 
We have focused on the Higgs-flare function $\mathcal{F}$ that controls the derivative couplings of two Goldstone bosons $\omega_i$ to any number of Higgs bosons. 
We have exhaustively studied this flare function $\mathcal{F}$ and particularly addressed the existence of its key zero at a symmetric point in the $(\omega_i,h)$ field space. 

We extend previous results concerning the expression of a few coefficients of this function in terms of the $c_{H\Box}$ Wilson coefficient of SMEFT;  
in our work we have addressed a larger number of such coefficients, we have employed the Warsaw basis, and we have proceeded to the next  order ($1/\Lambda^4$) in the SMEFT expansion. 

Further, we have completely eliminated the Wilson coefficient and obtained correlations that are intrinsic to HEFT and can be used to falsify SMEFT itself from the wrapping theory. This is the central result of a companion letter submitted together with this manuscript that provides {\it in extenso} discussion. 
We provide simple recipes that analysts following upcoming and future experimental data  can follow to test the framework of SMEFT.

With the latest ATLAS and CMS bounds on the $a_1$ (also known as $\kappa_{2V}$) and $a_2$ coefficients we have explicitly given 95\% confidence intervals for a few $a_i$, $i>2$ ones, that if exceeded would automatically rule out SMEFT, at least to the orders here considered, and point out to the need of extending the SMEFT framework.

It may be useful for those analysis to have explicit expressions of the $\omega\omega\to n h$ amplitudes in HEFT and we provide their leading order in perturbation theory.
In future work, we will additionally address cross-sections with the physical experimental cuts so that the parameter space can directly be compared with data.
These processes allow to access $\mathcal{F}$ order by order; we also observe that if a future collider could substantially increase the temperature of the collision environment, the entire function could be accessed from the $p_T$-spectrum of created Higgs bosons.

In the Appendix \ref{coeffspotential} and also in the companion letter we have also given similar correlations that we have extracted among the coefficients of the $V(h)$ nonderivative Higgs potential. This is attractive because it does not require the Equivalence Theorem (the process $\omega\omega\to nh$ needs to be extracted from $W_LW_L\to nh$ data and corrections are needed at low-energy, while $V(h)$ does not involve the Goldstone bosons) and is already accessible at LHC energies. Interestingly, it is affected by the properties of $\mathcal{F}$ since it is this function which controls the change of variable between HEFT and SMEFT, $h_1\to h$.

The same reasoning applies to the Yukawa fermion-Higgs couplings which is another interesting alley of investigation for future works where activity is ongoing~\cite{Burgess:2021ylu}.

\section*{Acknowledgments}
Supported by Spanish MICINN PID2019-108655GB-I00 grant, and Universidad Complutense de Madrid under research group 910309 and the IPARCOS institute; 
ERC Starting Grant REINVENT-714788; UCM CT42/18-CT43/18;
the Fondazione Cariplo and Regione Lombardia, grant 2017-2070. A.S.B. thanks Professor Juan Ferrera-Cuesta for pointing out the relevance of using Schwarz's lemma. We also thank Carlos Quezada-Calonge for discussions on $WW\to hh$ scattering.

\appendix
\section{Schwarz's lemma in complex analysis and its corollaries}
This brief appendix provides a short overview of Schwarz's lemma, which we quickly use to demonstrate the corollary of interest for subsection~\ref{subsec:Schwarz}. 
Again, the idea is whether the image set under $\mathcal{F}$ includes or not a disk around $\mathcal{F}=1$ large enough
to encompass the origin $\mathcal{F}=0$. The second corollary below gives a sufficient condition for this to be true. 
The point $h^*$ which is the preimage of $\mathcal{F}=0$ is the symmetric point around which the SMEFT expansion can be constructed. If the conditions of the second corollary are met, we know that $\mathcal{F}$ will be analytic in a region broad enough to guarantee the power-expansion.

To start, take a disk $D_z(0,1)$  around the origin in the preimage complex space (in our application, the extension $h\to z$ of the singlet Higgs field to be a complex variable). 
Second, $D_{f}(0,1)$ is a disk in the image complex space (also extending $\mathcal{F}\to f \in \mathbb{C}$), both disks having radius 1 and being centered around 0 as the notation indicates. We can then state the lemma {\cite{analisis}} as follows.

\subsection{Schwarz's Lemma} 

Let $f\;:\;D_z(0,1)\;\to \; D_f(0,1)$ be holomorphic with $f(0)=0$. Then $|f(z)|\leq |z| $ and  $|f'(0)|\leq 1$. 
Furthermore, if $|f(z_0)|= |z_0| $ for some $z_0\in D_z(0,1)$, then $|f(z)|=1 $   $\forall z\in D(0,1).$\\

\paragraph{Proof} 

Given those $f(0)$ and $f'(0)$, write $f(z)=zg(z)$:  $g$ is also holomorphic. 
Take $r<1$, if $|z|=r$ we have that $|g(z)|=\frac{|f(z)|}{r}$ and hence $|g(z)|= \frac{|f(z)|}{r}\leq \frac{1}{r}$ (since the image of $f$ is $D_f(0,1)$). The inequality $|g(z)|\leq \frac{1}{r}$  is satisfied for all $z\in  \bar{D}(0,r)$ thanks to the \textit{Maximum modulus principle} (if $f$ is a holomorphic function, then the modulus $|f|$ cannot exhibit a strict local maximum that is in the interior of the domain of $f$). This means that if $g(z_0)=1/r$, that is, it reaches its maximum for some $z_0$ satisfying $|z_0|< r$, then the function $g$ must be a constant (and the maximum is reached at the boundary anyway).  Now, taking the limit $r\to 1$ from the left we obtain $|g(z)|\leq 1$ and consequently $|f(z)|\leq |z|$ for all $z\in D_z(0,1)$. 

Noticing that $f'(z)=g(z)+zg'(z)$ it is immediate to prove that $|f'(0)|\leq 1$.\\

\subsection{Corollaries}
\paragraph{First Corollary} 

Let $f\;:\;D_z(0,1)\;\to \; D_f(0,M)$ analytic such that $f(0)=0$ and {$|f'(0)|=1$}. Then we will have that $M\geq 1$ and $D_f(0,\frac{\sqrt{2}}{\left(\sqrt{2}+1\right) \left(\sqrt{2}+2\right)M})\subset f(D_z(0,1))$ ( the open disc of radius $\frac{\sqrt{2}}{\left(\sqrt{2}+1\right) \left(\sqrt{2}+2\right)M}$ is contained in the image through $f$ of the open unit disc).\\
\paragraph{Proof} Thanks to Schwarz's lemma we know that $M\geq 1$ since otherwise $|f'(0)|<1$. We can then write $f$ as
\begin{equation}f(z):=z+\sum_{n=2}^\infty a_n z^n\end{equation}

 The triangular inequality of the complex norm yields
\begin{align}
|z|=|f(z)-\sum_{n=2}^\infty a_n z^n|\leq |f(z)|+|\sum_{n=2}^\infty a_n z^n|\ .
\end{align}

Choosing to evaluate with $|z_M|:=\frac{1}{\alpha M}$ with $\alpha> 1$, we find 
\begin{align}
|f(z_M)|&\geq |z_M|-|\sum_{n=2}^\infty a_n z_M^n|\geq |z_M|-\sum_{n=2}^\infty |a_n| |z_M|^n=\frac{1}{\alpha M}-\sum_{n=2}^\infty \frac{|a_n| }{(\alpha M)^n}\ .
\end{align}

Thanks to Cauchy's estimates (cancelling factorials) we know that for all $r<1$, $|a_n|\leq \frac{M}{r^n}$ and hence $|a_n|\leq M$; we may take the worst bound with $r\to 1$. Then,
\begin{align}
|f(z_M)| &\geq\frac{1}{\alpha M}-M\sum_{n=2}^\infty \frac{1}{(\alpha M)^n} =\frac{1}{\alpha M}-\frac{M}{(\alpha M)^2}\frac{1}{1-\frac{1}{\alpha M}}=\frac{1}{\alpha M}\Big(1-\frac{M}{\alpha M-1}\Big)\;,
\end{align}
(having reconstructed the geometric series).
Now, since $M\geq 1$ and we have taken $\alpha>1$ we have that
\begin{align}
  \frac{1}{\alpha M}\Big(1-\frac{1}{\alpha -1/M}\Big)\geq   \frac{1}{\alpha M}\Big(1-\frac{1}{\alpha-1}\Big)=\frac{1}{M}\Big(\frac{\alpha-2}{\alpha^2-\alpha}\Big),
\end{align}
giving the highest lower bound for $\alpha = 2+\sqrt{2}$.
Hence, taking  $|z_M|=\frac{1}{(2+\sqrt{2})M}$, we have
\begin{align}
|f(z_M)|&\geq \frac{\sqrt{2}}{\left(\sqrt{2}+1\right) \left(\sqrt{2}+2\right)M}\;.
\end{align}

We now proceed to prove that the image of the disk in the Higgs field
$f(D_z(0,1))$ contains the disk of the $\mathcal{F}$ function, namely $D_f(0,\frac{\sqrt{2}}{\left(\sqrt{2}+1\right) \left(\sqrt{2}+2\right)M})$.

Use for this an auxiliary 
$w_f\in D_f(0,\frac{\sqrt{2}}{\left(\sqrt{2}+1\right) \left(\sqrt{2}+2\right)M})$,
that is, $|w_f|\leq |f(z_0)|$;
the function $g(z)=w_f-f(z)$ verifies then
\begin{equation}|f(z_M)+g(z_M)|= |w|< \frac{\sqrt{2}}{\left(\sqrt{2}+1\right) \left(\sqrt{2}+2\right)M}\leq |f(z_M)|\;\;\; \text{for }\;\;|z_M|=\frac{1}{(2+\sqrt{2})M}\;.
\end{equation}
Now we make use of Rouché's theorem to state that $f$ and $g$ have the same number of zeroes in $D(0,\frac{1}{(2+\sqrt{2})M})$, i.e. at least one by definition (because by hypothesis $f(0)=0$). As a consequence there exists $z_0\in D(0,1)$ such that $f(z_0)=w_f$, in other words $f(D_z(0,1))$ contains $D_f(0,\frac{\sqrt{2}}{\left(\sqrt{2}+1\right) \left(\sqrt{2}+2\right)M})$.

\paragraph{Second Corollary} 
Let $g \;:\; D(0,R) \;\to\;D_g(0,M)$ analytic, such that $g(0)=0$ and $|g'(0)|=\mu>0$. Then $g(D(0,R))$ contains
another disk where $g$ is analytic,
\begin{equation}  \label{2nddisk}
g(D(0,R)) \supset
D_g(0,\frac{\sqrt{2}R^2\mu^2}{\left(\sqrt{2}+1\right) \left(\sqrt{2}+2\right)M})
\end{equation}
The $R$ appearing there is what can be tested to guarantee that the image 
includes a disk that in turn includes 0 (and therefore, $\exists h_* | \mathcal{F}(h_*)=0$ ), and the function is analytic between the vacuum and that symmetric point.

\paragraph{Proof} It follows in a relatively straightforward manner by applying the first corollary to the auxiliary function  
\begin{equation}  
f(z):=\frac{1}{Rg'(0)}g(R z)\;\;\text{ for }\;\; |z|<1\;.
\end{equation}


\section{From SMEFT to HEFT}

In this appendix we show the two alternative (complementary) procedures employed to obtain the effective Lagrangian simplifications discussed along the article: first by means of Higgs field redefinitions; in a second subsection, we show how to get this same simplifications of the effective action by means of partial integration and the use of the field equation of motion for the classical fields.      
 
These two approaches for the simplifications of the effective theory are but two sides of the same coin: field redefinitions in the integration of the generating functional modifies the classical action in the exponential $e^{iS}$, introducing new operators that are proportional to the classical EoM (this is, through an appropriate field redefinition one can remove operators in $S$ that are proportional to the EoM, while keeping the same generating functional); on the other hand, at tree-level the effective action $\Gamma$ coincides with the classical action $S$ evaluated in the (quantum) classical field, where, by construction, any operator in $\Gamma$ proportional to the EoM is deemed to vanish. Further details can be found, e.g., in App.~B in Ref.~\cite{Giudice:2007fh}, App.~A in Ref.~\cite{Pich:2013fea}, and Ref.~\cite{Guo:2015isa}.

\subsection{Simplification of the action by field redefinitions}

We will take the SMEFT Lagrangian as our starting point, which written in the modulus-phase form has the structure:
\begin{eqnarray}
\mL_{\rm SMEFT}&=& \frac{v^2}{4}\left(1+\frac{h}{v}\right)^2  \langle D_\mu U^\dagger D^\mu U\rangle 
\nonumber\\&& \qquad
\,+\frac{1}{2} \left(1+ (v+h)^2 B(h)\right)   (\partial_\mu h)^2 
\,\,\, -\,\,\, V_{SM}(h)  - \Delta V_{\text{dim-6}}(h)\, , 
\end{eqnarray}
which, up to $\mO(\Lambda^{-2})$, is given by $B(h)= 
\, -\,  \frac{2 c_{H\Box}}{\Lambda^2}   $.

We want to make the Higgs kinetic term canonical, as it is customary in the HEFT Lagrangian. Thus, one has the relation between the SMEFT Higgs modulus field $h$ and the HEFT Higgs singlet $h_1$ is provided by,
\begin{eqnarray}
\left(1+(v+h)^2B(h)\right)^{1/2} \partial_\mu  h \, =\ \partial_\mu h_1 
\quad \Longrightarrow \quad 
\left(1+(v+h)^2B(h)\right)^{1/2} \frac{dh}{dh_1}\,=\, 1\,  ,  
\end{eqnarray}
that leads to
\begin{eqnarray}
dh_1 = \left(1+(v+h)^2B(h)\right)^{1/2} dh
\quad \Longrightarrow \quad 
h_1 =\int_0^h\left(1+(v+h)^2B(h)\right)^{1/2} \,  dh\, .
\nonumber
\label{eq:general-h-h1-relation}
\end{eqnarray}
With this integration range prescription, one recovers the LO result $h_1=h$.~\footnote{Here we are not performing the necessary Higgs field shift required to take into account that the potential minimum of $V_{\rm HEFT}(h)$ is not the same as the minimum for the tree-level SM potential ($\langle h\rangle =0$). This can be always preformed in a second field redefinition afterwards if needed. } 
If we go up to NLO in $1/\Lambda^2$ we can compute this redefinition in a perturbative way: first, use the pertubative expansion of the integrand in Eq.~(\ref{eq:general-h-h1-relation}), e.g.,  $\left(1+(v+h)^2B(h)\right)^{1/2}= 1  -c_{H\Box}(v+h)^2/\Lambda^2$   
up to NLO in SMEFT; second, trivially integrate each term up to the considered perturvative order. This two steps lead to the $h_1=h_1(h)$ relation up to the desired perturbative order where, e.g., Eq.~(\ref{eq:h-from-h1-NLO}) in the text provides the expression up to $\mO(\Lambda^{-2}$ in the SMEFT expansion.   

Finally, through an iterative perturbative procedure, it is possible to invert this relation and extract the relation $h=h(h_1)$ as an expansion in powers of $1/\Lambda^2$, as it was shown in Eq.~(\ref{eq:h-from-h1-NLO}) for SMEFT at NLO. 

Under this field redefinition the SMEFT Lagrangian then becomes: 
\begin{eqnarray}
\mL_{SMEFT}&=& \frac{v^2}{4}\left(1+\frac{h(h_1)}{v}\right)^2  \langle D_\mu U^\dagger D^\mu U\rangle  
\,+\frac{1}{2}    (\partial_\mu h_1)^2 
\,\,\, -\,\,\, V_{SM}(h(h_1))  - \Delta V_{\text{dim-6}}(h(h_1)) 
\nonumber\\ 
&&= \frac{v^2}{4}\mF(h_1)  \langle D_\mu U^\dagger D^\mu U\rangle  
\,+\frac{1}{2}    (\partial_\mu h_1)^2 \,\, -\,\, V_{\rm HEFT} (h_1)\, .
\end{eqnarray}
with the flare function $\mF(h_1)$ provided in Eq.~(\ref{expandF}).

\subsection{Alternative route: partial integration and equations of motion on the field $h$.}
The previous transformations to cast a SMEFT Lagrangian into a HEFT, particularly Eq.~(\ref{eq:A-simpl}) for the partial integration in terms of $H$ fields and Eq.~(\ref{integral1}) changing the variable from $h$ (SMEFT's) to $h_1$ (HEFT's), lay a path that can be traded for a different one. We can easily start with the quantum effective action $\Gamma$ in terms of $h$, then apply partial integration (Green's theorem) and finally use the classical Higgs equation of motion (EoM) of $h$. The resulting effective action is then valid only up to a given order, and the field appearing therein should be interpreted as $h_1$.  
This will allow us to transform and remove any operator with only Higgs fields and two derivatives on the same field,
for example to express $\mathcal{O}^m_2:=h^m \partial^2 h$
in terms of $\mathcal{O}^n_1= h^n(\partial_\mu h)^2$.
First, we will show that the former ones can always be converted in the second ones up to a total derivative. For this, we will once more use partial integration (Leibniz's rule) to rewrite a Lagrangian term of this form as
\begin{eqnarray}
\mathcal{O}^n_1
& =&   -\, n   \, \mathcal{O}^n_1 \, -\, \mathcal{O}^{n+1}_2     \,\,\, +\,\,\, \partial^\mu(h^{n+1} \partial_\mu h)\, ,
\end{eqnarray}
which, as $n>0$ entails $n\neq -1$, and up to the now omitted total divergence, can be recast in the form 
\begin{eqnarray}
\mathcal{O}^n_1 = h^n(\partial_\mu h)^2 \,\,\,  &=& \,\,\, -\, \frac{1}{n+1}\mathcal{O}^{n+1}_2  
=\,\,\,  -\, \frac{1}{n+1} h^{n+1}\partial^2 h 
.
\end{eqnarray}

Thus, we find that all the possible $\mathcal{O}^n_1$ operators of our Lagrangian ($n\geq 0$) can be always rewritten in the $\mathcal{O}^{m}_2$ form (up to a total derivative) and viceversa. Specifically, up to a total derivative, 
\begin{eqnarray}
(h+v)^m \, \partial_\mu h\, \partial^\mu h 
&=& 
-\, \frac{\left[(v+h)^{m+1} -v^{m+1}\right]}{m+1}\, \partial^2 h\, ,
\end{eqnarray}
This is equivalent to Eq.~(\ref{convierteops}) but written in terms of the singlet and now for operators with arbitrarily large $n$ powers of $h$;
we can use it to rewrite more  general SMEFT operators in the HEFT form.~\footnote{Another useful relation for this type of partial integration simplifications is given by the identity: 
$\Delta \mathcal{L} =A \, \partial_\mu B\, \partial^\mu C = \frac{1}{2}\left((\partial^2A)\,B\, C - A \, (\partial^2 B)\, C - A\, B\, (\partial^2 C)\right)  +\partial_\mu\chi^\mu$, 
with the irrelevant total derivative term given by $\chi^\mu=((\partial^\mu A)\, B\, C -A\, (\partial^\mu B)\, C -A\, B\, (\partial^\mu C))/2$. This relation is essentially the position-space representation of the relation 
$p_B p_C=\frac{1}{2}(p_A^2 -p_B^2 -p_C^2)$, which is the square of the momentum conservation equation with all $p_{A,B,C}$ incoming, $p_B+p_C=-P_A$. By means of it one also can rewrite the $\mathcal{O}^n_1$ Lagrangian terms into $\mathcal{O}^m_2$ operators up to a total derivative (although it is a little less efficient than the simplifications in the main text).   }

Moving on, we transform and remove this second type of operators including $\partial^2 h$ that are generated in SMEFT; we will make use of the EoM of the Higgs field, which causes a difference that is pushed to higher orders in the expansion that are not included anyway, so that at fixed order they are equivalent. The EoM (within the electroweak sector alone) reads~\cite{Guo:2015isa,Buchalla:2013rka,Alonso:2015fsp,Alonso:2016oah}
\begin{eqnarray}
\partial^2h&=& \frac{(v+h)}{2}\,\langle D_\mu U^\dagger D^\mu U\rangle \, -\, V'(h)\, , 
\end{eqnarray}
with Higgs potential (ignorable at high $s$) given by $V(h)=V_{SM}(h)= 
m_h^2\left(-\frac{v^2}{8}+ \frac{h^2}{2} +\frac{h^3}{2v}+\frac{h^4}{8v^2}\right)$  at lowest order in the $1/\Lambda^2$ SMEFT  expansion, i.e., the tree-level SM potential,
and $V'_0(h)= 
m_h^2\left(h  +\frac{3h^2}{2v}+\frac{h^3}{2v^2}\right) $. For further detail, see Eqs.~(3) and~(10) in~\cite{Guo:2015isa}, where one can indeed see that we also get additional fermion operators whose discussion is beyond the scope of this article. 
Thus, we can make use of the EoM of the classical Higgs field to simplify the operators in the quantum effective action $\Gamma$ (which amounts to appropriate field redefinitions in the generating functional). 

The construction derived from a $\mathcal{O}_{H\Box}$-type operator can then be transformed as, up to a total derivative,  
\begin{eqnarray}
&&(v+h)^n (\partial_\mu h)^2 = \,-\, \frac{1}{n+1} \, [ (v+h)^{n+1}-v^{n+1}] \,  \partial^2 h 
\\ 
&&\qquad = \, 
\, - \, \frac{(v+h_1) \, [ (v+h_1)^{n+1}-v^{n+1}]}{2(n+1)}\,    \langle D_\mu U^\dagger D^\mu U\rangle  \, +\,  \frac{1}{n+1}\, [ (v+h_1)^{n+1}-v^{n+1}]\,  \, V' 
(h_1) \, ,  
\nonumber 
\end{eqnarray}
with $h_1$ now understood as the HEFT field. 
In the simplest non-trivial case, $n=2$, the term $\Delta \mathcal{L}=c_{H\Box} \mathcal{O}_{H\Box}$ yields, up to a total derivative,
\begin{eqnarray}
&&\Delta \mL=\frac{c_{H\Box}}{\Lambda^2} \mathcal{O}_{H\Box} =  -\, \frac{c_{H\Box}}{\Lambda^2} (v+h)^2 (\partial_\mu h)^2 
\nonumber\\
&& \qquad =  
\frac{c_{H\Box} (v+h) }{6\Lambda^2}  \, [ (v+h)^3\, - \, v^3]\,  \langle D_\mu U^\dagger D^\mu U\rangle   
\, \, \, -\,\,\,  \frac{c_{H\Box} }{3\Lambda^2}  \, [ (v+h)^3\, - \, v^3]\,   V' 
(h) \, .
\end{eqnarray}

With this, the SMEFT Lagrangian up to dimension-6 can be rewritten in the form 
\begin{eqnarray}
&&\mathcal{L}_{\rm SMEFT}= \frac{v^2}{4}\left(1+\frac{h_1}{v}\right)^2  \langle D_\mu U^\dagger D^\mu U\rangle 
\,+\frac{1}{2}\left( 1  - \frac{2 c_{H\Box} (h_1+v)^2}{\Lambda^2}\right)  (\partial_\mu h_1)^2 
\,\,\, -\,\,\, V(h_1)  
\nonumber\\ 
&&\qquad =
 \frac{v^2}{4} \mF(h_1) \langle D_\mu U^\dagger D^\mu U\rangle 
 +\frac{1}{2}  (\partial_\mu h_1)^2  -   V(h)   - \frac{c_{H\Box} \, [ (v+h_1)^3\, - \, v^3]}{3\Lambda^2} V'(h_1)\, .
\end{eqnarray}
with $\mF$ the flare function shown in the main text, given at this order by  
\begin{eqnarray}
\mathcal{F}(h_1)=  1+\frac{2h_1}{v} \left(1+ \frac{c_{H\Box}v^2 }{\Lambda^2} \right) 
+ \frac{h_1^2}{v^2} \left(1+ \frac{4c_{H\Box}v^2 }{\Lambda^2} \right) 
+  \frac{h_1^3}{v^3} \left(\frac{8c_{H\Box}v^2 }{3\Lambda^2} \right)
+ \frac{h_1^4}{v^4} \left(\frac{2c_{H\Box}v^2 }{3\Lambda^2} \right) \, . 
\nonumber\\
\label{eq:F-in-SMEFT}
\end{eqnarray} 
Satisfactorily, this result is in agreement with earlier investigations~\cite{Agrawal:2019bpm,Sanz-Cillero:2017jhb}, although now we are extending the relation up to $\mathcal{O}(h_1^4)$. Terms of $\mathcal{O}(h_1^5)$ in $\mathcal{F}(h_1)$ and higher start at $\mathcal{O}(\Lambda^{-4})$ or above, and they are thus suppressed in the SMEFT counting. 

\section{Restrictions on the coefficients of the HEFT potential $V(h_1)$ required for valid SMEFT} \label{coeffspotential}

Though a full phenomenological analysis will be left for other work (see~\cite{Taliercio:2022maa} for a precis), we here want to call attention on restrictions over the coefficients of the generic HEFT Higgs potential $V(h)$ that are required for SMEFT to be applicable, and are thus of interest to falsify it too, beyond those for $\mathcal{F}$ that we have concentrated over for most of the manuscript. These relations may be interesting at lower $E<0.5$ TeV which appears to make them more attractive; but overall, the physics is more contrived due to the important corrections that the equivalence theorem takes, so that distinguishing $\omega$ and $W_L$  becomes necessary. A full standard model analysis of the relations is needed, which we will not address now. In contrast, the equivalence theorem allows a much cleaner analysis of $\mathcal{F}$ in the TeV region (at the prize of exacting experimental conditions).

Again, as in subsection~\ref{subsec:correlations}, a proper expansion of the SMEFT Lagrangian that is analytic, and may thus be approximated by a polynomial of the quadratic singlet $H^\dagger H$, requires to avoid any singularities in passing to it from HEFT.
After the transformation to SMEFT coordinates as in Eq.~(\ref{transformaSMEFT}),
we must recover a SMEFT potential usable in Eq.~(\ref{SMEFTEWlagrangianV}),
 \begin{equation}
 V_{\rm SMEFT} = {\rm const.}\times   (H^\dagger H)   + {\rm const'}\times (H^\dagger H)^2   + {\rm const^{''}}\times (H^\dagger H)^3 + \dots
\end{equation}

In a nutshell, when the change of variable from HEFT to SMEFT is attempted, making the Higgs kinetic term canonical involves $\mathcal{F}$ intruding in the relation between the potentials as per Eq.~(\ref{Finverses}), 
\begin{equation}
    V_{\rm SMEFT}(H^\dagger H)  = V_{\rm HEFT}( h_1(z))  =  V_{\rm HEFT}(\mathcal{F}^{-1}(z))\ ,
\end{equation} 
with $z :=\sqrt{ 2 H^\dagger H/v^2}$ as in Eq.~(\ref{eq:Fm1}).

Given a generic $V_{\rm HEFT}$ and any coordinate transformation $F$, odd powers of $z$ appear in $V_{\rm SMEFT}$ upon expanding in this variable around $z\simeq 0$.
But, as such odd-power terms are nonanalytic (odd powers of the squared root $\sqrt{H^\dagger H}$) and cannot appear in SMEFT, which is a Taylor expansion, we need to demand that they vanish. 

Thus, for SMEFT to be valid, the expansion of $V_{\rm HEFT}(h_1)$ can only contain even terms in its $(h_1-h_\ast)^n$ expansion. This mimicks our discussion of $\mathcal{F}$, and thus the nonderivative potential density $V_{\rm HEFT}(h_1)$
must be an even function when expanded around the symmetric point $h_\ast$; ignoring the potential's zero-point value, that expansion starts at order two, namely
\begin{equation} \label{HEFTnonderpotential}
   V_{\rm HEFT}(h_1) = \frac{m_h^2 v^2}{2}  \left[ \frac{v_2^\ast}{v^2}(h_1-h_\ast)^2  + \frac{v_4^\ast}{v^4} (h_1-h_\ast)^4 +    \frac{v_6^\ast}{v^6} (h_1-h_\ast)^6   + \dots \right]
\end{equation}
This is completely analogous to the properties of $\mathcal{F}$ found in subsection~\ref{subsec:noSMEFT}, that upon expanding around the symmetric point where $\mathcal{F}(h_\ast)=0$ needed to have an expansion with only even powers of $h_1-h_\ast$.

\begin{equation}
\mathcal{F}(h_1) =
 \frac{1}{v^2} (h_1-h_\ast)^2    +   \frac{a^*_4}{v^4}  (h_1-h_\ast)^4     +       \frac{a^*_6}{v^6}  (h_1-h_\ast)^6     +\dots  \end{equation}

The global factor  $\frac{1}{2} m_h^2 v^2$ that has been extracted in Eq.~(\ref{HEFTnonderpotential}) helps us make a direct comparison between the correlations constraining the potential constants and the very same correlations that we find for the $a_i$ coefficients of the flare function $\mathcal{F}$.

That is, the trilinear, quadrilinear, pentalinear $\dots$ coefficients of the potential expanded around the physical vacuum,
\begin{equation}
    V_{\rm HEFT}= \frac{m_h^2 v^2}{2}   \left[  \left(\frac{h_1}{v}\right)^2 + v_3   \left(\frac{h_1}{v}\right)^3 + v_4    \left(\frac{h_1}{v}\right)^4 + \dots \right]
\end{equation}
are determined by the $v_{4}^\ast$, $v_{6}^\ast\dots$, coefficients around the symmetric vacuum, in analogy to Eq.~(\ref{matchcoeffs}) where the $a_i$ were determined by the $a_i^\ast$.

One difference among the two discussions is that, while $\mathcal{F}$ in the Standard Model Eq.~(\ref{SMF}) (and hence, the zeroth order SMEFT) is only of order $h^2$, the SM potential $V$ reaches
order $h^4$.
Likewise,  SMEFT at one more order takes $\mathcal{F}$ up to $\mathcal{O}(h^4)$ while the potential $V$ reaches $\mathcal{O}(h^6)$ and so on.
Translated in terms of the expansion coefficients of $\mathcal{F}$, the SM limit implies $a_2^\ast\neq 0$ while   $a_4^\ast\ ,\ a_4^\ast \dots = 0$; but those of $V_{\rm HEFT}(h_1)$ are one higher order,
$v_2^\ast\ ,\ v_4^\ast \neq 0$ with $v_6^\ast\ , \ v_8^\ast \dots=0$.

Truncating both series around the symmetric and the physical vacua to order $h_1^6$ we can solve for the first three coefficients around the physical vacuum, yielding (where $h_\ast/v\to h_\ast$ absorbs the normalization for visibility)  
\begin{equation}
    \begin{pmatrix}v_2^\ast \\ v_4^\ast \\ v_6^\ast  \end{pmatrix} = 
    -  \frac{1}{8h_\ast^4}
    \left(
\begin{array}{ccc}
\frac{15}{2} h_\ast^{3}   &7 h_\ast^4 & 3 h_\ast^5 \\
\frac{5}{2} h_\ast       
 &  5 h_\ast^2 &  3 h_\ast^3 \\
 \frac{1}{2}h_\ast^{-1} & 1 & h_\ast \\
\end{array}
\right)   
\begin{pmatrix}
v_1  \\ v_2 \\ v_3 
\end{pmatrix}\,,   
\end{equation}
with $v_1=0$ and $v_2=1$ at the potential minimum. This can then be employed to solve the relation, at the same order, for $v_4$, $v_5$, $v_6$ that then yields
\begin{equation}
    \begin{pmatrix} v_4\\ v_5 \\ v_6      \end{pmatrix}   
    =   
    \frac{1}{4h_\ast^4}
    \left(
\begin{array}{ccc}
-\frac{5}{2} h_\ast  & -5 h_\ast^2 & -6 h_\ast^3 \\
\frac{3}{2}  
 & 3 h_\ast   
 & 3 h_\ast^2  \\
 -\frac{1}{4} h_\ast^{-1}   & -\frac{1}{2} & -\frac{h_\ast}{2} \\
\end{array}
\right)
     \begin{pmatrix} v_1 \\ v_2  \\ v_3       \end{pmatrix} \,.    
\end{equation}
  These relations allow expressing $v_4$, $v_5$ and $v_6$ in terms of $v_1=0$, $v_2=1$, the trilinear coupling $v_3$ and the symmetric point position $h_\ast$. 

In SMEFT the symmetric point is always set at $\sqrt{2}|H|=(h+v)=0$. In the absence of derivative operators of the $c_{H\Box}^{(j)}$ type, the SMEFT radial coordinate $\sqrt{2}|H|=(h+v)$ coincides with the HEFT field combination $(h_1+v)$. Thus, $h=h_1=0$ at the potential minimum $|H|=v/\sqrt{2}$ and $h=h_1=h_\ast=-v$ at the symmetric point $H=0$. Note that $v$ is given by the minimum of the $V$ potential at the given SMEFT order (see Eq.~(\ref{eq:V_SMEFT}) and below).  
Setting $h_\ast/v=-1$ we have the three correlations in Table~\ref{tab:corV}:    
\begin{table}[!h] 
\setlength{\arrayrulewidth}{0.3mm} 
\setlength{\tabcolsep}{0.3cm}  
\renewcommand{\arraystretch}{2.4} 
    \centering
 \caption{ {\small Correlations among the HEFT Higgs potential $V(h)$ around the physical vacuum (thus, directly tree-level observables) derived from assuming that the flare function $\mathcal{F}(h)$ has a zero at $h^\ast=-v$; the field is normalized with the physical $v$ corresponding to the observation and not the bare Lagrangian parameter, additionally, $c_{H\Box}=0$. If these correlations are violated, SMEFT needs to be extended to a more general HEFT. }}
    \begin{tabular}{|c|c|}  \hline 
          $\displaystyle{v_4 = \frac{1}{4}(-5+6v_3)  }  $     & $\displaystyle{\Delta v_4=\frac{3}{2}\Delta v_3}$ \\[2ex]
         $\displaystyle{v_5 = \frac{3}{4}( v_3-1)}$& 
         $\displaystyle{\Delta v_5=\frac{3}{4}\Delta v_3} $\\ [2ex]
         $\displaystyle{v_6 = \frac{1}{8}(v_3-1)}$ &  $ \displaystyle{\Delta v_6=\frac{1}{8}\Delta v_3}$\\[2ex]
         \hline
    \end{tabular} 
    \label{tab:corV}
\end{table}

If we take the particular value $v_3\to 1$, we recover $v_4=\frac{1}{4}$ and $v_5=0=v_6$ that imply no new physics.

This result can be easily generalized to the case with SMEFT operators of the form $c_{H\Box}^{(6)}$, $c_{H\Box}^{(8)}$, etc. One must simply substitute the expression for $h_\ast$ at $\mO(\Lambda^{-2})$  (at $\mO(\Lambda^{-4})$) in Eq.~(\ref{sympoint2}) (in Eq.~(\ref{eq:SMEFTd8-sym-point})), instead of the value $h_\ast=-v$ employed in Table~\ref{tab:corV}.

\section{Zeroes of the Flare function in SMEFT} \label{appendixD}

\noindent
The flare function to order $1/\Lambda^{2}$ is:
\begin{align}  
    \mathcal{F}(h) = & 1 +   2\frac{h}{v} \underbrace{ \left( 1 + \frac{c_{H\Box} v^2}{\Lambda^2} \right)}_{a}
 +
 \left(\frac{h}{v} \right)^2 \underbrace{ \left( 1 + \frac{4 c_{H\Box} v^2}{\Lambda^2} \right)}_{b}
 +
    \left(\frac{h}{v} \right)^3 \underbrace{\left( \frac{8 c_{H\Box}  v^2}{3\Lambda^2}\right)}_{c}
+
    \left(\frac{h}{v}\right)^4 \underbrace{\left( \frac{2 c_{H\Box} v^2}{3 \Lambda^2} \right)}_{d}  
    \nonumber \\ 
    \nonumber
    & = 
    \left(1 +   \frac{h}{v} \right)^2 +  \left( \frac{h}{v} \right) \left( \frac{ 2 c_{H\Box} v^2}{\Lambda^2} \right)
 +
 \left(\frac{h}{v} \right)^2  \left(  \frac{4 c_{H\Box} v^2}{\Lambda^2} \right)
 +
    \left(\frac{h}{v} \right)^3 \left( \frac{8 c_{H\Box}  v^2}{3\Lambda^2}\right) +
    \left(\frac{h}{v}\right)^4 \left( \frac{2 c_{H\Box} v^2}{3 \Lambda^2} \right) 
    \nonumber  
    \\  & =  
       \left(1 +   \frac{h}{v} \right)^2 +  \left( \frac{c_{H\Box} v^2}{\Lambda^2} \right) 
       \left[  
      2 \left(  \frac{h}{v} \right) +
      4   \left(\frac{h}{v} \right)^2   +
      \frac{8}{3} \left(\frac{h}{v} \right)^3 +
      \frac{2}{3} \left(\frac{h}{v}\right)^4   
    \right] 
    \label{eq:flarefunction}
\end{align}
We can see that, at this order in $\Lambda$, both $h = -v $ and $h = -v + x /\Lambda^2 $ are possible zeroes:
\begin{align}
    \mathcal{F}(-v + x /\Lambda^2) = &
  = 
      \left( \frac{x }{\Lambda^2} \right)^2 + 
      \left(  \frac{ c_{H\Box} v^2}{\Lambda^2} \right)
      \left[ 
       \underbrace{-2 + 4 - 8/3 + 2/3}_{= 0}  + \op \left(\frac{x}{\Lambda^2} \right) \right] 
\end{align}
This is, the coefficients in front of $c_{H \Box}$ conspire to make the fixed point of $\mathcal{F}(h)$ be the one of the SM. \textit{I.e.} this cancellation would not hold if $a, b, c, d$ wouldn't be related the way they are in Eq.~\eqref{eq:flarefunction}, as $(-2, 4, -8/3, 2/3)$. 
Using the SMEFT basis, it is only at quadratic level that a shift is allowed. The flare function at order $\op(\rm{dim-6}, \Lambda^{-4})$ is given in Eq.~\eqref{F_SMEFT8}.
We can test again the solution $ h \to -v  $:
\begin{align}
\mathcal{F}(-v ) = &  
\left( -1  \right)  
\left[   2
  \left(\frac{c_{H\Box}  v^2}{\Lambda^2} + 
   \frac{3 c_{H\Box}^2 v^4}{2\Lambda^4} \right)  
    -
    \left( \frac{4 c_{H\Box} v^2}{\Lambda^2} + 
   \frac{12 c_{H\Box}^2 v^4)}{\Lambda^4} \right)  \right.   
   \nonumber
   + 
       \left( \frac{8 c_{H\Box} v^2}{3 \Lambda^2} + \frac{56 c_{H\Box}^2 v^4}{3 \Lambda^4} \right) 
  \\ & \nonumber
    -  \left( \frac{2 c_{H\Box} v^2}{3\Lambda^2} + 
   \frac{44 c_{H\Box}^2 v^4}{3\Lambda^4} \right) 
   + \left.  \left(
 \frac{88 c_{H\Box}^2 v^4}{15\Lambda^4} \right) 
- 
\left( \frac{44 c_{H\Box}^2 v^4}{45\Lambda^4} \right) \right] =  \\ & 
= \left( - \frac{c_{H\Box}  v^2}{\Lambda^2} \right)  
  \underbrace{ \left[  2 - 4 + \frac{8}{3} - \frac{2}{3}  \right]  }_{ = 0}
-    \left(\frac{c_{H\Box}  v^2}{\Lambda^2} \right)^2   
  \underbrace{ \left[  \frac{3}{2} - 12 + \frac{56}{3} - \frac{44}{3} + \frac{88}{15} - \frac{44}{45}      \right]  }_{ = -1/9}
\end{align}
in this case we do observe a shift of the solution $ \Delta = \left( \frac{c_{H \Box } v }{3 \Lambda^2}\right)^2$. Coincidentally similar to the solution $h_* = -v + \frac{c_{H\Box} v^3}{3 \Lambda^2} $, given in Eq.~\eqref{eq:SMEFTd8-sym-point}.


\bibliography{references}
\end{document}